\documentclass[reqno,12pt]{amsart}
\usepackage{fullpage}
\usepackage{amsfonts}
\usepackage{amssymb}
\usepackage{lscape}

\numberwithin{equation}{section}
\def\i{{\rm i}}
\def\X{{\rm X}}
\def\Y{{\rm Y}}

\def\arcsinh{{\rm arcsinh}}
\def\arccosh{{\rm arccosh}}

\def\pr{{\rm pr}}
\def\const{{\rm const}}
\def\Rnum{\mathbb{R}}
\def\Re{{\rm Re}}

\def\trans{\rm{trans.}}
\def\scal{\rm{scal.}}
\def\inver{\rm{inver.}}
\def\phas{\rm{phas.}}

\def\SO{{\rm SO}}
\def\U{{\rm U}}

\def\ie/{i.e.}

\newtheorem{prop}{Proposition}
\newtheorem{thm}{Theorem}

\newtheorem{lem}{Lemma}

\def\propref#1{Proposition~\ref{#1}}
\def\thmref#1{Theorem~\ref{#1}}

\def\Ref#1{Ref. \cite{#1}}

\def\endallowdisplaybreaks{}

\begin{document}

\title{Group-invariant solutions of semilinear Schr\"odinger equations in multi-dimensions}

\author{
Stephen C. Anco$^1$ \lowercase{\scshape{and}}
Wei Feng$^{1,2}$\\\\
${}^1$
Department of Mathematics\\
Brock University\\
St. Catharines, ON L2S3A1, Canada \\
\lowercase{\scshape{sanco@brocku.ca}}\\
${}^2$
Department of Mathematics\\
Zhejiang University of Technology\\
Hangzhou 310014, China\\
\lowercase{\scshape{fengwei9999@gmail.com}}
}

\begin{abstract}
Symmetry group methods are applied to obtain
all explicit group-invariant radial solutions to a class of semilinear
Schr\"odinger equations in dimensions $n\neq 1$.
Both focusing and defocusing cases of a power nonlinearity are considered,
including the special case of the pseudo-conformal power $p=4/n$ relevant for
critical dynamics.
The methods involve, firstly,
reduction of the Schr\"odinger equations to
group-invariant semilinear complex 2nd order ODEs
with respect to an optimal set of one-dimensional point symmetry groups,
and secondly, use of inherited symmetries, hidden symmetries,
and conditional symmetries to solve each ODE by quadratures.
Through Noether's theorem, all conservation laws arising from these
point symmetry groups are listed.
Some group-invariant solutions are found to exist for values of $n$ 
other than just positive integers, 
and in such cases an alternative two-dimensional form of 
the Schr\"odinger equations involving an extra modulation term 
with a parameter $m=2-n\neq 0$ 
is discussed. 
\end{abstract}

\maketitle

\section{\large Introduction}
\label{intro}

Semilinear Schr\"odinger equations with power nonlinearities
for $u(t,x)$, $x\in\Rnum^n$,
\begin{equation}\label{nls}
\i u_t=\Delta u +k|u|^p u, \quad p\neq 0, \quad k=\const\neq 0
\end{equation}
provide models of many interesting physical phenomena \cite{Sul},
such as
propagation of laser beams in nonlinear media,
slow oscillations of plasma waves,
motion of water waves at the free surface of an ideal fluid,
dynamics of imperfect Bose condensates,
and continuous-limits for mesoscopic molecular structure.
In multi-dimensions $n>1$,
solutions exhibit very rich types of behaviour \cite{Cav,Sul},
particularly radial similarity solutions
and more general group-invariant radial solutions
which are important for investigating collapse or blow-up behaviour,
dispersive behaviour, critical dynamics and asymptotic attractors,
as well as for testing numerical solution methods.

To-date in the literature,
only a few explicit $n$-dimensional radial solutions $u(t,|x|)$
are apparently known (e.g. see \Ref{PolZai}).
Most of the work on explicit solutions to
Schr\"odinger equations with power nonlinearities \eqref{nls}
has concentrated on systematically applying symmetry methods \cite{Olv,BluAnc}
to classify all possible types of group-invariant solutions
\cite{NikPop,FusSer,GagWin,GagWin2},
including radial and cylindrical as well as other less geometric types,
only for the pseudo-conformal power $p=4/3$ in dimension $n=3$
in addition to the lowest even powers $p=2,4$ in dimensions $n=2,3$.
These are the cases of greatest relevance for physical applications.

The present paper, in contrast, will be devoted to
deriving all group-invariant radial solutions $u(t,|x|)$ of
the Schr\"odinger equation \eqref{nls}
in all dimensions $n\neq 1$ and for all powers $p\neq0$,
including the case of the general pseudo-conformal power $p=4/n$
relevant for understanding critical dynamics in arbitrary dimensions. 
Both the focusing case $k>0$ and the defocusing case $k<0$ will be considered.

Radial solutions $u(t,|x|)$ of the Schr\"odinger equation \eqref{nls} satisfy 
the corresponding radial equation
\begin{equation}\label{radial-nls}
\i u_t=u_{rr} +(n-1)r^{-1}u_r +k|u|^p u, \quad n\neq 1
\end{equation}
with $r=|x|$. 
This equation \eqref{radial-nls} describes a general model 
for the slow modulation of $n$-dimensional radial waves in
weakly nonlinear, dispersive, isotropic media.
In particular, 
the amplitude for harmonic waves $\exp(\i(\kappa r-\omega t))$ 
with a dispersion relation $\omega=\omega(\kappa)$ in such a medium 
is given by \cite{Sul} $u(\epsilon t, \epsilon(r-\omega_\kappa t))$ 
in terms of a small parameter $|\epsilon|\ll 1$. 

Group-invariant solutions $u(t,r)$ of equation \eqref{radial-nls} 
arise from reductions by 
one-dimensional groups of point transformations on $r,t,u,\bar u$
that leave the equation invariant.
Each such symmetry reduction yields a complex 2nd order semilinear ODE
\begin{equation}\label{U-ode}
U''=f(\xi,U,\bar U,U',\bar U')
\end{equation}
formulated in terms of the invariants
\begin{equation}\label{invs}
\xi=\Xi(t,r,u,\bar u), \quad
U=\Upsilon(t,r,u,\bar u), \quad
\bar U=\bar \Upsilon(t,r,u,\bar u)
\end{equation}
determined by a given symmetry transformation group,
provided that this system \eqref{invs} can be inverted (at least implicitly)
to obtain both of the dependent variables $u,\bar u$,
and one of the independent variables $r,t$
in terms of $\xi,U,\bar U$, and the other independent variable
(which will be the case \cite{Olv} 
whenever the orbits of the transformation group 
acting on the variables $(r,t,u,\bar u)$ are one-dimensional 
and have a projectable regular action on $(r,t)$). 
Then each solution of the ODE \eqref{U-ode} for $U(\xi)$
will yield a group-invariant solution of the radial Schr\"odinger equation
\eqref{radial-nls} for $u(t,r)$.

If two groups of point symmetry transformations are related by
conjugation with respect to some point transformation
in the full symmetry group of the radial Schr\"odinger equation \eqref{radial-nls},
then the action of this point transformation on solutions $u(t,r)$
will map the group-invariant solutions determined by the two symmetry groups
into each other.
Consequently, for the purpose of finding all group-invariant radial solutions,
it is sufficient to work with a maximal set of
one-dimensional point symmetry groups that are conjugacy inequivalent.
For each such group,
once the resulting group-invariant solutions have been found,
the full symmetry group of the radial Schr\"odinger equation can be applied
on these solutions to obtain the group-invariant solutions determined by
all other one-dimensional point symmetry groups in the same conjugacy class.

Interestingly, group-invariant solutions $u(t,r)$ exist for values of $n$
other than just positive integers. 
In such cases the equation \eqref{radial-nls}
can be interpreted alternatively as modeling 
the slow modulation of $2$-dimensional radial waves in
a planar, weakly nonlinear, dispersive, isotropic medium
with a point-source disturbance at the origin. 
Specifically, this equation can be written in the equivalent form 
\begin{equation}\label{2D-radial-nls}
\i u_t=u_{rr} +(1-m)r^{-1}u_r +k|u|^p u
\end{equation}
with a parameter $m=2-n$ which is applicable for any value of $n$.
The term $m r^{-1}u_r$ has a natural interpretation through 
the net modulation defined by the $2$-dimensional integral 
\begin{equation}\label{modul}
\mathcal{C}(t) = \int_0^\infty u \;r dr
\end{equation}
on the radial domain. 
If $u$ is regular at the origin (\ie/ $\lim_{r\rightarrow 0} u_r =0$),
then this integral formally satisfies
\begin{equation}
\frac{d\mathcal{C}}{dt} = -(S+m\lim_{r\rightarrow 0} \i u),
\end{equation}
where
\begin{equation}
S(t) = \i k\int_0^\infty u |u|^p \;r dr
\end{equation}
measures the net amount of modulation produced by the nonlinear term
in the equation \eqref{2D-radial-nls}.
Note that $\mathcal{C}$ is conserved, $\dfrac{d\mathcal{C}}{dt}=0$,
whenever both the modulation term and the nonlinear term are absent,
$m=k=0$ (describing a linear, non-dispersive, planar medium).
Thus, when $m\neq 0$,
the modulation term $mr^{-1}u_r$ models the effect of a point-source disturbance
at the origin $r=0$,
which alters the slow modulation of harmonic waves
in a planar, weakly nonlinear, dispersive medium.

For a given point symmetry reduction of 
the radial Schr\"odinger equation \eqref{radial-nls}, 
or the equivalent $2$-dimensional equation \eqref{2D-radial-nls},
the task of solving the complex 2nd order semilinear ODE \eqref{U-ode} 
for $U(\xi)$ is typically very non-trivial. 
One systematic approach is the method of reduction of order \cite{Olv,BluAnc},
which relies essentially
on the algebraic structure of the group of point symmetries admitted by this ODE
and on the whether the ODE admits a Lagrangian structure or not,
depending on the values of the parameters $p$ and $n$ (or $m=2-n$). 
Reduction of order is simplest to carry out using 
an equivalent polar system of 2nd order real ODEs 
for the amplitude and phase of the variable
\begin{equation}\label{U-polar}
U=A\exp(\i\Phi).
\end{equation}
Firstly,
equation \eqref{radial-nls} together with its complex-conjugate equation 
are the respective Euler--Lagrange equations
$\delta L/\delta\bar u=0$ and $\delta L/\delta u=0$
of the radial Schr\"odinger Lagrangian
\begin{equation}\label{radial-Lagr}
L=( \tfrac{1}{2}\i\bar u u_t - \tfrac{1}{2}\i\bar u_t u
+ \bar u_r u_r - \tfrac{2}{p+2}k\bar u^{1+p/2} u^{1+p/2} )r^{n-1}.
\end{equation}
As a consequence if this Lagrangian is invariant
under the given point symmetry transformations used for the reduction,
then the polar system for $A(\xi)$ and $\Phi(\xi)$ will inherit a Lagrangian
obtained by reduction of the radial Euler--Lagrange structure.
Secondly,
the phase rotation transformations
\begin{equation}\label{phasegroup}
u\rightarrow e^{\i\phi}u, \quad
\bar u\rightarrow e^{-\i\phi}\bar u
\end{equation}
comprise a $\U(1)$ group of point symmetries of equation \eqref{radial-nls}.
This group is readily shown to commute with the full group of point symmetries
admitted by equation \eqref{radial-nls}.
Therefore, if the given point symmetry reduction starts from
any other one-dimensional group of point symmetries,
then the polar system for $A(\xi)$ and $\Phi(\xi)$ will inherit
a $\U(1)$ group of point symmetry transformations on $\xi,A,\Phi$
obtained by reduction of the phase rotation transformations on $u,\bar u$.
Moreover, this system will also inherit all point symmetries that belong to
the normalizer subgroup of the given point symmetry used for the reduction.
An important remark is that the full group of point symmetries
admitted by the system can possibly contain ``hidden'' point symmetries 
in addition to those point symmetries inherited via reduction.

In the case when a given point symmetry reduction of
the radial Schr\"odinger equation \eqref{radial-nls}
inherits a Lagrangian structure,
the polar system for $A(\xi)$ and $\Phi(\xi)$
can be solved explicitly (up to quadratures) if its Lagrangian is invariant
under a two-dimensional group of point symmetry transformations.
In the alternative case when no Lagrangian structure exists,
a four-dimensional point symmetry group with a solvable Lie algebra structure
is needed to obtain the explicit solutions (up to quadratures)
for $A(\xi)$ and $\Phi(\xi)$. 
In all cases, each one-dimensional group of point symmetry transformations
admitted by the polar system for $A(\xi)$ and $\Phi(\xi)$,
or equivalently by the complex semilinear ODE for $U(\xi)$,
can be used to find a single solution that is invariant
under the admitted symmetry transformations on $\xi,A,\Phi$. 

A different reduction method is applicable in the Lagrangian cases.
If a given polar system for $A(\xi)$ and $\Phi(\xi)$ inherits
a $\U(1)$-invariant Lagrangian obtained by reduction of both
the radial Schr\"odinger Lagrangian \eqref{radial-Lagr}
and the $\U(1)$ group of phase rotation symmetries \eqref{phasegroup},
then the 2nd order ODE for $\Phi(\xi)$ can be explicitly integrated
so that the system reduces to a 2nd order ODE for $A(\xi)$ alone,
with $\Phi(\xi)$ given by an integral in terms of $A(\xi)$
containing an arbitrary constant.
This reduced ODE for $A(\xi)$ has two useful features.
First, all nonlinear terms in the reduced ODE vanish if
the nonlinearity power is $p=-1$
and the arbitrary constant in integral for $\Phi(\xi)$ is zero.
In this case, the resulting linear ODE for $A(\xi)$ can be solved explicitly
in terms of special functions.
Second, a Lagrangian structure can be derived for the reduced ODE
starting from the $\U(1)$-invariant Lagrangian of the original polar system.
In the nonlinear case this ODE can be solved (up to quadratures)
if its Lagrangian is invariant under
a one-dimensional group of point symmetry transformations on $\xi,A$.
In all cases the reduced ODE also can be solved (up to quadratures)
if it admits a two-dimensional group of
non-variational point symmetry transformations.
More generally, any one-dimensional group of point symmetries
admitted by the reduced ODE can be used to find a single invariant solution
for $A(\xi)$.
Each solution found for $A(\xi)$ determines a solution for $\Phi(\xi)$
and hence yields a solution to the polar system. 
These solutions will turn out to differ from those solutions obtained
via the previous reduction method if the group of point symmetries
used to obtain $A(\xi)$ is a ``hidden'' group which is not inherited
under reduction from any point symmetries admitted by
the polar system for $A(\xi)$ and $\Phi(\xi)$. 
In particular, a ``hidden'' symmetry group can arise from
point transformations on $\xi,A$ that leave invariant
the integral for $\Phi(\xi)$.

The rest of this paper is organized as follows.

Section~\ref{symms-conslaws} contains some preliminaries
on symmetries and Noether's theorem.
We first state the full point symmetry structure for both
the $n$-dimensional Schr\"odinger equation \eqref{nls}
and the radial Schr\"odinger equation \eqref{radial-nls}.
In addition we summarize the conservation laws that arise from
the Lagrangian structure of these two equations via Noether's theorem.
Next for the radial Schr\"odinger equation \eqref{radial-nls} we present
a maximal set of one-dimensional point symmetry groups that are conjugacy inequivalent.
Modulo phase rotations,
the symmetry groups in this set consist of time translations, scalings,
and a particular combination of inversions (pseudo-conformal transformations)
and time translations.
These three symmetry groups will be referred to as
the optimal subgroups for symmetry considerations.
We then write down the complex 2nd order semilinear ODEs \eqref{U-ode}
given by reduction under each point symmetry subgroup in the optimal set,
and we summarize the Lagrangian structure admitted by each of the ODEs,
depending on $p$ and $n$.

Section~\ref{methods}
explains the reduction of order method in detail
for solving complex 2nd order semilinear ODEs \eqref{U-ode}
by use of point symmetries.
In particular,
we show how to streamline the standard reduction steps in an efficient way
by combining the Lagrangian and non-Lagrangian cases
through the use of canonical coordinates
determined by any admitted one-dimensional point symmetry group.
We also explain details of the alternative reduction method
in the $\U(1)$-invariant Lagrangian cases, which we carry out by again using
canonical coordinates to streamline the steps.

Sections~\ref{translation-group}, ~\ref{scaling-group}, ~\ref{conformal-group}
apply these reduction methods to derive
solutions $U(\xi)$ for each ODE \eqref{U-ode} arising from
the three optimal subgroups of point symmetries
for the radial Schr\"odinger equation \eqref{radial-nls}.
In particular, we are able to obtain explicit solutions in terms of
elementary functions.

The main results of the paper are presented in section~\ref{solns}.
We first list all of the group-invariant radial solutions $u(t,r)$
determined by applying the full group of point symmetries of
the radial Schr\"odinger equation \eqref{radial-nls}
to each of the group-invariant solutions $U(\xi)$ derived from
the three optimal subgroups of point symmetries.
We next discuss a few analytical features of these solutions $u(t,r)$,
including cases that involve non-integer values of $n$.

Finally, section~\ref{concl} has some concluding remarks 
as well as comments on group-invariant radial solutions pertaining to
blow-up behaviour of $u(t,r)$ for powers $p\geq 4/n$.

\section{\large Symmetries and Conservation Laws}
\label{symms-conslaws}

For the Schr\"odinger equation \eqref{nls} in $\Rnum^n$,
a {\em point symmetry} is a one-dimensional Lie group of transformations
acting on the variables $(t,x,u,\bar u)$
such that the prolongation $\pr\X$ of its infinitesimal generator
\begin{equation}\label{symmX}
\X=
\tau(t,x,u,\bar u)\partial/\partial t
+ \zeta(t,x,u,\bar u)\cdot\partial/\partial x
+ \eta(t,x,u,\bar u)\partial/\partial u
+ \bar\eta(t,x,u,\bar u)\partial/\partial\bar u
\end{equation}
satisfies $\pr\X(\i u_t-\Delta u -k|u|^p u)=0$
for all formal solutions $u(t,x)$ of equation \eqref{nls}.
This is the condition for equation \eqref{nls} to be infinitesimally invariant
under the transformation group generated by $\X$.
Each such generator acting on solutions $u(t,x)$
has an equivalent {\em characteristic form}
\begin{equation}\label{charX}
\hat\X=
P(t,x,u,\bar u,u_t,\bar u_t,\nabla u,\nabla\bar u) \partial/\partial u
+ \bar P(t,x,u,\bar u,u_t,\bar u_t,\nabla u,\nabla\bar u) \partial/\partial\bar u
\end{equation}
with
\begin{equation}\label{pointP}
P=\eta-\tau u_t-\zeta\cdot\nabla u
\end{equation}
satisfying
\begin{equation}\label{Pdeteq}
\i D_t P - D_x\cdot D_x P -k(1+p/2)\bar u^{p/2} u^{p/2} P
-k(p/2)\bar u^{p/2-1} u^{1+p/2}\bar P
=0
\end{equation}
as given by applying $\pr\hat\X$ to equation \eqref{nls}
and then eliminating $u_t$, $\bar u_t$, and $x$-derivatives of $u_t$, $\bar u_t$
through the equation \eqref{nls}, its complex conjugate equation, 
and differential consequences of these equations.
Here $D_t$ and $D_x$ denote total derivatives with respect to $t$ and $x$.
Thus, a solution $u=f(t,x)$ of the Schr\"odinger equation \eqref{nls}
is {\em group-invariant} under a one-dimensional point symmetry group
with a generator \eqref{symmX}
if (and only if) it satisfies the additional equation
\begin{equation}\label{invsoln}
\eta(t,x,u,\bar u)-\tau(t,x,u,\bar u) u_t-\zeta(t,x,u,\bar u) \cdot\nabla u
=0.
\end{equation}

It is straightforward to solve equation \eqref{Pdeteq} to determine
all point symmetry generators \eqref{symmX}.
In particular,
because $\tau$, $\zeta$, $\eta$ do not depend on any derivatives of $u$ and $\bar u$,
the equation \eqref{Pdeteq} splits with respect to
$x$-derivatives of $u$ and $\bar u$,
yielding an overdetermined linear system of PDEs on
the functions $\tau(t,x,u,\bar u)$, $\zeta(t,x,u,\bar u)$, $\eta(t,x,u,\bar u)$.
This system leads to the following well-known result
\cite{NikPop,PopKunEsh}.
(Here $\odot$ and $\wedge$ will respectively denote 
the symmetric and antisymmetric parts of an outer product.)

\begin{thm}\label{thm:pointsymms}
The point symmetries of the Schr\"odinger equation \eqref{nls} are generated by
\begin{subequations}\label{nls-pointsymms}
\begin{align}
\text{phase~rotation} \quad
& \X_{\rm 1}=
\i u\partial/\partial u-\i\bar{u}\partial/\partial{\bar{u}},
\label{nls-Phas}
\\
\text{time~translation} \quad
& \X_{\rm 2}=
\partial/\partial t,
\label{nls-tTrans}
\\
\text{space~translations} \quad
& \X_{{\rm 3}(l)}=
e_{(l)} \cdot \partial/\partial x,
\quad l=1,\ldots,n,
\label{nls-xTrans}
\\
\text{Galilean~boosts} \quad
& \X_{{\rm 4}(l)}=
2te_{(l)}\cdot \partial/\partial x
-\i (e_{(l)}\cdot x)(u\partial/\partial u-\bar{u}\partial/\partial {\bar{u}}),
\label{nls-Gal}\\&\qquad
l=1,\ldots,n,
\nonumber
\\
\text{space~rotations} \quad
& \X_{{\rm 5}(l,l')}=
(e_{(l)}\cdot x)e_{(l')}\cdot\partial/\partial x
-(e_{(l')}\cdot x)e_{(l)}\cdot\partial/\partial x,
\label{nls-Rot}\\&\qquad
l=1,\ldots,n-1; l'=l+1,\ldots n,
\nonumber
\\
\text{scaling} \quad
& \X_{\rm 6}=
2t \partial/\partial t+x \cdot \partial/\partial x
-(2/p)u\partial/\partial u-(2/p)\bar{u}\partial/\partial {\bar{u}},
\label{nls-Scal}
\\
\text{inversion} \quad
& \X_{\rm 7}=
t^2\partial/\partial t+tx\cdot \partial/\partial x
-(2t/p +\i |x|^2/4)u\partial/\partial u
\nonumber\\&\qquad
-(2t/p -\i |x|^2/4)\bar{u} \partial/\partial {\bar{u}}
\quad \text{only for}\quad p=4/n,
\label{nls-Inver}
\end{align}
\end{subequations}
where $\{e_{(1)},\ldots,e_{(n)}\}$ is any orthonormal basis for $\Rnum^n$.
The corresponding transformation groups acting on solutions
$u=f(t,x)$ of the Schr\"odinger equation \eqref{nls} are given by
\begin{subequations}\label{nls-family-sol}
\begin{align}
& u=\exp{(\i\phi)}f(t,x),
\label{nls-Phas-group}
\\
& u=f(t-\epsilon,x),
\label{nls-tTrans-group}
\\
& u=f(t,x-\epsilon e_{(l)}),
\quad l=1,\ldots,n,
\label{nls-xTrans-group}
\\
& u=\exp\left(-\i (2\epsilon e_{(l)}\cdot x-\epsilon^2 t)/4\right) f(t,x-\epsilon te_{(l)}),
\quad l=1,\ldots,n,
\label{nls-Gal-group}
\\
& u=f(t,x+(\cos(\phi)-1)(e_{(l)}\odot e_{(l)}+e_{(l')}\odot e_{(l')})\cdot x
+2\sin(\phi)(e_{(l)}\wedge e_{(l')})\cdot x),
\label{nls-Rot-group}
\\&\qquad
l=1,\ldots,n-1; l'=l+1,\ldots n,
\nonumber
\\
& u=\lambda^{-2/p}f(\lambda^{-2}t,\lambda^{-1}x),
\label{nls-Scal-group}
\\
& u=(1+\epsilon t)^{-2/p}\exp\left(-\i\epsilon |x|^2/(4+4\epsilon t)\right) f(t/(1+\epsilon t),x/(1+\epsilon t))
\label{nls-Inver-group}
\\&\qquad
\text{only for}\quad p=4/n, 
\nonumber
\end{align}
\end{subequations}
with group parameters
$-\infty<\epsilon<\infty$, $0<\lambda<\infty$, $0 \leq \phi < 2\pi$.
\end{thm}

The special power $p=4/n$ for which the inversion group exists
is commonly called the {\em pseudo-conformal power}.
In the case $p\neq 4/n$,
the transformations \eqref{nls-Phas-group}--\eqref{nls-Scal-group}
comprise a semi-direct product of a scaling group acting on
a central extension of the Galilean group,
which has the Lie algebra structure
\begin{align}
&
[\X_{\rm 1},\X_{\rm 2}]=[\X_{\rm 1},\X_{{\rm 3}(l)}]= [\X_{\rm 1},\X_{{\rm 4}(l)}]=[\X_{\rm 1},\X_{{\rm 5}(l,l')}]= [\X_{\rm 1},\X_{\rm 6}]=0,
\\
&
[\X_{\rm 2},\X_{{\rm 3}(l)}]= 0,
\;
[\X_{\rm 2},\X_{{\rm 4}(l)}]= 2\X_{{\rm 3}(l)},
\;
[\X_{\rm 2},\X_{{\rm 5}(l,l')}]=0,
\\
&
[\X_{{\rm 3}(k)},\X_{{\rm 4}(l)}]= -\delta_{kl}\X_{\rm 1},
\;
[\X_{{\rm 3}(k)},\X_{{\rm 5}(l,l')}]= \delta_{kl}\X_{{\rm 3}(l')}- \delta_{kl'}\X_{{\rm 3}(l)},
\\
&
[\X_{{\rm 4}(k)},\X_{{\rm 5}(l,l')}]= \delta_{kl}\X_{{\rm 4}(l')}- \delta_{kl'}\X_{{\rm 4}(l)},
\\
&
[\X_{\rm 6},\X_{\rm 2}]=-2\X_{\rm 2},
\;
[\X_{\rm 6},\X_{{\rm 3}(k)}]= -\X_{{\rm 3}(k)},
\;
[\X_{\rm 6},\X_{{\rm 4}(k)}]= 2\X_{{\rm 4}(k)},
\;
[\X_{\rm 6},\X_{{\rm 5}(l,l')}]=0.
\end{align}
In the case $p=4/n$,
the inversion transformation \eqref{nls-Inver-group} intertwines non-trivially
with the previous group, as given by the commutator structure
\begin{align}
&
[\X_{\rm 7},\X_{\rm 1}]=[\X_{\rm 7},\X_{{\rm 4}(l)}]= [\X_{\rm 7},\X_{{\rm 5}(l,l')}]= 0,
\\
&
[\X_{\rm 7},\X_{\rm 2}]=-\X_{\rm 6},
\;
[\X_{\rm 7},\X_{{\rm 3}(l)}]= -\tfrac{1}{2}\X_{{\rm 4}(l)},
\;
[\X_{\rm 7},\X_{\rm 6}]=-2\X_{\rm 7}.
\end{align}

For any domain $\Omega\subseteq\Rnum^n$,
the Schr\"odinger equation \eqref{nls} has the variational formulation
\begin{equation}\label{extremaL}
\frac{\delta\mathcal L}{\delta \bar u} =0
\end{equation}
given by the Lagrangian functional
\begin{equation}\label{nls-L}
\mathcal L= \int_{t_0}^{t_1}\int_{\Omega} L(u,\bar u,u_t,\bar u_t,\nabla u,\nabla\bar u) \;d^n xdt ,\quad
L=\tfrac{1}{2}\i \bar{u}u_t-\tfrac{1}{2}\i u\bar{u}_t
+|\nabla u|^2-\tfrac{2}{p+2}k|u|^{2+p}.
\end{equation}
A {\em variational point symmetry} of this functional \eqref{nls-L}
is an infinitesimal point transformation \eqref{symmX} on $(t,x,u,\bar u)$
under which $\mathcal L$ is invariant
up to spatial boundary terms at $\partial\Omega$
and temporal boundary terms at $t=t_0$ and $t=t_1$.
This invariance condition holds if and only if
the Lagrangian satisfies
\begin{equation}\label{invL}
\pr\hat\X L=D_t A+D_x\cdot B
\end{equation}
for some functions $A$ and $B$ of $t$, $x$, $u$, $\bar u$,
and derivatives of $u$, $\bar u$ with respect to $t$ and $x$,
where $\hat\X$ is the characteristic form \eqref{charX}--\eqref{pointP} of
the generator \eqref{symmX}.
An equivalent condition on the Lagrangian is that $\pr\hat\X L$ is annihilated
by the variational derivatives with respect to $u$ and $\bar u$.

Since invariance of $\mathcal L$ implies that its extrema \eqref {extremaL}
are preserved,
every variational point symmetry of the Lagrangian functional \eqref{nls-L}
for the Schr\"odinger equation \eqref{nls} is thereby
a point symmetry of the Schr\"odinger equation itself such that
\begin{equation}\label{varsymmX}
\frac{\delta(\pr\hat\X L)}{\delta \bar u} =0 ,\quad
\frac{\delta(\pr\hat\X L)}{\delta u} =0.
\end{equation}
This provides a straightforward way to determine
all of the variational point symmetries
starting from \thmref{thm:pointsymms},
which yields the following result.

\begin{thm}\label{thm:varsymms}
The variational point symmetries of the Schr\"odinger equation \eqref{nls}
are generated by the symmetries \eqref{nls-Phas}--\eqref{nls-Rot} 
for all powers $p$ 
and additionally by the symmetries \eqref{nls-Scal}--\eqref{nls-Inver}
for the pseudo-conformal power $p=4/n$.
\end{thm}

Thus, among all of the point symmetries listed in \thmref{thm:pointsymms},
the only non-variational symmetry is the scaling \eqref{nls-Scal} for $p\neq 4/n$. 

Variational symmetries give rise to conservation laws
for the Schr\"odinger equation \eqref{nls} by means of Noether's theorem
as follows.
The invariance condition \eqref{invL} combined with
the variational identity
\begin{equation}\label{noetherid}
\pr\hat\X L=
\frac{\delta L}{\delta u}P
+ \frac{\delta L}{\delta\bar u}\bar P
+D_t \Big(\frac{\partial L}{\partial u_t}P + \frac{\partial L}{\partial\bar u_t}\bar P\Big)
+D_x \cdot\Big(\frac{\partial L}{\partial\nabla u}P + \frac{\partial L}{\partial\nabla\bar u}\bar P\Big)
\end{equation}
yields the multiplier equation
\begin{equation}\label{noetherthm}
\Re((\i u_t-\Delta u -k|u|^p u)\bar P)
= D_t T +D_x\cdot X
\end{equation}
with
\begin{equation}\label{noetherrel}
T =A -\frac{\partial L}{\partial u_t}P  -\frac{\partial L}{\partial\bar u_t}\bar P ,
\quad
X =B -\frac{\partial L}{\partial\nabla u}P -\frac{\partial L}{\partial\nabla\bar u}\bar P.
\end{equation}
On all formal solutions $u(t,x)$ of the Schr\"odinger equation \eqref{nls},
the multiplier equation \eqref{noetherthm} then produces
a {\em conservation law}
\begin{equation}\label{conslaw}
D_t T + D_x\cdot X=0,
\end{equation}
where the {\em conserved density} $T$ and {\em flux} $X$ are given
by the Noether relation \eqref{noetherrel}.
In particular, $A$ and $B$ can be shown to have the specific form
\begin{equation}
A= -\tau L,
\quad
B= -\zeta L
\end{equation}
as derived from the invariance condition \eqref{invL}.
Therefore,
the conserved density and flux are given explicitly by the simple formulas
\begin{align}
T= -\tau L +\Re(\i u\bar P),
\quad
X= -\zeta L -2\Re(\bar P\nabla u)
\end{align}
in terms of $L$ and $P=\eta-\tau u_t-\zeta\cdot\nabla u$.
This result together with \thmref{thm:varsymms} yields
all of the conservation laws generated by
the variational point symmetries of the Schr\"odinger equation \eqref{nls},
as shown in Table~\ref{conslaws} using the notation
$\nabla_{(l)}=e_{(l)}\cdot \nabla$
where $\{e_{(1)},\ldots,e_{(n)}\}$ is any orthonormal basis for $\Rnum^n$.

\begin{table}
\begin{center}
\begin{tabular}{|c|c|c|c|}
\hline
& $T$
& $X$
&
\\
\hline
1
& $-|u|^2$
& $\i(u\nabla \bar{u}-\bar{u}\nabla u)$
& $L^2$ norm
\\
\hline
2
& $|\nabla u|^2-\tfrac{2}{p+2}k|u|^{2+p}$
& $-u_t\nabla \bar{u}-\bar{u}_t\nabla u$
& energy
\\
\hline
3
& $\tfrac{1}{2}\i (u\nabla_{(l)}\bar{u}-\bar{u}\nabla_{(l)}u)$
& $\begin{aligned}&
(|\nabla u|^2-\tfrac{2}{p+2}k|u|^{2+p})e_{(l)} \\&
+ \tfrac{1}{2}\i (\bar{u}u_t -u\bar{u}_t) e_{(l)} \\&
-\nabla_{(l)}u\nabla \bar{u} -\nabla_{(l)}\bar{u}\nabla u
\end{aligned}$
& $\begin{aligned} &\text{momentum}\\  &l=1,\ldots,n\end{aligned}$
\\
\hline
4
& $\begin{aligned}&
\i t(u\nabla_{(l)}\bar{u}-\bar{u}\nabla_{(l)}u) \\&
+ (e_{(l)}\cdot x)|u|^2
\end{aligned}$
& $\begin{aligned}&
-2t((\nabla \bar{u})(\nabla_{(l)}u)+(\nabla u)(\nabla_{(l)}\bar{u})) \\&
+2t(|\nabla u|^2-\tfrac{2}{p+2}k|u|^{2+p})e_{(l)} \\&
+\i t(\bar{u}u_t-u\bar{u}_t)e_{(l)} \\&
-\i (e_{(l)}\cdot x)(u\nabla \bar{u}-\bar{u}\nabla u)
\end{aligned}$
& $\begin{aligned} &\text{Galilean} \\& \text{momentum}\\  &l=1,\ldots,n\end{aligned}$
\\
\hline
5
& $\begin{aligned}&
\tfrac{1}{2}\i (e_{(l')}\cdot x)(\bar{u}\nabla_{(l)}u -u\nabla_{(l)}\bar{u}) \\&
-\tfrac{1}{2}\i (e_{(l)}\cdot x)(\bar{u}\nabla_{(l')}u -u\nabla_{(l')}\bar{u})
\end{aligned}$
& $\begin{aligned}&
(e_{(l')}\cdot x) (\nabla_{(l)}u\nabla\bar{u}+\nabla_{(l)}\bar{u}\nabla u) \\&
-(e_{(l)}\cdot x)(\nabla_{(l')}u\nabla\bar{u}+\nabla_{(l')}\bar{u}\nabla u) \\&
+\i (e_{(l)}\wedge e_{(l')})\cdot x(\bar{u} u_t-u\bar{u}_t) \\&
+2(e_{(l)}\wedge e_{(l')})\cdot x(|\nabla u|^2-\tfrac{2}{p+2}k|u|^{2+p})
\end{aligned}$
& $\begin{aligned} &\text{angular}\\& \text{momentum}\\  &l,l'=1,\ldots,n; \\& l\neq l' \end{aligned}$
\\
\hline
6
& $\begin{aligned}&
2t(|\nabla u|^2-\tfrac{2}{p+2}k|u|^{2+p}) \\&
+\tfrac{1}{2}\i((x \cdot \nabla\bar{u})u-(x\cdot \nabla u)\bar{u})
\end{aligned}$
& $\begin{aligned}&
(|\nabla u|^2-\tfrac{2}{p+2}k|u|^{2+p})x \\&
+\tfrac{1}{2}\i (\bar{u}u_t-u\bar{u}_t)x
-\tfrac{2}{p}\nabla (|u|^2) \\&
-2t(u_t\nabla \bar{u}+\bar{u}_t\nabla u) \\&
-(x\cdot\nabla u)\nabla \bar{u} -(x\cdot\nabla\bar{u})\nabla u
\end{aligned}$
& $\begin{aligned} &\text{dilation}\\&\text{energy}\\ &p=4/n\end{aligned}$
\\
\hline
7
& $\begin{aligned}&
t^2(|\nabla u|^2-\tfrac{2}{p+2}k|u|^{2+p}) \\&
+\tfrac{1}{2}\i t ((x\cdot \nabla \bar{u})u-(x\cdot \nabla u)\bar{u}) \\&
+\tfrac{1}{4}|x|^2|u|^2
\end{aligned}$
& $\begin{aligned}&
t(|\nabla u|^2-\tfrac{2}{p+2}k|u|^{2+p})x \\&
-\tfrac{1}{2}\i t(\bar{u}u_t-u\bar{u}_t)x -\tfrac{2}{p}t\nabla(|u|^2) \\&
-2t((x\cdot \nabla u)\nabla \bar{u}+(x\cdot \nabla \bar{u})\nabla u) \\&
-t^2(u_t\nabla \bar{u}+\bar{u}_t\nabla u) \\&
+\tfrac{1}{4}\i |x|^2(\bar{u}\nabla u-u\nabla \bar{u})
\end{aligned}$
& $\begin{aligned} &\text{pseudo-conformal}\\&\text{energy}\\ &p=4/n\end{aligned}$
\\
\hline
\end{tabular}
\end{center}
\caption{Conservation laws of the semilinear Schr\"odinger equation \eqref{nls} in $\Rnum^n$}
\label{conslaws}
\end{table}

\subsection{Radial reduction}

A space rotation generator \eqref{nls-Rot} acts as an infinitesimal rotation
in the 2-plane specified by a pair of basis vectors
$e_{(l)}$ and $e_{(l')}$ in $\Rnum^n$ with $l\neq l'$.
The corresponding group of transformations
\begin{equation}
(t,x,u)
\rightarrow
(t,x+(\cos(\phi)-1)(e_{(l)}\odot e_{(l)}+e_{(l')}\odot e_{(l')})\cdot x
+2\sin(\phi)(e_{(l)}\wedge e_{(l')})\cdot x,u)
\end{equation}
is an $\SO(2)$ Lie group.
Composition of all such transformations acting in the
$n(n-1)/2$ distinct 2-planes determined by an orthonormal basis
$\{e_{(1)},\ldots,e_{(n)}\}$ of $\Rnum^n$
produces an $\SO(n)$ Lie group of rotations,
whose invariants consist of functions of $t$, $|x|=r$, $u$, $\bar u$.

Reduction of the Schr\"odinger equation \eqref{nls} in $\Rnum^n$
under this point symmetry group of rotations
gives the radial Schr\"odinger equation \eqref{radial-nls}
where $u(t,r)$ is a group-invariant solution.
The radial Lagrangian \eqref{radial-Lagr} arises naturally from
this reduction of the Lagrangian functional \eqref{nls-L} in $\Rnum^n$,
as given by
\begin{equation}\label{nls-radL}
{\mathcal L}_{\rm rad.} =
\int_{t_0}^{t_1}\int_{0}^{\infty} L(r,u,\bar u,u_t,\bar u_t,u_r,\bar u_r) \;drdt
\end{equation}
in the radial domain $0\leq r<\infty$.

From general results on symmetry reductions \cite{Olv},
a point symmetry generator \eqref{symmX} of the Schr\"odinger equation \eqref{nls}
in $\Rnum^n$ will admit a radial reduction
if and only if it belongs to the normalizer algebra of
the $\mathfrak{so}(n)$ Lie subalgebra of space rotations \eqref{nls-Rot}
in the Lie algebra of all point symmetry generators \eqref{nls-pointsymms}.
Likewise, through Noether's theorem,
a conservation law \eqref{conslaw} of the Schr\"odinger equation \eqref{nls}
will admit a radial reduction
if and only if its corresponding variational point symmetry generator
belongs to this same normalizer algebra.

Because the set of all formal radial solutions $u(t,r)$ is contained strictly
as a subset in the set of all formal solutions $u(t,x)$
to the Schr\"odinger equation \eqref{nls} in $\Rnum^n$,
the radial Schr\"odinger equation \eqref{radial-nls} could possibly admit
additional (``hidden'') point symmetries and conservation laws 
other than those inherited through radial reduction.
Consequently, the only way to find all radial point symmetries,
as well as all radial conservation laws corresponding to
variational point symmetries,
is by directly solving the determining equations for their generators.

For the radial Schr\"odinger equation \eqref{radial-nls},
the generator of a point symmetry acting on the variables $(t,r,u,\bar u)$
is given by
\begin{equation}\label{radial-symmX}
\X=
\tau(t,r,u,\bar u)\partial/\partial t
+ \rho(t,r,u,\bar u)\partial/\partial r
+ \eta(t,r,u,\bar u)\partial/\partial u
+ \bar\eta(t,r,u,\bar u)\partial/\partial\bar u
\end{equation}
such that $\pr\X(\i u_t-u_{rr}-(n-1)r^{-1}u_r -k|u|^p u)=0$
holds for all formal solutions $u(t,r)$ of equation \eqref{radial-nls}.
The characteristic form for each such generator
\begin{equation}\label{radial-charX}
\hat\X=
P(t,r,u,\bar u,u_t,\bar u_t,u_r,\bar u_r) \partial/\partial u
+ \bar P(t,r,u,\bar u,u_t,\bar u_t,u_r,\bar u_r) \partial/\partial\bar u
\end{equation}
with
\begin{equation}\label{radial-pointP}
P=\eta-\tau u_t-\rho u_r
\end{equation}
satisfies
\begin{equation}\label{radial-Pdeteq}
\i D_t P - D_r{}^2 P -(n-1)r^{-1}D_r P -k(1+p/2)\bar u^{p/2} u^{p/2} P
-k(p/2)\bar u^{p/2-1} u^{1+p/2}\bar P
=0,
\end{equation}
where $u_t$, $\bar u_t$, and $r$-derivatives of $u_t$, $\bar u_t$
are eliminated through the equation \eqref{radial-nls} 
and the complex-conjugate equation. 
Here $D_t$ and $D_r$ denote total derivatives with respect to $t$ and $r$.
This determining equation \eqref{radial-Pdeteq}
splits with respect to $r$-derivatives of $u$ and $\bar u$,
giving an overdetermined linear system of PDEs on
the functions $\tau(t,r,u,\bar u)$, $\rho(t,r,u,\bar u)$, $\eta(t,r,u,\bar u)$.
After a straightforward integrability analysis,
the system reduces to the PDEs
\begin{subequations}\label{radial-symm-eqns}
\begin{align}
&\tau_r=\tau_u=\tau_{\bar{u}}=\rho_u=\rho_{\bar{u}}=\eta_{\bar u}=\bar\eta_u=0,
\label{det-eqn-1}
\\
&\eta_{uu}=\bar\eta_{\bar u\bar u}=(pn-4)\tau_{tt}=\tau_{ttt}=0,
\label{det-eqn-2}
\\
& \rho=\tfrac{1}{2} r\tau_t, \quad
\eta_{ut}=\bar\eta_{\bar{u}t} ,\quad
\eta_{ur}-\bar\eta_{\bar{u}r} =-\i\rho_t ,\quad
\eta_u +\bar\eta_{\bar u} = -\tfrac{4}{p}\rho_r,
\label{det-eqn-3}
\\
&\bar u\eta+u\bar\eta=\bar{u}u(\eta_u +\bar\eta_{\bar u}),
\label{det-eqn-4}
\end{align}
\end{subequations}
which are easily solved to obtain the following result.

\begin{thm}\label{thm:radial-pointsymms}
The point symmetries of the radial Schr\"odinger equation \eqref{radial-nls}
are generated by
\begin{subequations}\label{radial-pointsymms}
\begin{alignat}{2}
\text{phase~rotation} \quad
&\X_{\phas} && =
\i u\partial/\partial u-\i \bar{u} \partial/\partial {\bar{u}},
\label{radial-Phas}
\\
\text{time~translation} \quad
&\X_{\trans} && =
\partial/\partial t,
\label{radial-Trans}
\\
\text{scaling} \quad
&\X_{\scal} && =
2t\partial/\partial t+r\partial/\partial r
-(2/p)u\partial/\partial u-(2/p)\bar{u}\partial/\partial {\bar{u}},
\label{radial-Scal}
\\
\text{inversion} \quad
&\X_{\inver} &&=
t^2\partial/\partial t +tr\partial/\partial r
-(2t/p +\i r^2/4)u \partial/\partial u
\nonumber\\ &&&\qquad
-(2t/p -\i r^2/4)\bar{u}\partial/\partial {\bar{u}}
\quad \text{only for}\quad p=4/n, 
\label{radial-Inver}
\end{alignat}
\end{subequations}
with the Lie algebra structure
\begin{align}
&
[\X_{\phas},\X_{\trans}]=[\X_{\phas},\X_{\scal}]=[\X_{\phas},\X_{\inver}]=0,
\\
&
[\X_{\trans},\X_{\scal}]=2\X_{\trans},
\;
[\X_{\trans},\X_{\inver}]=\X_{\scal},
\;
[\X_{\scal},\X_{\inver}]=2\X_{\inver}.
\label{radial-commutators}
\end{align}
The corresponding transformation groups acting on solutions $u=f(t,r)$
are given by
\begin{subequations}\label{radial-family-sol}
\begin{align}
&u=\exp{(\i\phi)}f(t,r),
\label{radial-Phas-group}
\\
&u=f(t-\epsilon,r),
\label{radial-Trans-group}
\\
&u=\lambda^{-2/p}f(\lambda^{-2}t,\lambda^{-1}r),
\label{radial-Scal-group}
\\
&u=(1+\epsilon t)^{-2/p}\exp{(-\i\epsilon r^2/(4+4\epsilon t))} f(t/(1+\epsilon t),r/(1+\epsilon t))
\quad \text{only for}\quad p=4/n, 
\label{radial-Inver-group}
\end{align}
\end{subequations}
with group parameters
$-\infty<\epsilon<\infty$, $0<\lambda<\infty$, $0 \leq \phi < 2\pi$.
\end{thm}

Thus,
the only point symmetries admitted by the radial Schr\"odinger equation \eqref{radial-nls}
are the ones it inherits from reduction of the Schr\"odinger equation \eqref{radial-nls} in $\Rnum^n$ under the $\SO(n)$ group of rotations.

The Lagrangian functional \eqref{nls-radL}
for the radial Schr\"odinger equation \eqref{radial-nls}
is invariant up to spatial boundary terms (at $r=0$ and $r\rightarrow\infty$)
and temporal boundary terms (at $t=t_0$ and $t=t_1$)
under an infinitesimal point transformation \eqref{radial-symmX} on $(t,r,u,\bar u)$
if and only if
\begin{equation}\label{radial-invL}
\pr\hat\X L=D_t A+D_r B
\end{equation}
holds for some functions $A$ and $B$ of $t$, $r$, $u$, $\bar u$,
and derivatives of $u$, $\bar u$ with respect to $t$ and $r$.
This is the condition for $\X$ to be a variational point symmetry of
the Lagrangian functional \eqref{nls-radL}.
Since the transformation group generated by a variational point symmetry
necessarily preserves the extrema $\delta{\mathcal L}_{\rm rad.}/\delta\bar u=0$ 
of this functional ${\mathcal L}_{\rm rad.}$, 
every such symmetry generator $\X$ is thereby
a point symmetry of the radial Schr\"odinger equation itself
for which the variational condition \eqref{radial-invL} holds,
or equivalently for which the equation
\begin{equation}\label{radial-varsymmX}
\frac{\delta(\pr\hat\X L)}{\delta \bar u} 
= \frac{\delta(\pr\hat\X L)}{\delta u} =0
\end{equation}
is satisfied.
Condition \eqref{radial-varsymmX} can be shown to reduce to
the symmetry determining equations \eqref{radial-symm-eqns}
plus the additional equation
\begin{equation}\label{radial-varsymm-cond}
\tau_t +(n-1)r^{-1}\rho -(1+4/p)\rho_r=0.
\end{equation}
It is straightforward to determine which of the point symmetries
from \thmref{thm:radial-pointsymms} satisfy this equation \eqref{radial-varsymm-cond},
which yields the following result.

\begin{thm}\label{thm:radial-varsymms}
The variational point symmetries of the radial Schr\"odinger equation \eqref{radial-nls}
are generated by the symmetries \eqref{radial-Phas}--\eqref{radial-Trans}
when $p\neq 4/n$ 
and by all of the symmetries \eqref{radial-Phas}--\eqref{radial-Inver}
when $p=4/n$. 
\end{thm}

From Noether's theorem,
these variational symmetries yield the multiplier equation
\begin{equation}\label{radial-noetherthm}
\Re((\i u_t-u_{rr}-(n-1)r^{-1}u_r -k|u|^p u)\bar P)
= D_t T +D_r X,
\end{equation}
which produces a conservation law
\begin{equation}\label{radial-conslaw}
D_t T + D_r X=0
\end{equation}
holding on all formal solutions $u(t,r)$ of the radial Schr\"odinger equation \eqref{radial-nls}.
Here $P$ is the function \eqref{radial-pointP} given by
the characteristic form of a symmetry generator \eqref{radial-charX},
while the conserved density $T$ and flux $X$ are given by the simple formulas
\begin{align}
T= -\tau L +\Re(\i u\bar P) ,
\quad
X= -\rho L -2\Re(\bar P u_r),
\end{align}
which can be derived from the variational condition \eqref{radial-invL}
in terms of the radial Lagrangian \eqref{radial-Lagr}.
The resulting conservation laws generated by
all of the variational point symmetries in \thmref{thm:radial-varsymms}
are shown in Table~\ref{radial-conslaws}.

\begin{table}[h!]
\begin{center}
\begin{tabular}{|c|c|c|c|}
\hline
& $T$
& $X$
&
\\
\hline
1
& $-r^{n-1}|u|^2$
& $\i r^{n-1}(u\bar{u}_r-\bar{u}u_r)$
& $L^2$ norm (charge)
\\
\hline
2
& $r^{n-1}(|u_r|^2-\tfrac{2}{p+2}k|u|^{2+p})$
& $-r^{n-1}(u_t\bar{u}_r+\bar{u}_t u_r)$
& energy
\\
\hline
3
& $\begin{aligned}&
2tr^{n-1}(|u_r|^2-\tfrac{2}{p+2}k|u|^{2+p}) \\&
+\tfrac{1}{2} \i r^{n}(u\bar{u}_r-\bar{u}u_r)
\end{aligned}$
& $\begin{aligned}&
-r^n(|u_r|^2+\tfrac{2}{p+2}k|u|^{2+p}) \\&
+\tfrac{1}{2}\i r^n(\bar{u}u_t-u\bar{u}_t) \\&
-\tfrac{2}{p}r^{n-1}(u\bar{u}_r+\bar{u}u_r) \\&
-2tr^{n-1}(u_t\bar{u}_r+\bar{u}_t u_r)
\end{aligned}$
& $\begin{aligned} &\text{dilation energy}\\ &p=4/n\end{aligned}$
\\
\hline
4
& $\begin{aligned}&
t^2r^{n-1}(|u_r|^2-\tfrac{2}{p+2}k|u|^{2+p}) \\&
+\tfrac{1}{2}\i tr^n(u\bar{u}_r-\bar{u}u_r) \\&
+\tfrac{1}{4}r^{n+1}|u|^2
\end{aligned}$
& $\begin{aligned}&
-\tfrac{2}{p}tr^{n-1}(u\bar{u}_r+\bar{u}u_r) \\&
-t^2r^{n-1}(u_t\bar{u}_r+\bar{u}_tu_r) \\&
-tr^n(|u_r|^2+\tfrac{2}{p+2}k|u|^{2+p}) \\&
+\tfrac{1}{2}\i tr^n(\bar{u}u_t-u\bar{u}_t) \\&
+\tfrac{1}{4}\i r^{n+1}(\bar{u}u_r-u\bar{u}_r)
\end{aligned}$
& $\begin{aligned} &\text{pseudo-conformal energy}\\ &p=4/n\end{aligned}$
\\
\hline
\end{tabular}
\end{center}
\caption{Conservation laws of the semilinear radial Schr\"odinger equation \eqref{radial-nls}\label{radial-conslaws}}
\end{table}

\subsection{Group invariance of radial solutions}

A solution $u=f(t,r)$ of the radial Schr\"odinger equation \eqref{radial-nls}
is {\em group-invariant} under a one-dimensional point symmetry group
if (and only if) it satisfies $\hat\X u\big|_{u=f(t,r)}=0$
where $\hat\X$ is the symmetry generator in characteristic form
\eqref{radial-charX}--\eqref{radial-pointP}.
To find all group-invariant solutions, it is sufficient to consider
a maximal set of one-dimensional point symmetry subgroups that are
conjugacy inequivalent in the full Lie group of point symmetries admitted
by the radial Schr\"odinger equation \eqref{radial-nls}.
This is most easily carried out at the Lie algebra level by exhibiting
a maximal set of one-dimensional subalgebras that are
conjugacy inequivalent in the Lie algebra of point symmetry generators.
Such a set, called {\em optimal symmetry generators},
can be straightforwardly determined by the methods in Refs.~\cite{Olv,WinPat}
starting from the point symmetry algebra given in \thmref{thm:radial-pointsymms}.

\begin{lem}\label{lemma:optimalsymms}
The Lie algebra of point symmetry generators \eqref{radial-pointsymms}
for the radial Schr\"odinger equation \eqref{radial-nls} is isomorphic to
$\mathfrak{u}(1)\oplus A(2)$ when $p\neq 4/n$
and $\mathfrak{u}(1)\oplus \mathfrak{sl}(2,\Rnum)$ when $p=4/n$.
(Here $A(2)$ denotes a 2-dimensional, non-abelian Lie algebra,
which is unique up to isomorphism.)
An optimal set of symmetry generators $\X$ consists of
\begin{subequations}\label{radial-optimal-symms}
\begin{gather}
\X_{\phas},
\label{optimal-Phas}\\
\X_{\trans}+\nu \X_{\phas},
\label{optimal-TransPhas}\\
\X_{\scal}+\mu \X_{\phas}, 
\label{optimal-ScalPhas}
\end{gather}
with parameters $-\infty<\nu<\infty$, $-\infty<\mu<\infty$,
and also
\begin{equation}
\X_{\trans}+\X_{\inver}+\kappa \X_{\phas} \quad\text{ for }\quad
p=4/n
\label{optimal-TransInverPhas}
\end{equation}
\end{subequations}
with parameter $-\infty<\kappa<\infty$.
\end{lem}

These generators \eqref{radial-optimal-symms} define group-invariant solutions
whose form is determined by the respective characteristic equations
\begin{subequations}\label{radial-optimal-inv-eqns}
\begin{align}
& u =0,
\label{Phas-inv}\\
& \i\nu u - u_t  =0,
\label{TransPhas-inv}\\
& (\tfrac{2}{p}-\i\mu)u +2t u_t +r u_r =0,
\label{ScalPhas-inv}\\
& (\tfrac{2}{p}t+\i(\tfrac{1}{4}r^2-\kappa))u +(t^2+1) u_t +tr u_r =0 ,
\quad p=4/n,
\label{TransInverPhas-inv}
\end{align}
\end{subequations}
which come from $\hat\X u=0$,
where $\hat\X$ denotes a generator \eqref{radial-optimal-symms}
in characteristic form \eqref{radial-charX}--\eqref{radial-pointP}.
The corresponding one-dimensional point transformation subgroups are given by
\begin{subequations}\label{radial-optimal-groups}
\begin{align}
& t\rightarrow t, r\rightarrow r, u\rightarrow\exp(\i\phi)u,
\label{Phas-group}\\
& t\rightarrow t+\epsilon, r\rightarrow r, u\rightarrow\exp(\i\nu\epsilon)u,
\label{TransPhas-group}\\
& t\rightarrow \lambda^2 t, r\rightarrow \lambda r,
u\rightarrow\exp(\i\mu\ln\lambda)\lambda^{-2/p}u,
\label{ScalPhas-group}\\
& t\rightarrow (\sin\phi +t\cos\phi)/(\cos\phi -t\sin\phi),
r\rightarrow r/(\cos\phi -t\sin\phi),
\nonumber\\&
u\rightarrow \exp\big(\i\kappa\phi -\i\tfrac{1}{4}r^2\sin\phi/(\cos\phi -t\sin\phi)\big)(\cos\phi -t\sin\phi)^{2/p} u,
\quad p=4/n, 
\label{TransInverPhas-group}
\end{align}
\end{subequations}
with group parameters
$-\infty<\epsilon<\infty$, $0<\lambda<\infty$, $0 \leq \phi < 2\pi$.

Invariance under the phase rotation symmetry generator \eqref{optimal-Phas}
yields only a trivial group-invariant solution \eqref{Phas-inv}.
Hence, for finding non-trivial group-invariant solutions,
only the remaining symmetry generators
\eqref{optimal-TransPhas}--\eqref{optimal-TransInverPhas}
in the optimal set need to be considered.
The specific form for these solutions is given by integration of 
the characteristic equations \eqref{TransPhas-inv}--\eqref{TransInverPhas-inv},
which can be expressed directly in terms of the invariants 
$\xi=\Xi(t,r)$, $U=\Upsilon(t,r,u)$, $\bar U=\bar \Upsilon(t,r,\bar u)$
of the corresponding generators $\X$.

\subsection{Radial scaling reductions}
The symmetry group of scaling transformations \eqref{ScalPhas-group}
with the generator \eqref{optimal-ScalPhas} 
has invariants 
\begin{equation}\label{invs:Scal+Phas}
\xi = t/r^2,
\quad
U= r^{2/p}\exp(-\i\mu\ln r)u,
\quad
\bar U = r^{2/p}\exp(\i\mu\ln r)\bar u, 
\end{equation}
depending on a parameter $-\infty<\mu<\infty$.
Hence the corresponding form for group-invariant solutions \eqref{ScalPhas-inv} 
is given by 
\begin{equation}\label{u:Scal+Phas}
u(t,r)=r^{- 2/p}\exp(\i \mu\ln r) U(\xi) . 
\end{equation}
The radial Schr\"odinger equation \eqref{radial-nls} thereby reduces to
the complex semilinear 2nd order ODE
\begin{align}
&
4\xi^2 U'' + \big( (8-2n+8/p)\xi - \i(1+4\mu \xi) \big)U'
\nonumber\\&\quad
+ \big((4-2n)/p+4/p^2-\mu^2 +\i \mu (n-2-4/p) \big)U + k |U|^p U = 0.
\label{Scal+Phas-ODE}
\end{align}
For $\mu=0$, this reduction describes radial similarity solutions.

From \thmref{thm:radial-varsymms},
the symmetry group \eqref{ScalPhas-group} leaves invariant
the Lagrangian functional \eqref{nls-radL} for the radial Schr\"odinger equation
iff $p=4/n$.
Hence, in the case of the pseudo-conformal power,
ODE \eqref{Scal+Phas-ODE} has a variational formulation given by
reduction of the radial Schr\"odinger Lagrangian \eqref{radial-Lagr},
\begin{equation}\label{L:Scal+Phas}
L =
-4 \xi^2 |U'|^2 + (\tfrac{4(p-1)}{p^2}-\mu^2) |U|^2
+ \tfrac{2}{p+2}k|U|^{2+p}
+ \i(\tfrac{1}{2}+2\mu\xi)(U\bar{U}'- \bar{U}U')
\text{ for } p=4/n.
\end{equation}

\subsection{Radial time-translation reductions}
The symmetry group of time translation transformations \eqref{TransPhas-group}
with the generator \eqref{optimal-TransPhas}
has invariants 
\begin{equation}\label{invs:Trans+Phas}
\xi=r,
\quad
U = \exp{(-\i\nu t)}u, 
\quad
\bar U = \exp{(\i\nu t)}\bar u, 
\end{equation}
depending on a parameter $-\infty<\nu<\infty$.
Hence
\begin{equation}\label{u:Trans+Phas}
u(t,r)= \exp{(\i\nu t)}U(\xi)
\end{equation}
yields the corresponding form for group-invariant solutions \eqref{TransPhas-inv}. 
The radial Schr\"odinger equation \eqref{radial-nls} thereby reduces to
the complex semilinear 2nd order ODE
\begin{equation}\label{Trans+Phas-ODE}
U'' + (n-1)\xi^{-1}U' + \nu U +k |U|^p U = 0.
\end{equation}
For $\nu=0$, this reduction describes radial static solutions,
which are invariant under the symmetry subgroup of
time translation transformations
\begin{equation}
t\rightarrow t+\epsilon, r\rightarrow r, u\rightarrow u.
\label{Trans-group}
\end{equation}
When $\nu\neq0$, the reduction \eqref{u:Trans+Phas} describes 
radial standing wave solutions. 

Since, by \thmref{thm:radial-varsymms},
the Lagrangian functional \eqref{nls-radL} for the radial Schr\"odinger equation
is invariant under the symmetry group \eqref{TransPhas-group},
ODE \eqref{Trans+Phas-ODE} has a variational formulation given by
reduction of the radial Schr\"odinger Lagrangian \eqref{radial-Lagr},
\begin{equation}\label{L:Trans+Phas}
L = \xi^{n-1}( -|U'|^2+\nu |U|^2+ \tfrac{2}{p+2}k|U|^{2+p}).
\end{equation}

\subsection{Radial pseudo-conformal reductions}
The symmetry group of combined inversion and time translation transformations \eqref{TransInverPhas-group}
with the generator \eqref{optimal-TransInverPhas}
for $p=4/n$
has invariants 
\begin{equation}\label{invs:Trans+Inver+Phas}
\begin{aligned}
&
\xi=(1+t^2)/r^2,
\quad
U = r^{2/p}\exp\big(\i \kappa\arctan(1/t) + \i r^2t/(4(1+t^2)) \big) u,
\\&
\bar U = r^{2/p}\exp\big(-\i \kappa\arctan(1/t) -\i r^2t/(4(1+t^2)) \big)\bar u,
\end{aligned}
\end{equation}
depending on a parameter $-\infty<\kappa<\infty$.
Hence
\begin{equation}\label{u:Trans+Inver+Phas}
u(t,r)= r^{-2/p}\exp\big(-\i \kappa\arctan(1/t) - \i r^2t/(4(1+t^2)) \big) U(\xi)
\end{equation}
yields the corresponding form for group-invariant solutions \eqref{TransInverPhas-inv}. 
The radial Schr\"odinger equation \eqref{radial-nls} thereby reduces to
the complex semilinear 2nd order ODE
\begin{equation}\label{Trans+Inver+Phas-ODE}
4 \xi^2 U'' + 8\xi U' + (\kappa\xi^{-1}-\xi^{-2}/4+ n(1-n/4)) U + k |U|^{4/n} U = 0.
\end{equation}

From \thmref{thm:radial-varsymms},
since the Lagrangian functional \eqref{nls-radL} for the radial Schr\"odinger equation
is invariant under the symmetry group \eqref{TransInverPhas-group},
ODE \eqref{Trans+Inver+Phas-ODE} has a variational formulation given by
reduction of the radial Schr\"odinger Lagrangian \eqref{radial-Lagr} with $p=4/n$,
\begin{equation}\label{L:Trans+Inver+Phas}
L =
-4\xi^2|U'|^2 + (\kappa\xi^{-1} - \tfrac{1}{4}\xi^{-2} + n(1-\tfrac{1}{4}n)) |U|^2 + \tfrac{n}{n+2}k|U|^{2+4/n}).
\end{equation}

\subsection{Group invariance of optimal radial reductions}

The ODEs \eqref{Scal+Phas-ODE}, \eqref{Trans+Phas-ODE}, \eqref{Trans+Inver+Phas-ODE}
arising by the reduction of the radial Schr\"odinger equation
under its optimal point symmetry subgroups
\eqref{optimal-ScalPhas}, \eqref{optimal-TransPhas}, \eqref{optimal-TransInverPhas}
have the following symmetry structure.

\begin{prop}\label{prop:U-ODE-symms}
(i) The point symmetries admitted by
the scaling-group ODE \eqref{Scal+Phas-ODE}
and the pseudo-conformal-group ODE \eqref{Trans+Inver+Phas-ODE}
are generated only by phase rotations
\begin{equation}\label{U-phas-symm}
\Y_{\phas}= \i U \partial/\partial U - \i \bar{U} \partial/\partial \bar{U}.
\end{equation}
(ii) The point symmetries admitted by
the translation-group ODE \eqref{Trans+Phas-ODE}
are generated by
\begin{align}
\text{scaling}
\quad
\Y_{\scal} &
= \xi \partial/ \partial \xi
- (2/p) U \partial/\partial U - (2/p) \bar{U} \partial/\partial \bar{U}
\quad\text{ for }\quad \nu=0,
\label{Trans+Phas-U-scalsymm}
\\
\text{dilation}
\quad
\Y_{\rm{dil.}} &
= \xi^{2-n}(\xi/(n-2) \partial/ \partial \xi
- U \partial/\partial U - \bar{U} \partial/\partial \bar{U} )
\label{Trans+Phas-U-dilsymm}
\\&\nonumber
\qquad \text{ for }\quad \nu=0,\quad p=2(3-n)/(n-2)\neq 0,
\end{align}
in addition to phase rotations \eqref{U-phas-symm}.
\end{prop}

In the case $\nu=0$,
the translation-group ODE \eqref{Trans+Phas-ODE} determines
all static solutions of the radial Schr\"odinger equation \eqref{radial-nls},
\begin{equation}\label{Trans-ODE}
U'' + (n-1)\xi^{-1}U' +k |U|^p U = 0
\quad\text{with}\quad 
U=u, \xi=r.
\end{equation}
The scaling symmetry \eqref{Trans+Phas-U-scalsymm} of this ODE is inherited
from the scaling invariance of the radial Schr\"odinger equation,
due to the commutator structure $[\X_{\trans},\X_{\scal}]=2\X_{\trans}$.
In contrast, the dilation symmetry \eqref{Trans+Phas-U-dilsymm} is
not inherited from any invariance of the radial Schr\"odinger equation
and thus describes a hidden symmetry arising for static solutions.

\section{\large Quadrature of complex 2nd order semilinear ODEs}
\label{methods}

Details of the two reduction of order methods outlined in section~\ref{intro}
for solving complex 2nd order semilinear ODEs of the general form
\eqref{Scal+Phas-ODE}, \eqref{Trans+Phas-ODE}, \eqref{Trans+Inver+Phas-ODE}
will now be presented.

\subsection{Reduction by point symmetries}
\label{method-1}

Consider a complex 2nd order semilinear ODE
\begin{equation}\label{U-ODE}
\alpha(\xi) U'' + \beta(\xi) U' + \gamma(\xi)U +k|U|^p U = 0
\end{equation}
with a real independent variable $\xi$ and a complex dependent variable $U$.
Here $\alpha$, $\beta$, $\gamma$ are allowed to be complex functions of $\xi$,
while $k$ and $p$ are assumed to be non-zero real constants.
Every such ODE \eqref{U-ODE} is invariant under
a $\U(1)$ group of phase rotation symmetries given by the generator
\eqref{U-phas-symm}.
Now suppose an ODE \eqref{U-ODE} admits another
one-dimensional group of point symmetries,
with a generator of the form
\begin{equation}\label{U-symm}
\Y= \zeta(\xi) \partial/ \partial\xi + \Omega(\xi) U \partial/\partial U
+ \bar\Omega(\xi) \bar{U} \partial/\partial \bar{U},
\end{equation}
whereby
\begin{equation}
[\Y_{\phas},\Y]=0.
\end{equation}
If the ODE has a variational formulation for which both $\Y_{\phas}$ and $\Y$
are variational symmetries,
then these two symmetries can be used to reduce the ODE to quadratures
by means of first integrals.
However, the standard reduction steps \cite{Olv} would require
finding the Lagrangian functional and checking its invariance under the two symmetries,
which can be cumbersome to carry out.
A simpler, more direct way to accomplish the same reduction is to utilize
canonical coordinates and integrating factors associated to the symmetries
$\Y_{\phas}$ and $\Y$ as follows.

Make a change of variables from $(\xi,U,\bar U)$ to $(z,V,\bar V)$ given by
the canonical coordinates of the generator \eqref{U-symm},
\begin{equation}\label{zVcoords}
\Y z= 1 ,
\quad
\Y V =0 ,
\quad
\Y \bar{V} =0,
\end{equation}
where
\begin{equation}
z=\int(1/\zeta)d\xi ,
\quad
V/U=\exp\Big(-\int(\Omega/\zeta)d\xi\Big)
\end{equation}
are functions only of $\xi$.
In terms of these variables,
the generator \eqref{U-symm} takes the form of a $z$-translation
\begin{equation}\label{V-trans}
\Y= \partial/ \partial z,
\end{equation}
while the phase rotation generator \eqref{U-phas-symm} is given by
\begin{equation}\label{V-phas}
\Y_{\phas}= \i V \partial/\partial V - \i \bar{V} \partial/\partial \bar{V}.
\end{equation}
Consequently, the ODE \eqref{U-ODE} is transformed into
\begin{equation}\label{V-ODE}
a V'' + b V' + cV+ k|V|^p V = 0
\end{equation}
for $V(z)$, where the coefficients $a,b,c$ are constants.
The following result is straightforward to prove by considering
the characteristic form of the symmetry generators \eqref{V-trans}--\eqref{V-phas},
\begin{equation}
\hat\Y= -V' \partial/\partial V - \bar{V}' \partial/\partial \bar{V} ,
\quad
\hat\Y_{\phas}= \i V \partial/\partial V - \i \bar{V} \partial/\partial \bar{V}.
\end{equation}

\begin{lem}\label{lemma:V-quadrature}
(i)
An ODE \eqref{V-ODE} has a non-trivial invariant solution
with respect to $z$-translations
iff $\bar c=c$ and $c/k<0$ 
(unless $p$ is a rational number having an odd numerator). 
Then $V'=0$ yields
\begin{equation}
|V|=(-c/k)^{1/p},
\end{equation}
which determines $V$ up to an arbitrary constant phase.
\newline
(ii)
An ODE \eqref{V-ODE} has integrating factors $\bar V'$ and $-\i\bar V$
iff $\bar a=a$, $\bar b=-b$, $\bar c=c$.
Then the first integrals are respectively given by
\begin{align}
&
a|V'|^2 + c|V|^2 + 2k\int |V|^{p+1} d|V| = C_1,
\label{V-integral-1}\\
&
a(\bar V V' -\bar V' V) + b |V|^2 = \i 2C_2,
\label{V-integral-2}
\end{align}
from which $V(z)$ can be determined by quadratures,
where $C_1$ and $C_2$ are arbitrary real constants.
\end{lem}

The solutions arising from Lemma~\ref{lemma:V-quadrature}
can be easily obtained in a more explicit form in terms of polar variables
\begin{equation}
V =A\exp(\i\Phi).
\end{equation}
In particular,
the first integrals \eqref{V-integral-1}--\eqref{V-integral-2} yield
\begin{align}
&
a {A'}^2 = -a A^2{\Phi'}^2 -c A^2 - 2k\int A^{p+1} dA + C_1,
\\
&
a\Phi'= \tilde b +  C_2/A^2,
\quad
\tilde b = \i 2b.
\end{align}
Combining these two differential equations, we get the quadratures
\begin{subequations}\label{V-quadrature}
\begin{align}
&
\int\frac{dA}{\sqrt{H(A)}} = \pm z +C_3,
\\
&
\Phi= (\tilde b/a)z  \pm(C_2/a) \int\frac{dA}{A^2 \sqrt{H(A)}} + C_4, 
\end{align}
\end{subequations}
assuming $A'\neq 0$,
where
\begin{equation}
H(A) =
(aC_1-2\tilde b C_2)/a^2 - ((\tilde b/a)^2 +c/a) A^2 -(C_2/a)^2 A^{-2}
- (2k/a)\int A^{p+1} dA.
\end{equation}
The quadratures \eqref{V-quadrature}
determine the general solution $V(z)$ of ODE \eqref{V-ODE}
in the case $\bar a=a$, $\bar b=-b$, $\bar c=c$.
Moreover, these conditions on $a,b,c$ are necessary and sufficient for
the ODE \eqref{V-ODE} to have an Euler--Lagrange structure
\begin{equation}
\delta L/\delta\bar V =a V'' + b V' +c V + k|V|^p V=0 ,
\quad
\bar a=a, \bar b=-b, \bar c=c
\end{equation}
given by the Lagrangian
\begin{equation}
L =
-a\bar V' V' +\tfrac{1}{2}b(\bar V V' - \bar V' V) +\tfrac{1}{2}c |V|^2
+ 2k \int |V|^{p+1} d|V|,
\end{equation}
for which both the $z$-translation generator \eqref{V-trans}
and the phase rotation generator \eqref{V-phas}
are variational symmetries.
Noether's theorem thereby asserts
\begin{align}
&
\pr\hat\Y L= -\frac{dL}{dz}
= -V'\frac{\delta L}{\delta V} -\bar V'\frac{\delta L}{\delta\bar V}
+\frac{d}{dz} \Big( (a\bar V'-\tfrac{1}{2}b\bar V)V' + (a V' +\tfrac{1}{2}b V)\bar V' \Big),
\\
&
\pr\hat\Y_{\phas} L= 0
= \i V\frac{\delta L}{\delta V} -\i\bar V\frac{\delta L}{\delta\bar V}
+\frac{d}{dz} \Big( (-a\bar V'+\tfrac{1}{2}b\bar V)\i V + (a V' +\tfrac{1}{2}b V)\i \bar V \Big),
\end{align}
giving a derivation of the first integrals \eqref{V-integral-1}--\eqref{V-integral-2}.

In the general case without conditions on $a,b,c$,
the $z$-translation invariant solution $V=\const$ of ODE \eqref{V-ODE}
is given by
\begin{equation}
A=(-c/k)^{1/p} =\const,
\quad
\Phi=\const.
\end{equation}

Finally, changing variables $(z,V,\bar V)$ back into $(\xi,U,\bar U)$
in the original complex semilinear ODE \eqref{U-ODE} yields
(i) the group-invariant solution with respect to the symmetry generator \eqref{U-symm};
(ii) the general solution under necessary and sufficient conditions
for the generator \eqref{U-symm} to be a variational symmetry.

\subsection{Reduction by hidden conditional symmetries}
\label{method-2}

Consider a real 1st-order Lagrangian of the form
\begin{equation}\label{U-Lagr}
L =
\varpi(\xi)\Big(
-\tilde\alpha(\xi) \bar U' U' +\i\tilde\beta(\xi)(\bar U' U- U' \bar U)
+ \tilde\gamma(\xi)|U|^2 +2k\int |U|^{p+1} d|U|
\Big),
\end{equation}
where $\varpi$, $\tilde\alpha$, $\tilde\beta$, $\tilde\gamma$
are real functions of $\xi$,
and $k\neq0$, $p\neq 0$ are real constants.
The Euler--Lagrange equation $\delta L/\delta\bar U=0$
yields a complex 2nd order semilinear ODE of the form \eqref{U-ODE}
whose coefficients $\alpha$, $\beta$, $\gamma$ are given by
\begin{equation}\label{U-ODE-coeffs}
\alpha= \tilde\alpha ,
\quad
\beta = \tilde\alpha' + \tilde\alpha \varpi'/\varpi -\i 2\tilde\beta ,
\quad
\gamma = \tilde\gamma -\i (\tilde\beta' + \tilde\beta \varpi'/\varpi).
\end{equation}
Both the Lagrangian \eqref{U-Lagr}
and the Euler--Lagrange ODE \eqref{U-ODE}, \eqref{U-ODE-coeffs}
are invariant under the generator \eqref{U-phas-symm}
of phase rotation symmetries on $U,\bar U$.
Hence $\Y_{\phas}$ is a variational symmetry to which
Noether's theorem can be applied to obtain a reduction of this ODE.

The simplest way to carry out the reduction is by use of polar variables
\begin{equation}
U =A\exp(\i\Phi).
\end{equation}
In particular, the ODE \eqref{U-ODE}, \eqref{U-ODE-coeffs}
gets converted into a coupled semilinear system of real ODEs
\begin{align}
&
\tilde\alpha A'' +((\tilde\alpha\varpi)'/\varpi) A'
-\tilde\alpha (\Phi'-\tilde\beta/\tilde\alpha)^2 A
+ (\tilde\gamma+\tilde\beta^2/\tilde\alpha) A  + kA^{p+1} =0,
\label{A-EL-ODE}\\
&
\tilde\alpha \Phi''A + 2\tilde\alpha (\Phi'-\tilde\beta/\tilde\alpha) A'
+((\tilde\alpha\varpi)'/\varpi)A \Phi' - ((\tilde\beta\varpi)'/\varpi)A =0,
\label{Phi-EL-ODE}
\end{align}
which are seen to be the Euler--Lagrange equations
$\delta L/\delta A=0$ and $A^{-1}\delta L/\delta\Phi=0$
of the Lagrangian \eqref{U-Lagr} expressed in polar variables
\begin{equation}\label{A-Phi-Lagr}
L =
\varpi(\xi)\Big(
-\tilde\alpha(\xi) {A'}^2 -\tilde\alpha(\xi) A^2 (\Phi'-\tilde\beta(\xi)/\tilde\alpha(\xi))^2
+ (\tilde\gamma(\xi) +\tilde\beta(\xi)^2/\tilde\alpha(\xi)) A^2 +2k\int A^{p+1} dA
\Big).
\end{equation}
The phase rotation generator \eqref{U-phas-symm} thereby becomes
\begin{equation}\label{Phi-shift-symm}
\Y_{\phas}= \partial/\partial\Phi,
\end{equation}
producing a group of shift transformations $\Phi\rightarrow \Phi+\epsilon$
with group parameter $-\infty<\epsilon<\infty$.
Then the invariance $\pr\hat\Y_{\phas} L=0$ yields the multiplier equation
\begin{equation}\label{shift-multiplier}
\varpi A\big(
\tilde\alpha \Phi''A + 2\tilde\alpha (\Phi'-\tilde\beta/\tilde\alpha) A'
+((\tilde\alpha\varpi)'/\varpi) \Phi' - ((\tilde\beta\varpi)'/\varpi)A
\big)
=\frac{d}{d\xi}\Big( \varpi A^2 (\alpha \Phi'-\tilde\beta) \Big),
\end{equation}
which produces the first integral
\begin{equation}\label{Phi-integral}
\varpi A^2 (\tilde\alpha\Phi'-\tilde\beta) =C_1
\end{equation}
for the system \eqref{A-EL-ODE}--\eqref{Phi-EL-ODE}.
Through equation \eqref{Phi-integral},
$\Phi'$ can be eliminated in terms of $A$, so that
the system reduces to a single real semilinear ODE
\begin{equation}\label{A-ODE}
\tilde\alpha A'' +((\tilde\alpha\varpi)'/\varpi) A'
+ (\tilde\gamma+\tilde\beta^2/\tilde\alpha) A
-({C_1}^2/(\tilde\alpha \varpi^2)) A^{-3} + kA^{p+1} =0.
\end{equation}
This ODE \eqref{A-ODE} is the Euler--Lagrange equation
$\delta \tilde L/\delta A=0$ of the modified Lagrangian
\begin{equation}\label{A-Lagr}
\begin{aligned}
\tilde L = &
\Big( L +2(\Phi'-\tilde\beta(\xi)/\tilde\alpha(\xi))C_1 \Big)
\Big|_{\displaystyle \tilde\alpha(\xi)\Phi'-\tilde\beta(\xi) =(C_1/\varpi(\xi))A^{-2}}
\\
= &
\varpi(\xi)\Big(
-\tilde\alpha(\xi) {A'}^2
+ (\tilde\gamma(\xi) +\tilde\beta(\xi)^2/\tilde\alpha(\xi)) A^2
+ ({C_1}^2/(\tilde\alpha(\xi)\varpi(\xi)^2)) A^{-2} +2k\int A^{p+1} dA
\Big),
\end{aligned}
\end{equation}
where $C_1$ is treated as a constant parameter.

There are two cases, depending on $C_1$ and $p$,
for which the ODE \eqref{A-ODE} can be solved.
Firstly, if $C_1=0$ and $p=-1$
then the ODE becomes linear
\begin{equation}\label{single-A-linear-ODE}
\tilde\alpha A'' +((\tilde\alpha\varpi)'/\varpi) A'
+ (\tilde\gamma+\tilde\beta^2/\tilde\alpha) A  +k =0
\end{equation}
and its general solution will be given by special functions.
Secondly, if $C_1\neq0$ or $p\neq -1$
then the ODE will reduce to quadratures by means of integrating factors
arising from variational symmetries of the Lagrangian \eqref{A-Lagr}.

Note that the set of solutions of the ODE \eqref{A-ODE} for each value of $C_1$
corresponds to the subset of solutions of the system 
\eqref{A-EL-ODE}--\eqref{Phi-EL-ODE}
defined by the resulting level set of $C_1$.
Suppose a point symmetry generator
\begin{equation}\label{A-Phi-symm}
\Y= \zeta(\xi)\partial/\partial\xi + \Upsilon(\xi)A\partial/\partial A
+ \omega(\xi)\partial/\partial\Phi
\end{equation}
admitted by system \eqref{A-EL-ODE}--\eqref{Phi-EL-ODE}
leaves invariant the level set of $C_1$, so that 
$\pr\hat\Y C_1 =
2\varpi A(\Upsilon A-\zeta A')(\tilde\alpha \Phi' -\tilde\beta)
+ \varpi\tilde\alpha A^2(\omega-\zeta \Phi')' 
=0$
when 
$\Phi'=(\tilde\beta +(C_1/\varpi)A^{-2})/\tilde\alpha$. 
Then, by reduction, the level-set ODE \eqref{A-ODE} will inherit
a corresponding point symmetry generator
\begin{equation}\label{A-symm}
\tilde\Y= \zeta(\xi)\partial/\partial\xi + \Upsilon(\xi)A\partial/\partial A.
\end{equation}
Such symmetries are equivalent to conditional
point symmetries of the form \eqref{U-symm}
for the original complex 2nd order semilinear ODE \eqref{U-ODE}
such that the generator preserves the first integral \eqref{Phi-integral}
given in terms of $U,\bar U$ by
\begin{equation}
C_1= \varpi(\tfrac{1}{2}\i\tilde\alpha (U\bar U'- U'\bar U) -\tilde\beta U\bar U).
\end{equation}
Any variational symmetries that are inherited in this way
by the level-set ODE \eqref{A-ODE} will yield the same integrating factors
derived from the previous reduction method in section \ref{method-1}.

Now suppose that the level-set ODE \eqref{A-ODE} admits
a variational point symmetry that is not inherited from any point symmetry
admitted by the system \eqref{A-EL-ODE}--\eqref{Phi-EL-ODE}.
Such a variational symmetry will be a hidden conditional symmetry which
yields an additional integrating factor for the ODE \eqref{A-ODE}.
The resulting reduction is most easily carried out
by utilizing canonical coordinates associated to the symmetry generator
as follows.

For a symmetry of the form \eqref{A-symm},
we change variables from $(\xi,A)$ to canonical coordinates $(z,B)$
given by
\begin{equation}\label{zBcoords}
\tilde\Y z= 1 ,
\quad
\tilde\Y B =0,
\end{equation}
where
\begin{equation}\label{zB}
z=\int(1/\zeta)d\xi ,
\quad
B=\exp\Big(-\int(\Upsilon/\zeta)d\xi\Big) A.
\end{equation}
The symmetry generator \eqref{A-symm} thereby becomes a $z$-translation
\begin{equation}\label{B-trans}
\tilde\Y= \partial/ \partial z
\end{equation}
and thus the ODE \eqref{A-ODE} is transformed into
\begin{equation}\label{B-ODE}
a B'' + b B' + cB+ qB^{-3}+k B^{p+1} = 0
\end{equation}
for $B(z)$, where the coefficients $a,b,c,q$ are real constants.
Similarly, the Lagrangian \eqref{A-Lagr} becomes
(modulo a total $z$-derivative)
\begin{equation}\label{B-Lagr}
\zeta\tilde L = \hat L =
\exp((b/a)z)\Big( -a {B'}^2 + cB^2 - qB^{-2} +2k\int B^{p+1}dB \Big),
\end{equation}
where $1/\zeta=\dfrac{dz}{d\xi}$ is the Jacobian of the transformation
for the independent variable.
Then Noether's theorem applied to the $z$-translation generator \eqref{B-trans}
in characteristic form
\begin{equation}
\widehat{\tilde\Y}= -B' \partial/\partial B
\end{equation}
yields the multiplier equation
\begin{equation}\label{B-multiplier}
\begin{aligned}
\pr\widehat{\tilde\Y}(\hat L)
=& -\exp((b/a)z)\frac{d}{dz}\Big( -a {B'}^2 + cB^2 - qB^{-2} +2k\int B^{p+1}dB \Big)
\\
= & -B'\frac{\delta\hat L}{\delta B}
+\frac{d}{dz} \Big( 2(\exp((b/a)z)a B') B' \Big).
\end{aligned}
\end{equation}
This equation \eqref{B-multiplier} shows that $b=0$ is
a necessary and sufficient condition for the $z$-translation generator \eqref{B-trans} to be a variational symmetry of the ODE \eqref{A-ODE}.
As an immediate consequence, the following result holds.

\begin{lem}\label{lemma:B-firstintegr}
An ODE \eqref{B-ODE} has integrating factor $B'$ iff $b=0$.
The resulting first integral is given by
\begin{equation}\label{B-integral}
a{B'}^2 + cB^2 -qB^{-2} +2k\int B^{p+1} dB = C_2,
\end{equation}
from which $B(z)$ is determined by the quadrature
\begin{equation}\label{B-quadrature}
\int\frac{dB}{\sqrt{H(B)}} = \pm z +C_3,
\end{equation}
where
\begin{equation}
H(B) =
C_2/a - (c/a) B^2 +(q/a) B^{-2} - (2k/a)\int B^{p+1} dB.
\end{equation}
\end{lem}

When $b=0$,
the quadrature \eqref{B-quadrature} combined with the change of variable \eqref{zB}
will yield the general solution of the ODE \eqref{A-ODE} for $A(\xi)$.
In this case the first integral \eqref{Phi-integral} provides
a quadrature for $\Phi(\xi)$, thereby yielding the general solution of
the system \eqref{A-EL-ODE}--\eqref{Phi-EL-ODE}.
These quadratures for $A(\xi)$ and $\Phi(\xi)$ can be written down
in a fairly simple way.
First, we find that the change of variable \eqref{zB} relating
the Lagrangians \eqref{B-Lagr} and \eqref{A-Lagr} gives
\begin{equation}\label{ABrelation}
A^2/B^2= \exp\Big(2\int(\Upsilon/\zeta)d\xi\Big)
= (a\zeta/(\varpi\tilde\alpha)) \exp\Big((b/a)\int(1/\zeta)d\xi\Big).
\end{equation}
Hence, from the quadrature \eqref{B-quadrature} together with the variables \eqref{zB},
we see
\begin{subequations}\label{A-Phi-quadrature}
\begin{equation}
\int\frac{dB}{\sqrt{H(B)}} = \pm \int(1/\zeta)d\xi +C_3
\label{A-B-quadrature}
\end{equation}
and
\begin{equation}
A=(a\zeta/(\varpi\tilde\alpha))^{1/2} B
\end{equation}
for $b=0$.
Last, changing variables in the first integral \eqref{Phi-integral}
and using the other first integral \eqref{B-integral},
we obtain
\begin{equation}
\Phi= \pm (C_1/a)\int\frac{dB}{B^2 \sqrt{H(B)}} + \int (\tilde\beta/\tilde\alpha)d\xi + C_4.
\label{Phi-B-quadrature}
\end{equation}
\end{subequations}

The integrals \eqref{A-Phi-quadrature} yield the general solution of
the system \eqref{A-EL-ODE}--\eqref{Phi-EL-ODE}
for $A(\xi)$ and $\Phi(\xi)$ under necessary and sufficient conditions for
the reduced ODE \eqref{A-ODE} to admit a variational symmetry \eqref{A-symm}.
This determines the corresponding solution $U(\xi)=A(\xi)\exp(\i\Phi(\xi))$ to
the original complex semilinear Euler--Lagrange ODE
\eqref{U-ODE}, \eqref{U-ODE-coeffs}.

Finally, suppose the level-set ODE \eqref{A-ODE} admits a hidden point symmetry
that is not a variational symmetry.
Such a hidden conditional symmetry can be used to obtain
an invariant solution of the ODE \eqref{A-ODE},
which will determine a corresponding solution of
the system \eqref{A-EL-ODE}--\eqref{Phi-EL-ODE}
through the first integral \eqref{Phi-integral} as follows.

\begin{lem}\label{lemma:B-invsoln}
An ODE \eqref{A-ODE} has a non-trivial invariant solution
with respect to $z$-translations
iff $c\neq 0$ or $q\neq 0$ when $p\neq -4$, 
or $c\neq 0$ and $q\neq -k$ when $p=-4$. 
Then $B'=0$ yields
\begin{equation}
0=c+qB^{-4} + kB^p,
\end{equation}
which is an algebraic equation determining $B=\const\neq 0$.
\end{lem}

Since $B'=0$, relation \eqref{ABrelation} yields 
\begin{equation}\label{A-invsoln}
A=(a\zeta/(\varpi\tilde\alpha))^{1/2} B \exp\Big((b/2a)\int(1/\zeta)d\xi\Big).
\end{equation}
Then, after changing variables from $(\xi,A)$ to $(z,B)$
in the first integral \eqref{Phi-integral}, we find
$\dfrac{d\Phi}{dz} = (C_1/aB^2)\exp(-(b/a)z)+ \zeta\tilde\beta/\tilde\alpha$
from which we obtain the quadrature
\begin{equation}
\Phi=
-C_1/(bB^2) \exp\Big((-b/a)\int(1/\zeta)d\xi\Big)
+ \int (\tilde\beta/\tilde\alpha)d\xi + C_2.
\label{Phi-invsoln}
\end{equation}
These expressions \eqref{A-invsoln}--\eqref{Phi-invsoln} yield a solution of
the system \eqref{A-EL-ODE}--\eqref{Phi-EL-ODE},
and hence $U(\xi)=A(\xi)\exp(\i\Phi(\xi))$ gives a solution to
the original Euler--Lagrange ODE \eqref{U-ODE}, \eqref{U-ODE-coeffs},
corresponding to an invariant solution of the level-set ODE \eqref{A-ODE}
under a hidden conditional point symmetry \eqref{A-symm}.

\section{\large Solutions to the optimal translation-group ODE}
\label{translation-group}

The reduction of order methods from section~\ref{methods}
will now be applied to the translation-group ODE \eqref{Trans+Phas-ODE}
arising by the reduction of the radial Schr\"odinger equation
under its optimal subgroup of point symmetries \eqref{optimal-TransPhas}.

As a preliminary step,
we use polar variables $U=A\exp(\i\Phi)$ to convert
this $\U(1)$-invariant ODE \eqref{Trans+Phas-ODE}
into a semilinear system of real ODEs
\begin{subequations}\label{APhi-ODEs:Trans+Phas}
\begin{align}
&
A'' - A\Phi'^2 + (n-1)\xi^{-1}A' + \nu A + k A^{p+1} =0,
\\
&
\Phi'' + 2A^{-1}A'\Phi' + (n-1)\xi^{-1}\Phi' =0.
\end{align}
\end{subequations}
The ODEs in this system \eqref{APhi-ODEs:Trans+Phas}
are the respective Euler--Lagrange equations
$\delta L/\delta A=0$ and $A^{-1}\delta L/\delta\Phi=0$
of the $\U(1)$-invariant Lagrangian \eqref{L:Trans+Phas}
expressed in polar variables,
\begin{equation}\label{APhi-L:Trans+Phas}
L = \xi^{n-1}( -A'^2 - A^2\Phi'^2 + \nu A^2 + \tfrac{2}{2+p}kA^{2+p} ).
\end{equation}

In polar form,
the point symmetries of ODE \eqref{Trans+Phas-ODE} listed in Proposition~\ref{prop:U-ODE-symms}
consist of
\begin{subequations}
\begin{align}
\text{phase~rotation}
\quad
\Y_{\phas} & =
\partial/\partial \Phi,
\label{APhi-phas:Trans+Phas}
\\
\text{scaling}
\quad
\Y_{\scal} & =
\xi \partial/ \partial \xi - (2/p) A \partial/\partial A
\quad\text{for~}\quad
\nu=0,
\label{APhi-scal:Trans+Phas}
\\
\text{dilation}
\quad
\Y_{\rm{dil.}} & =
\xi^{2-n} \left( \xi \partial/ \partial \xi + (2-n) A \partial/\partial A \right)
\label{APhi-dil:Trans+Phas}
\\&\nonumber
\qquad\text{for~}\quad
\nu=0,
\quad
p=2(3-n)/(n-2).
\end{align}
\end{subequations}
Invariance of the polar Lagrangian \eqref{APhi-L:Trans+Phas}
under $\Y_{\phas}$ produces a first integral \eqref{Phi-integral},
yielding
\begin{equation}\label{Phi-integral:Trans+Phas}
\Phi' = C_1 \xi^{1-n} A^{-2}.
\end{equation}
The polar system \eqref{APhi-ODEs:Trans+Phas} thereby reduces to
a single real semilinear ODE
\begin{equation}\label{A-ODE:Trans+Phas}
A'' +(n-1) \xi^{-1}A' + \nu A + k A^{1+p} - C_1^2 \xi^{2-2n} A^{-3} = 0,
\end{equation}
which is the Euler--Lagrange equation
$\delta \tilde L/\delta A=0$ of a modified Lagrangian \eqref{A-Lagr},
given by
\begin{equation}\label{A-L:Trans+Phas}
L = \xi^{n-1} ( -A'^2 + \nu A^2 + \tfrac{2}{2+p}kA^{2+p} + C_1^2\xi^{2-2n}A^{-2} ).
\end{equation}
Solutions of the ODE \eqref{A-ODE:Trans+Phas} for $A(\xi)$
represent the level set $C_1=\const$ of solutions $(A(\xi),\Phi(\xi))$
to the polar system \eqref{APhi-ODEs:Trans+Phas},
or equivalently the level set
\begin{equation}\label{levelset:Trans+Phas}
C_1= \tfrac{1}{2}\i \xi^{n-1} (U\bar U'- U'\bar U)
=\const
\end{equation}
of solutions $U(\xi)$ to the translation-group ODE \eqref{Trans+Phas-ODE}.

Note that the level-set ODE \eqref{A-ODE:Trans+Phas} will be linear iff
$C_1=0$ and $p=-1$.

\begin{prop}\label{prop:hiddensymms:Trans+Phas}
In the nonlinear case $C_1 \neq 0$ or $p \neq -1$,
the level-set ODE \eqref{A-ODE:Trans+Phas} admits point symmetries
only when
\begin{equation}
\nu=0.
\end{equation}
For this case the admitted point symmetries consist of
\begin{subequations}\label{hiddensymms:Trans+Phas}
\begin{align}
\Y_1 =&
\xi \partial/\partial \xi - (2/p) A \partial/\partial A,
\quad\text{for}\quad
C_1=0 \text{~or~} p=4/(n-2),
\label{scalsymm:Trans+Phas}
\\
\Y_2 =&
\xi^{2 - n} (\xi /(n-2)\partial/\partial \xi -A \partial/\partial A),
\quad\text{for}\quad
p = 2(3-n)/(n-2),
\label{dilsymm:Trans+Phas}
\\
\Y_3 =&
2\left( (C_1^2/k) \xi^{1/3} - \xi\right) \partial/\partial \xi - A \partial/\partial A,
\quad\text{for}\quad
C_1\neq0, p=-4, n=4/3,
\label{symm3:Trans+Phas}
\\
\Y_4 =&
2( k -  C_1^2 \xi^2) \xi\partial/\partial \xi + (k - 4 C_1^2 \xi^2) A \partial/\partial A,
\quad\text{for}\quad
C_1\neq0, p=-4, n=0,
\label{symm4:Trans+Phas}
\\
\Y_5 =&
\xi^3 \partial/\partial \xi - 6( \xi^2 A - 16 /k ) \partial/\partial A,
\quad\text{for}\quad
C_1=0, p=1, n=16,
\label{symm5:Trans+Phas}
\\
\Y_6 =&
9\xi^{2/3} \partial/\partial \xi - \left( 12 \xi^{-1/3} A + (4/k) \xi^{-7/3} \right) \partial/\partial A,
\quad\text{for}\quad
C_1=0, p=1, n=13/3.
\label{symm6:Trans+Phas}
\end{align}
\end{subequations}
\end{prop}
Since the point symmetries \eqref{scalsymm:Trans+Phas} and \eqref{dilsymm:Trans+Phas}
are the only ones also admitted by the polar system \eqref{APhi-ODEs:Trans+Phas},
the remaining 4 point symmetries \eqref{symm3:Trans+Phas}--\eqref{symm6:Trans+Phas}
thus describe hidden conditional symmetries arising only for
level-set solutions of the polar system \eqref{APhi-ODEs:Trans+Phas}
in the case $\nu=0$, $C_1\neq 0$ or $p\neq -1$.

We now carry out the two reduction methods
(cf sections~\ref{method-1} and~\ref{method-2})
to obtain explicit solutions $U(\xi)$
for the translation-group ODE \eqref{Trans+Phas-ODE}.
The first reduction method can be applied only in the case $\nu=0$,
using the scaling and dilation symmetries
\eqref{Trans+Phas-U-scalsymm} and \eqref{Trans+Phas-U-dilsymm}.
The second reduction method is applicable to
the case $\nu=0$ by using the hidden conditional symmetries
\eqref{symm3:Trans+Phas}--\eqref{symm6:Trans+Phas},
and the case $\nu\neq 0$ by using the linearization
which holds when $C_1=0$, $p=-1$.

\subsection{Scaling-symmetry quadratures for the translation-group ODE}
The canonical coordinates of the scaling symmetry \eqref{Trans+Phas-U-scalsymm}
are
\begin{equation}\label{Trans+Phas-scal-zV}
z= \ln \xi,
\quad
V=\xi^{2/p}U
\end{equation}
with $\nu=0$.
Hence the translation-group ODE \eqref{Trans+Phas-ODE}
gets transformed into
\begin{equation}\label{Trans-ODE-scal-coords}
V''+(n-2-4/p) V'+(2(2-n)/p+4/p^2)V+k|V|^{p}V=0,
\end{equation}
which is equivalent to the ODE \eqref{Trans-ODE}
describing static solutions of the radial Schr\"odinger equation.
To apply the reduction Lemma~\ref{lemma:V-quadrature}
to ODE \eqref{Trans-ODE-scal-coords},
we note $a=1$, $b= n-2-4/p$, $c= 2(2-n)/p+(2/p)^2$.

Then from Lemma~\ref{lemma:V-quadrature}(i) we obtain the invariant solution
\begin{equation}\label{Trans-ODE-scal-V-invsol}
V=\big(2(n-2-2/p)/(kp)\big)^{1/p}
\exp{(\i\Phi)},\quad
\Phi=\const.
\end{equation}
This yields
\begin{equation}\label{Trans-ODE-scal-invsol}
U=\big(2(n-2-2/p)/(kp)\big)^{1/p}\xi^{-2/p}\exp{(\i\Phi)},\quad
\Phi=\const,
\end{equation}
which is the scaling-invariant solution of ODE \eqref{Trans-ODE}.

Next, by Lemma~\ref{lemma:V-quadrature}(ii), we can obtain the general solution
for $V(z)$ in the case when $n-2-4/p=0$
under which the ODE \eqref{Trans-ODE-scal-coords} becomes
\begin{equation}\label{Trans-ODE-scal-special-p}
V''-(2/p)^2 V +k|V|^{p}V=0, \quad
p=4/(n-2).
\end{equation}
The quadratures \eqref{V-quadrature} in polar variables $V=A\exp(\i\Phi)$
are then given by
\begin{equation}\label{Trans-ODE-scal-quadrature-A}
\int\frac{dA}{\sqrt{H(A)}} = \pm z +C_3 ,
\quad
\Phi= C_2 \int\frac{dA}{A^2 \sqrt{H(A)}} + C_4,
\end{equation}
(after renaming constants) where
\begin{equation}
H(A) = C_1 -C_2^2 A^{-2} +(2/p)^2 A^2 -2k\int A^{1+p}dA .
\end{equation}
These integrals \eqref{Trans-ODE-scal-quadrature-A}
cannot be evaluated generally to obtain explicit solutions
for $A(z)$ and $\Phi(z)$ in terms of elementary functions.
Some special cases where explicit solutions can be derived
are possible if we make a change of variables
\begin{equation}\label{Trans-ODE-A-quadrature-y}
\int\frac{dA}{\sqrt{H(A)}} = (1/s) \int\frac{dy}{\sqrt{H_1(y)}},
\quad
y=A^{s}
\end{equation}
and, when $C_2\neq 0$,
\begin{equation}\label{Trans-ODE-Phi-quadrature-y}
\int\frac{dA}{A^2 \sqrt{H(A)}} = (1/\tilde s) \int\frac{dy}{\sqrt{H_2(y)}},
\quad
y=A^{\tilde s}
\end{equation}
so that the respective expressions
\begin{equation}\label{Trans-ODE-scal-denom-A}
H_1(y) = y^{2-2/s} H(y^{1/s})
= C_1 y^{2-2/s} -C_2^2  y^{2-4/s} +(2/p)^2 y^2 -\tilde k y^{2+p/s}
\end{equation}
and
\begin{equation}\label{Trans-ODE-scal-denom-Phi}
H_2(y) = y^{2+2/\tilde s} H(y^{1/\tilde s})
= C_1 y^{2+2/\tilde s} -C_2^2  y^{2} +(2/p)^2 y^{2+4/\tilde s}-\tilde k y^{2+(4+p)/\tilde s}
\quad\text{ for } C_2\neq 0
\end{equation}
are quadratic polynomials in $y$ for some values of $s$ and $\tilde s$,
depending on $p,C_1,C_2$,
where
\begin{equation}
\tilde k=2k/(p+2),
\quad
p=4/(n-2)\neq -2.
\end{equation}
The required conditions from expression \eqref{Trans-ODE-scal-denom-A}
consist of
\begin{subequations}
\begin{align}
&
2+p/s=0,1,2 ;
\label{Trans-cond1}\\
&
2-2/s=0,1,2
\quad\text{ if } C_1\neq 0;
\label{Trans-cond2}\\
&
2-4/s=0,1,2
\quad\text{ if } C_2\neq 0.
\label{Trans-cond3}
\end{align}
From expression \eqref{Trans-ODE-scal-denom-Phi},
the required conditions are given by
\begin{align}
&
2+4/\tilde s=0,1,2 \text{ and } 2+(4+p)/\tilde s =0,1,2
\quad\text{ if } C_2\neq 0;
\label{Trans-cond4}\\
& 2+2/\tilde s=0,1,2
\quad\text{ if } C_1\neq 0, C_2\neq 0.
\label{Trans-cond5}
\end{align}
\end{subequations}
It is straightforward to solve these conditions for $p,s,\tilde s$.
For each value found for $p$, only a single value each for $s$ and $\tilde s$
is needed to evaluate the integrals
\eqref{Trans-ODE-A-quadrature-y} -- \eqref{Trans-ODE-Phi-quadrature-y}.
We thus find
\begin{align}
& C_1=0 \text{ and } C_2=0:
\quad
p\neq -2, s= -p/2;
\label{Trans-ODE-scal-y-quadratures-a}\\
& C_1\neq0 \text{ and } C_2=0:
\quad
p=-1, s=1;
\quad
p=-4, s=2;
\label{Trans-ODE-scal-y-quadratures-b}\\
& C_1=0 \text{ and } C_2\neq0:
\quad
p=-4, s=2, \tilde s=-2;
\label{Trans-ODE-scal-y-quadratures-c}\\
& C_1\neq0 \text{ and } C_2\neq0:
\quad
p=-4, s=2, \tilde s=-2;
\label{Trans-ODE-scal-y-quadratures-d}
\end{align}
with $n = 2+4/p$.
By scaling and shifting $y$, we can then match the integrals
\eqref{Trans-ODE-A-quadrature-y} -- \eqref{Trans-ODE-Phi-quadrature-y}
to one of the forms
\begin{gather}
\int\frac{dy}{\sqrt{y^2-1}} =\arccosh(y) ,
\quad
\int\frac{dy}{\sqrt{y^2+1}} =\arcsinh(y) ,
\label{special-integrals-cosh}\\
\int\frac{dy}{\sqrt{y^2}} =\ln|y| ,
\label{special-integrals-exp}\\
\int\frac{dy}{\sqrt{1-y^2}} =\arcsin(y),
\label{special-integrals-sin}\\
\int\frac{dy}{\sqrt{y}} =2\sqrt{y},
\label{special-integrals-sqrt}\\
\int dy =y.
\label{special-integrals-lin}
\end{gather}

More generally, a similar method can be used to obtain
additional explicit solutions for $A(z)$ and $\Phi(z)$ by requiring that
the expressions \eqref{Trans-ODE-scal-denom-A}--\eqref{Trans-ODE-scal-denom-Phi}
are either quartic polynomials in $y$,
which will yield elliptic functions,
or squares of quartic polynomials in $y$,
which will yield elementary functions.
(These solutions will be worked out elsewhere \cite{AncFen}.)

\subsubsection{Quadrature for $p\neq -2$}

From case \eqref{Trans-ODE-scal-y-quadratures-a}, we have
$s=-p/2$, $C_1=0$ and $C_2=0$.
The quadratures \eqref{V-quadrature} are thus given by
\begin{equation}
\Phi = C_4
\end{equation}
and
\begin{equation}
\pm z +C_3 =
(-2/p) \int\frac{dy}{\sqrt{(4/p^2)y^2 - 2k/(p+2)}} ,
\quad
y=A^{-p/2},
\end{equation}
which can be matched to the form \eqref{special-integrals-cosh}.
This yields
\begin{subequations}
\begin{align}
&
A^{-p/2} = p(2(p+2)/k)^{-1/2}\cosh(\pm z + C_3),
\quad
\Phi = C_4,
\quad
k/(p + 2)>0,
\\
&
A^{-p/2} = p(-2(p+2)/k)^{-1/2}\sinh(\pm z + C_3),
\quad
\Phi = C_4,
\quad
k/(p + 2)<0.
\end{align}
\end{subequations}
Hence we obtain two solutions of ODE \eqref{Trans-ODE-scal-special-p}
for $p\neq -2$:
\begin{subequations}
\begin{align}
&
V =
(2(p+2)/(kp^2))^{1/p}(\cosh(\pm z + C_3))^{-2/p}\exp(\i C_4),
\quad
k/(p + 2)>0,
\label{Trans-ODE-scal-V-a-sol1}\\
&
V =
(-2(p+2)/(kp^2))^{1/p}(\sinh(\pm z + C_3))^{-2/p}\exp(\i C_4),
\quad
k/(p + 2)<0.
\label{Trans-ODE-scal-V-a-sol2}
\end{align}
\end{subequations}
These solutions \eqref{Trans-ODE-scal-V-a-sol1}--\eqref{Trans-ODE-scal-V-a-sol2}
can be merged after we change variables \eqref{Trans+Phas-scal-zV},
giving the following result.

\begin{prop}\label{prop:Trans-ODE-scal-U-case-a}
For $p \neq -2$, 
the translation-group ODE \eqref{Trans-ODE} has a solution
\begin{equation}\label{Trans-ODE-scal-U-a-sol}
\begin{aligned}
& U = 
(\pm 8(p+2)/(kp^2))^{1/p}(\tilde C_3 \xi^2 \pm 1/\tilde C_3)^{-2/p}\exp(\i C_4),
\\&
\quad
n = 2+4/p, 
\quad
\pm k/(p+2)>0, 
\end{aligned}
\end{equation}
with real constants $\tilde C_3$, $C_4$.
\end{prop}

\subsubsection{Quadrature for $p=-1$}

From case \eqref{Trans-ODE-scal-y-quadratures-b}, we have
$s=1$, $C_1\neq 0$ and $C_2=0$.
The quadratures \eqref{V-quadrature} are thus given by
\begin{equation}
\Phi = C_4
\end{equation}
and
\begin{equation}
\pm z +C_3 =
\int\frac{dy}{\sqrt{4(y - k/4)^2 + C_1 - k^2/4}},
\quad
y=A,
\end{equation}
which can be matched to the forms
\eqref{special-integrals-cosh} and \eqref{special-integrals-exp}.
This yields
\begin{subequations}
\begin{align}
&
A = \tfrac{1}{4}\left(k + (4C_1 - k^2)^{1/2}\sinh(2(\pm z + C_3))\right),
\quad
\Phi = C_4,
\quad
C_1 >k^2/4,
\\
&
A = \tfrac{1}{4}\left(k + (k^2 - 4C_1)^{1/2}\cosh(2(\pm z + C_3))\right),
\quad
\Phi = C_4,
\quad
C_1 < k^2/4,
\\
&
A = \tfrac{1}{4}\left(k + 4\exp(2(\pm z + C_3))\right),
\quad
\Phi = C_4,
\quad
C_1 =k^2/4,
\\
&
A = \tfrac{1}{4}\left(k - 4\exp(2(\pm z + C_3))\right),
\quad
\Phi = C_4,
\quad
C_1 = k^2/4.
\end{align}
\end{subequations}
Hence we obtain four solutions of ODE \eqref{Trans-ODE-scal-special-p}
for $p=-1$:
\begin{subequations}
\begin{align}
&
V = \tfrac{1}{4}\left(k + (4C_1 - k^2)^{1/2}\sinh(2(\pm z + C_3))\right)\exp(\i C_4),
\quad
C_1 > k^2/4,
\label{Trans-ODE-scal-V-bb-sol1}
\\
&
V = \tfrac{1}{4}\left(k + (k^2 - 4C_1)^{1/2}\cosh(2(\pm z + C_3))\right)\exp(\i C_4),
\quad
C_1 < k^2/4,
\label{Trans-ODE-scal-V-bb-sol2}
\\
&
V = \tfrac{1}{4}\left(k \pm 4\exp(2(z + C_3))\right)\exp(\i C_4),
\quad
C_1 = k^2/4,
\label{Trans-ODE-scal-V-bb-sol3}
\\
&
V = \tfrac{1}{4}\left(k \pm 4\exp(2(-z + C_3))\right)\exp(\i C_4), \quad
C_1 = k^2/4.
\label{Trans-ODE-scal-V-bb-sol4}
\end{align}
\end{subequations}
These solutions \eqref{Trans-ODE-scal-V-bb-sol1}--\eqref{Trans-ODE-scal-V-bb-sol4}
give the following result after we change variables \eqref{Trans+Phas-scal-zV}.

\begin{prop}
For $p=-1$, 
the translation-group ODE \eqref{Trans-ODE} has solutions
\begin{gather}
\begin{aligned}
&
U = \tfrac{1}{4}\left(k\xi^2 + (\pm (C_1 - k^2/4))^{1/2}(\tilde C_3 \xi^4 \mp 1/\tilde C_3)\right)\exp(\i C_4) ,
\\&
\quad
C_1\neq k^2/4,
\quad
n=-2,
\end{aligned}
\label{Trans-ODE-scal-U-bb-sol1-sol2}\\
\begin{aligned}
&
U=\tfrac{1}{4}\left(k\xi^2 +\tilde C_3 \xi^4\right)\exp(\i C_4) ,
\quad
n=-2,
\end{aligned}
\label{Trans-ODE-scal-U-bb-sol3}\\
\begin{aligned}
&
U=\tfrac{1}{4}\left(k\xi^2 + \tilde C_3\right)\exp(\i C_4) ,
\quad
n=-2,
\end{aligned}
\label{Trans-ODE-scal-U-bb-sol4}
\end{gather}
with real constants $C_1$, $\tilde C_3$, $C_4$.
\end{prop}

\subsubsection{Quadrature for $p=-4$}

From cases \eqref{Trans-ODE-scal-y-quadratures-b}--\eqref{Trans-ODE-scal-y-quadratures-d},
we have
$s=2$, $C_1\neq 0$ and $C_2=0$;
$s=2$, $\tilde s=-2$, and $C_2\neq 0$.

In the first subcase,
the quadratures \eqref{V-quadrature} are given by
\begin{equation}
\Phi = C_4
\end{equation}
and
\begin{equation}\label{Trans-ODE-z-quadr-b}
\pm z +C_3 =
\int\frac{dy}{\sqrt{(y + 2C_1)^2 + 4(k-C_1^2)}}, \quad
y=A^2.
\end{equation}
Quadrature \eqref{Trans-ODE-z-quadr-b} can be matched to the forms
\eqref{special-integrals-cosh} and \eqref{special-integrals-exp}.
This yields
\begin{subequations}
\begin{align}
&
A^2 = -2C_1 + 2(k - C_1^2)^{1/2}\sinh(\pm z + C_3),
\quad
\Phi=C_4 ,
\quad
C_1^2<k,
\\
&
A^2 = -2C_1 + 2(C_1^2 - k)^{1/2}\cosh(\pm z + C_3),
\quad
\Phi=C_4 ,
\quad
C_1^2>k,
\\
&
A^2 = -2C_1 + \exp(\pm z + C_3),
\quad
\Phi=C_4 ,
\quad
C_1^2=k,
\\
&
A^2 = -2C_1 - \exp(\pm z + C_3),
\quad
\Phi=C_4 ,
\quad
C_1^2=k, C_1<0.
\end{align}
\end{subequations}
Hence we obtain four solutions of ODE \eqref{Trans-ODE-scal-special-p}
for $p=-4$:
\begin{subequations}
\begin{align}
&
V = \left(-2C_1 + 2(k - C_1^2)^{1/2}\sinh(\pm z + C_3)\right)^{1/2}\exp(\i C_4),
\quad
C_1^2<k,
\label{Trans-ODE-scal-V-b-sol1}
\\
&
V = \left(-2C_1 + 2(C_1^2 - k)^{1/2}\cosh(\pm z + C_3)\right)^{1/2}\exp(\i C_4),
\quad
C_1^2>k,
\label{Trans-ODE-scal-V-b-sol2}
\\
&
V = \left(-2C_1 + \exp(\pm z + C_3)\right)^{1/2}\exp(\i C_4),
\quad
C_1^2=k,
\label{Trans-ODE-scal-V-b-sol3}
\\
&
V = \left(-2C_1 - \exp(\pm z + C_3)\right)^{1/2}\exp(\i C_4),
\quad
C_1^2=k, C_1<0.
\label{Trans-ODE-scal-V-b-sol4}
\end{align}
\end{subequations}

In the second subcase,
the quadratures \eqref{V-quadrature} are given by
\begin{equation}\label{Trans-ODE-z-quadr-c}
\pm z +C_3 =
\int\frac{dy}{\sqrt{(y + 2C_1)^2 + 4(k - C_1^2 - C_2^2)}},
\quad
y=A^2
\end{equation}
and
\begin{equation}\label{Trans-ODE-Phi-quadr-c}
\Phi
= C_4 - C_2 \int\frac{dy}{\sqrt{4(k-C_2^2)y^2 + 4C_1 y +1}},
\quad
y=A^{-2}.
\end{equation}
Quadrature \eqref{Trans-ODE-z-quadr-c}
can be matched to the forms
\eqref{special-integrals-cosh} and \eqref{special-integrals-exp},
which yields
\begin{subequations}
\begin{align}
&
A^2 = -2C_1 + 2(k - C_1^2 - C_2^2)^{1/2}\sinh(\pm z + C_3),
\quad
C_1^2 + C_2^2 < k,
\label{Trans-ODE-scal-A-c-sol1}
\\
&
A^2 = -2C_1 + 2(C_2^2 + C_1^2 - k)^{1/2}\cosh(\pm z + C_3),
\quad
C_1^2 + C_2^2 > k,
\label{Trans-ODE-scal-A-c-sol2}
\\
&
A^2 = -2C_1 + \exp(\pm z + C_3),
\quad
C_1^2 + C_2^2 = k,
\label{Trans-ODE-scal-A-c-sol3}
\\
&
A^2 = -2C_1 - \exp(\pm z + C_3),
\quad
C_1^2 + C_2^2 = k, C_1<0.
\label{Trans-ODE-scal-A-c-sol4}
\end{align}
\end{subequations}
Quadrature \eqref{Trans-ODE-Phi-quadr-c}
can be matched to all of the forms
\eqref{special-integrals-cosh}--\eqref{special-integrals-lin}.
This yields
\begin{subequations}
\begin{align}
&\begin{aligned}
\Phi = &
C_4 + \tfrac{1}{2}( k/C_2^2 - 1)^{-1/2}
\arcsinh\left((k - C_2^2 - C_1^2)^{-1/2}(C_1+ 2(k - C_2^2)A^{-2})\right),
\\& C_1^2 + C_2^2 < k,
\end{aligned}
\\
&\begin{aligned}
\Phi = &
C_4 - \tfrac{1}{2}(k/C_2^2 - 1)^{-1/2}
\arccosh\left((C_2^2 + C_1^2 - k)^{-1/2}(C_1 + 2(k - C_2^2)A^{-2})\right),
\\&
C_2^2 < k < C_1^2 + C_2^2,
\end{aligned}
\\
& \begin{aligned}
\Phi = &
C_4 - \tfrac{1}{2}(1-k/C_2^2)^{-1/2}
\arcsin\left((C_2^2 + C_1^2 - k)^{-1/2}(C_1- 2(C_2^2 - k)A^{-2})\right),
\\&
C_2^2 >k,
\end{aligned}
\\
&\begin{aligned}
\Phi =
C_4 - (C_2/(2C_1))\ln|C_1A^{-2} + 1/2| ,
\quad
C_1^2 + C_2^2 =k,
\end{aligned}
\\
&\begin{aligned}
\Phi = C_4 - (C_2/C_1)(C_1A^{-2} + 1/4)^{1/2} ,
\quad
C_2^2=k,
\end{aligned}
\\
&\begin{aligned}
\Phi = C_4 - C_2A^{-2} ,
\quad
C_2^2=k, C_1=0.
\end{aligned}
\end{align}
\end{subequations}
Hence we obtain 7 more solutions of ODE \eqref{Trans-ODE-scal-special-p}
for $p=-4$:
\begin{subequations}
\begin{equation}
\begin{aligned}
V = & \left(-2C_1 +2(k - C_1^2 - C_2^2)^{1/2}\sinh(\pm z + C_3)\right)^{1/2}
\\&\times
\exp\bigg(\i C_4 - (\i/2)(k/C_2^2 - 1)^{-1/2}
\arcsinh\left(\frac{(k - C_1^2 - C_2^2)^{1/2} + C_1\sinh(\pm z + C_3)}
{-C_1 + (k - C_1^2 - C_2^2)^{1/2}\sinh(\pm z + C_3)}\right)\bigg) ,
\\&
\quad
C_1^2 + C_2^2 < k,
\end{aligned}
\label{Trans-ODE-scal-V-c-sol1}
\end{equation}
\begin{equation}
\begin{aligned}
V =& \left(-2C_1 + 2(C_2^2 + C_1^2 - k)^{1/2}\cosh(\pm z + C_3)\right)^{1/2}
\\&\times
\exp\bigg(\i C_4 - (\i/2)(k/C_2^2 - 1)^{-1/2}
\arccosh\left(\frac{(C_2^2 + C_1^2 - k)^{1/2} - C_1\cosh(\pm z + C_3)}
{C_1 - (C_2^2 + C_1^2 - k)^{1/2}\cosh(\pm z + C_3)}\right)\bigg) ,
\\&
\quad
C_2^2 < k < C_1^2 + C_2^2,
\end{aligned}
\label{Trans-ODE-scal-V-c-sol3}
\end{equation}
\begin{equation}
\begin{aligned}
V =& \left(-2C_1 + 2(C_2^2 + C_1^2 - k)^{1/2}\cosh(\pm z + C_3)\right)^{1/2}
\\& \times
\exp\bigg(\i C_4 - (\i/2)(1 - k/C_2^2)^{-1/2}
\arcsin\left(\frac{(C_2^2 + C_1^2 - k)^{1/2} - C_1\cosh(\pm z+C_3)}
{-C_1 + (C_2^2 + C_1^2 - k)^{1/2}\cosh(\pm z+C_3)}\right)\bigg) ,
\\&
\quad
C_2^2 >k,
\end{aligned}
\label{Trans-ODE-scal-V-c-sol2}
\end{equation}
\begin{equation}
\begin{aligned}
V = &\left(-2C_1 + \exp(\pm z + C_3)\right)^{1/2}
\exp\left(\i\tilde C_4 +\i(C_2/(2C_1))
\ln\big|2C_1\exp(\mp z - C_3) -1\big|\right) ,
\\&
C_1^2 + C_2^2 =k,
\end{aligned}
\label{Trans-ODE-scal-V-c-sol4}
\end{equation}
\begin{equation}
\begin{aligned}
V = & \left(-2C_1 - \exp(\pm z + C_3)\right)^{1/2}
\exp\left(\i\tilde C_4 +\i(C_2/(2C_1))
\ln\big|2C_1\exp(\mp z - C_3) +1\big|\right) ,
\\&
C_1^2 + C_2^2=k,
\end{aligned}
\label{Trans-ODE-scal-V-c-sol5}
\end{equation}
\begin{equation}
\begin{aligned}
V = & (-2C_1 + 2|C_1|\cosh(\pm z + C_3))^{1/2}
\exp\bigg(\i C_4 - \i(k^{1/2}/(2C_1))
\Big(\frac{C_1 + |C_1|\cosh(\pm z + C_3)}{-C_1 + |C_1|\cosh(\pm z + C_3)}\Big)^{1/2}\bigg) ,
\\&
\quad
C_2^2 =k,
\end{aligned}
\label{Trans-ODE-scal-V-c-sol6}
\end{equation}
\begin{equation}
\begin{aligned}
V = & \exp(\tfrac{1}{2}(\pm z + C_3))
\exp\left(\i C_4 - \i k^{1/2}\exp(\mp z - C_3)\right) ,
\quad
C_2^2 =k, C_1=0.
\end{aligned}
\label{Trans-ODE-scal-V-c-sol7}
\end{equation}
\end{subequations}

After we change variables \eqref{Trans+Phas-scal-zV},
these solutions \eqref{Trans-ODE-scal-V-b-sol1}--\eqref{Trans-ODE-scal-V-b-sol4}
and \eqref{Trans-ODE-scal-V-c-sol1}--\eqref{Trans-ODE-scal-V-c-sol7}
give the following result.

\begin{prop}
For $p=-4$, 
the translation-group ODE \eqref{Trans-ODE} has solutions
\begin{equation}
\begin{aligned}
U = & \left(-2C_1\xi + (\pm(k - C_1^2))^{1/2}(\tilde C_3 \xi^2 \mp 1/\tilde C_3)\right)^{1/2}\exp(\i C_4),
\quad
C_1^2\neq k,
\quad
n=1,
\end{aligned}
\label{Trans-ODE-scal-U-b-sol1-sol2}
\end{equation}
\begin{equation}
\begin{aligned}
U = & \left(\pm 2k^{1/2}\xi + \tilde C_3 \xi^2\right)^{1/2}\exp(\i C_4),
\quad
k>0,
\quad
n=1,
\end{aligned}
\label{Trans-ODE-scal-U-b-sol3}
\end{equation}
\begin{equation}
\begin{aligned}
U = & \left(\pm 2k^{1/2}\xi + \tilde C_3\right)^{1/2}\exp(\i C_4),
\quad
k>0,
\quad
n=1,
\end{aligned}
\label{Trans-ODE-scal-U-b-sol4}
\end{equation}
\begin{equation}
\begin{aligned}
U = & \left(-2C_1\xi + (k - C_1^2 - C_2^2)^{1/2}(\tilde C_3 \xi^2 - 1/\tilde C_3)\right)^{1/2}
\\&\times
\exp\bigg(\i C_4 - \tfrac{1}{2}\i (k/C_2^2 -1)^{-1/2}
\arcsinh\Big(\frac{2(k - C_1^2 - C_2^2)^{1/2}\xi + C_1(\tilde C_3 \xi^2 -1/\tilde C_3 )}{-2C_1\xi + (k - C_1^2 - C_2^2)^{1/2}(\tilde C_3 \xi^2 -1/\tilde C_3)}\Big)\bigg) ,
\\&
\quad
C_1^2 + C_2^2 < k,
\quad
n=1,
\end{aligned}
\label{Trans-ODE-scal-U-c-sol1}
\end{equation}
\begin{equation}
\begin{aligned}
U = & \left(-2C_1\xi + (C_2^2 + C_1^2 - k\right)^{1/2}
(\tilde C_3 \xi^2 + 1/\tilde C_3))^{1/2}
\\&\times
\exp\bigg(\i C_4 - \tfrac{1}{2}\i (k/C_2^2 - 1)^{-1/2}
\arccosh\Big(\frac{2(C_2^2 + C_1^2 - k)^{1/2}\xi - C_1 (\tilde C_3 \xi^2 + 1/\tilde C_3)}
{2C_1\xi - (C_2^2 + C_1^2 - k)^{1/2}(\tilde C_3 \xi^2 + 1/\tilde C_3)}\Big)\bigg) ,
\\&
\quad
C_2^2 < k < C_1^2 + C_2^2,
\quad
n=1,
\end{aligned}
\label{Trans-ODE-scal-U-c-sol3}
\end{equation}
\begin{equation}
\begin{aligned}
U =& \left(-2C_1\xi + (C_2^2 + C_1^2 - k)^{1/2}
(\tilde C_3 \xi^2 + 1/\tilde C_3)\right)^{1/2}
\\&\times
\exp\bigg(\i C_4 - \tfrac{1}{2}\i (1 - k/C_2^2)^{-1/2}
\arcsin\Big(\frac{2(C_2^2 + C_1^2 - k)^{1/2}\xi - C_1(\tilde C_3 \xi^2 + 1/\tilde C_3)}
{-2C_1\xi + (C_2^2 + C_1^2 - k)^{1/2}(\tilde C_3 \xi^2 + 1/\tilde C_3 )}\Big)\bigg) ,
\\&
\quad
C_2^2 > k,
\quad
n=1,
\end{aligned}
\label{Trans-ODE-scal-U-c-sol2}
\end{equation}
\begin{equation}
\begin{aligned}
U = & \left(-2C_1\xi + \tilde C_3 \xi^2\right)^{1/2}
\exp\left(\i\tilde C_4 + \i(C_2/(2C_1))
\ln|2C_1\xi^{-1} - \tilde C_3|\right) ,
\quad
C_1^2 + C_2^2 = k,
\quad
n=1,
\end{aligned}
\label{Trans-ODE-scal-U-c-sol4}
\end{equation}
\begin{equation}
\begin{aligned}
U = & \left(-2C_1\xi + \tilde C_3 \right)^{1/2}
\exp\left(\i \tilde C_4 + \i(C_2/(2C_1))
\ln |2C_1\xi - \tilde C_3|\right) ,
\quad
C_1^2 + C_2^2 = k,
\quad
n=1,
\end{aligned}
\label{Trans-ODE-scal-U-c-sol5}
\end{equation}
\begin{equation}
\begin{aligned}
U = & (C_1/\tilde C_3)^{1/2} (\tilde C_3\xi-1)
\exp\bigg(\i C_4 - \tfrac{1}{2}\i(k^{1/2}/C_1)
\Big(\frac{\tilde C_3 \xi + 1}{\tilde C_3 \xi - 1}\Big)\bigg) ,
\quad
k>0,
\quad
n=1,
\end{aligned}
\label{Trans-ODE-scal-U-c-sol6}
\end{equation}
\begin{equation}
\begin{aligned}
U = & \tilde C_3^{1/2} \xi
\exp\left(\i C_4 - \i (k^{1/2}/ \tilde C_3) \xi^{-1}\right) ,
\quad
k>0,
\quad
n=1,
\end{aligned}
\label{Trans-ODE-scal-U-c-sol7}
\end{equation}
\begin{equation}
\begin{aligned}
U = & \tilde C_3^{1/2}
\exp\left(\i C_4 - \i (k^{1/2} /\tilde C_3) \xi\right) ,
\quad
k>0, 
\quad
n=1,
\end{aligned}
\label{Trans-ODE-scal-U-c-sol8}
\end{equation}
with real constants $C_1$, $C_2$, $\tilde C_3$, $C_4$, $\tilde C_4$.
\end{prop}

\subsection{Dilation-symmetry quadratures for the translation-group ODE}
The canonical coordinates of the dilation symmetry \eqref{Trans+Phas-U-dilsymm}
are
\begin{equation}\label{Trans+Phas-dil-zV}
z= \xi^{n-2},
\quad
V= \xi^{n-2} U
\end{equation}
with $\nu=0$.
In this case the translation-group ODE \eqref{Trans-ODE}
gets transformed into
\begin{equation}\label{Trans-ODE-dil-coords}
V''+k(1+p/2)^2|V|^{p}V=0,
\quad p = 2(3-n)/(n-2), n\neq 2,3.
\end{equation}
To apply Lemma~\ref{lemma:V-quadrature} to this ODE \eqref{Trans-ODE-dil-coords},
we note $a=1/(1+p/2)^2$, $b=0$, $c=0$.

From Lemma~\ref{lemma:V-quadrature}(i) we see that the invariant solution is
trivial, since the only root of $F$ is $|V|=0$.
Thus ODE \eqref{Trans-ODE} has no non-trivial dilation-invariant solution.

Next, we can obtain the general solution for $V(z)$
from the quadratures \eqref{V-quadrature},
since the conditions of Lemma~\ref{lemma:V-quadrature}(ii)
are directly satisfied for the ODE \eqref{Trans-ODE-dil-coords}.
In polar variables $V=A\exp(\i\Phi)$, these quadratures
are given by
\begin{equation}\label{Trans-ODE-dil-quadrature-A}
\int\frac{dA}{\sqrt{H(A)}} = \pm z +C_3 ,
\quad
\Phi= C_2 \int\frac{dA}{A^2 \sqrt{H(A)}} + C_4,
\end{equation}
where
\begin{equation}
H(A) = C_1 - C_2^2 A^{-2} -2k(1+p/2)^2\int A^{1+p}dA.
\end{equation}
These integrals \eqref{Trans-ODE-dil-quadrature-A}
cannot be evaluated generally to obtain explicit solutions
for $A(z)$ and $\Phi(z)$ in terms of elementary functions.
Some special cases where explicit solutions can be derived are possible
if we change variables \eqref{Trans-ODE-A-quadrature-y}--\eqref{Trans-ODE-Phi-quadrature-y}
to make the expressions
\begin{align}
&
H_1(y) = y^{2-2/s} H(y^{1/s})
= C_1 y^{2-2/s} - C_2^2  y^{2-4/s}  -\tilde k y^{2+p/s},
\label{Trans-ODE-dil-denom-A}\\
&
H_2(y) = y^{2+2/\tilde s} H(y^{1/\tilde s})
= C_1 y^{2+2/\tilde s} - C_2^2  y^{2} -\tilde k y^{2+(4+p)/\tilde s}
\quad\text{ for } C_2\neq 0
\label{Trans-ODE-dil-denom-Phi}
\end{align}
into quadratic polynomials in $y$ for some values of $s$ and $\tilde s$,
depending on $p,C_1,C_2$,
where
\begin{equation}
\tilde k=2k/(p+2),
\quad
p = 2(3-n)/(n-2) \neq -2.
\end{equation}
The required conditions from expressions \eqref{Trans-ODE-dil-denom-A}
and \eqref{Trans-ODE-dil-denom-Phi} are respectively given by
\eqref{Trans-cond1}, \eqref{Trans-cond2}, \eqref{Trans-cond3}, \eqref{Trans-cond5}
and
\begin{equation}
2+(4+p)/\tilde s =0,1,2
\quad\text{ if } C_2\neq 0.
\label{Trans-cond6}
\end{equation}
Solving these conditions for $p,s,\tilde s$
in the same way as for the previous case,
we find
\begin{align}
& C_1=0 \text{ and } C_2=0:
\quad
p\neq -2, s=-p/2;
\label{Trans-ODE-dil-y-quadratures-a}\\
& C_1\neq0 \text{ and } C_2=0:
\quad
p=-1, s=1;
\quad
p=-4, s=2;
\label{Trans-ODE-dil-y-quadratures-b}\\
& C_1=0 \text{ and } C_2\neq0:
\quad
p=-4, s=2, \tilde s=-1;
\quad
p=-8, s=4, \tilde s=2;
\label{Trans-ODE-dil-y-quadratures-c}\\
& C_1\neq0 \text{ and } C_2\neq0:
\quad
p=-4, s=2, \tilde s=-1;
\label{Trans-ODE-dil-y-quadratures-d}
\end{align}
with $n = 2(p+3)/(p+2)$.

We can then match the integrals \eqref{Trans-ODE-A-quadrature-y}--\eqref{Trans-ODE-Phi-quadrature-y}
to one of the forms \eqref{special-integrals-cosh}--\eqref{special-integrals-lin}
after scaling and shifting $y$.

As noted previously, additional explicit solutions for $A(z)$ and $\Phi(z)$
can be derived by requiring that
the expressions \eqref{Trans-ODE-dil-denom-A}--\eqref{Trans-ODE-dil-denom-Phi}
are either quartic polynomials in $y$,
which will yield elliptic functions,
or squares of quartic polynomials in $y$,
which will yield elementary functions.
(These solutions will be worked out elsewhere \cite{AncFen}.)

\subsubsection{Quadrature for $p\neq -2$}

From case \eqref{Trans-ODE-dil-y-quadratures-a},
we have $s=-p/2$, $C_1=0$ and $C_2=0$.
The quadratures \eqref{V-quadrature} are thus given by
\begin{equation}
\Phi = C_4
\end{equation}
and
\begin{equation}
\pm z +C_3 =
-(2/p) \int\frac{dy}{\sqrt{-k(p + 2)/2}} ,
\quad
y=A^{-p/2},
\end{equation}
which can be matched to the form \eqref{special-integrals-lin}.
This yields
\begin{equation}
A^{-p/2} = (-k p^2 (p + 2)/8)^{1/2}(\pm z + C_3),
\quad
\Phi = C_4,
\quad
k(p+2)<0.
\end{equation}
Hence we obtain a solution of ODE \eqref{Trans-ODE-dil-coords}
for $p\neq -2$:
\begin{equation}\label{Trans-ODE-dil-V-a-sol}
V =
(-kp^2(p + 2)/8)^{-1/p}(\pm z + C_3)^{-2/p}\exp(\i C_4),
\quad
k(p + 2)<0.
\end{equation}
This solution \eqref{Trans-ODE-dil-V-a-sol} gives the
following result after we change variables \eqref{Trans+Phas-dil-zV}.

\begin{prop}\label{prop:Trans-ODE-dil-U-case-a}
For $p \neq -2$, 
the translation-group ODE \eqref{Trans-ODE} has a solution
\begin{equation}\label{Trans-ODE-dil-U-a-sol}
\begin{aligned}
& U =
(-kp^2(p + 2)/8)^{-1/p}(\xi + \tilde C_3\xi^{3-n})^{-2/p}\exp(\i C_4),
\\&
\quad
k(p + 2)<0,
\quad
n = 2(p + 3)/(p + 2), 
\end{aligned}
\end{equation}
with real constants $\tilde C_3$, $C_4$.
\end{prop}

\subsubsection{Quadrature for $p=-1$}

From case \eqref{Trans-ODE-dil-y-quadratures-b},
we have $s=1$, $C_1\neq0$ and $C_2=0$.
The quadratures \eqref{V-quadrature} are thus given by
\begin{equation}
\Phi = C_4
\end{equation}
and
\begin{equation}
\pm z +C_3 =
\int\frac{dy}{\sqrt{C_1 - (k/2)y}} ,
\quad
y=A,
\end{equation}
which can be matched to the form \eqref{special-integrals-sqrt}.
This yields
\begin{equation}
A = (2C_1/k) - (k/8)(\pm z + C_3)^2 ,
\quad
\Phi = C_4.
\end{equation}
Hence we obtain a solution of ODE \eqref{Trans-ODE-dil-coords}
for $p = -1$:
\begin{equation}\label{Trans-ODE-dil-V-bb-sol}
V =
\left((2C_1/k) - (k/8)(\pm z + C_3)^2\right)\exp(\i C_4).
\end{equation}
This solution \eqref{Trans-ODE-dil-V-bb-sol} gives the
following result after we change variables \eqref{Trans+Phas-dil-zV}.

\begin{prop}\label{prop:Trans-ODE-dil-U-case-bb}
For $p=-1$, 
the translation-group ODE \eqref{Trans-ODE} has a solution
\begin{equation}\label{Trans-ODE-dil-U-bb-sol}
U =
\left((2C_1/k)\xi^{-2} - (k/8)(\xi + \tilde C_3/\xi)^2\right)\exp(\i C_4),
\quad 
n=4,
\end{equation}
with real constants $C_1$, $\tilde C_3$, $C_4$.
\end{prop}

\subsubsection{Quadrature for $p=-4$}

From cases 
\eqref{Trans-ODE-dil-y-quadratures-b}--\eqref{Trans-ODE-dil-y-quadratures-d},
we have
$s=2$, $\tilde s=-1$ and $C_2\neq0$.

The quadratures \eqref{V-quadrature} are thus given by
\begin{equation}\label{Trans-ODE-dil-z-quadr-e}
\pm z +C_3 =
(1/2)\int\frac{dy}{\sqrt{C_1 y + k - C_2^2}} ,
\quad
y=A^2,
\end{equation}
which can be matched to the forms
\eqref{special-integrals-sqrt}--\eqref{special-integrals-lin},
and
\begin{equation}\label{Trans-ODE-dil-Phi-quadr-e}
\Phi =
C_4 - C_2\int\frac{dy}{\sqrt{C_1 + (k - C_2^2)y^2}} ,
\quad
y=A^{-1},
\end{equation}
which can be matched to all of the forms
\eqref{special-integrals-cosh}--\eqref{special-integrals-sin}.
This yields
\begin{subequations}
\begin{align}
&
A^2 = (C_2^2 - k)/C_1 + C_1(\pm z + C_3)^2,
\quad
C_1\neq 0,
\label{Trans-ODE-dil-A-e-sol}
\\
&
A^2 = 2(k - C_2^2)^{1/2}(\pm z + C_3) ,
\quad
C_2^2 < k, C_1 = 0, 
\label{Trans-ODE-dil-A-c-sol}
\end{align}
\end{subequations}
and
\begin{subequations}
\begin{align}
&\begin{aligned}
\Phi = &
C_4 - C_2(k - C_2^2)^{-1/2}\arcsinh\left(((k - C_2^2)/C_1)^{1/2}A^{-1}\right),
\quad
C_2^2 < k, C_1 > 0,
\end{aligned}
\\
& \begin{aligned}
\Phi = &
C_4 - C_2(C_2^2 - k)^{-1/2}\arcsin\left(((C_2^2 - k)/C_1)^{1/2}A^{-1}\right),
\quad
C_2^2 >k, C_1>0,
\end{aligned}
\\
&\begin{aligned}
\Phi = &
C_4 - C_2(k - C_2^2)^{-1/2}\arccosh\left(((C_2^2 - k)/C_1)^{1/2}A^{-1}\right),
\quad
C_2^2 < k, C_1 < 0,
\end{aligned}
\\
&\begin{aligned}
\Phi = &
C_4 - (C_2/C_1^{1/2})A^{-1} ,
\quad
C_2^2 = k, C_1>0,
\end{aligned}
\\
&\begin{aligned}
\Phi = &
C_4 - C_2 (k - C_2^2)^{-1/2} \ln |A^{-1}|,
\quad
C_2^2 < k, C_1 = 0.
\end{aligned}
\end{align}
\end{subequations}
Hence we obtain 5 more solutions of ODE \eqref{Trans-ODE-scal-special-p}
for $p=-4$:
\begin{subequations}
\begin{equation}
\begin{aligned}
V =& \left((C_2^2 - k)/C_1 + C_1(\pm z + C_3)^2\right)^{1/2}
\\&\times
\exp\bigg(\i C_4 - \i C_2(k - C_2^2)^{-1/2}
\arcsinh\Big((-1 + C_1^2(\pm z + C_3)^2/(k - C_2^2))^{-1/2}\Big)\bigg) ,
\\&
\quad
C_2^2 < k, C_1>0,
\end{aligned}
\label{Trans-ODE-dil-V-e-sol1}
\end{equation}
\begin{equation}
\begin{aligned}
V =& \left((C_2^2 - k)/C_1 + C_1(\pm z + C_3)^2\right)^{1/2}
\\&\times
\exp\bigg(\i C_4 - \i C_2(C_2^2 - k)^{-1/2}
\arcsin\Big((1 + C_1^2(\pm z + C_3)^2/(C_2^2 - k))^{-1/2}\Big)\bigg) ,
\\&
\quad
C_2^2 >k, C_1>0,
\end{aligned}
\label{Trans-ODE-dil-V-e-sol2}
\end{equation}
\begin{equation}
\begin{aligned}
V = & \left((C_2^2 - k)/C_1 + C_1(\pm z + C_3)^2\right)^{1/2}
\\&\times
\exp\bigg(\i C_4 - \i C_2(k - C_2^2)^{-1/2}
\arccosh\Big((1 + C_1^2(\pm z + C_3)^2/(C_2^2 - k))^{-1/2}\Big)\bigg) ,
\\&
C_2^2 < k, C_1<0,
\end{aligned}
\label{Trans-ODE-dil-V-e-sol3}
\end{equation}
\begin{equation}
\begin{aligned}
V = & C_1^{1/2} |\pm z + C_3|\exp\left(\i C_4 \mp \i (k^{1/2}/C_1)|\pm z + C_3|^{-1}\right),
\quad
C_2^2=k, C_1>0,
\end{aligned}
\label{Trans-ODE-dil-V-e-sol4}
\end{equation}
\begin{equation}
\begin{aligned}
V = & (4(k - C_2^2))^{1/4}(\pm z + C_3)^{1/2}
\exp\left(\i \tilde C_4 + \tfrac{1}{2} \i C_2(k - C_2^2)^{-1/2}\ln |\pm z + C_3|\right) ,
\\&
\quad
C_2^2 < k, C_1 = 0.
\end{aligned}
\label{Trans-ODE-dil-V-c-sol}
\end{equation}
\end{subequations}

After we change variables \eqref{Trans+Phas-dil-zV},
these solutions \eqref{Trans-ODE-dil-V-e-sol1}--\eqref{Trans-ODE-dil-V-c-sol}
give the following result.

\begin{prop}
For $p=-4$,
the translation-group ODE \eqref{Trans-ODE} has solutions
\begin{equation}
\begin{aligned}
U =& \left((C_2^2 - k)\xi^2/C_1 + C_1(1 + \tilde C_3\xi)^2\right)^{1/2}
\\&\times
\exp\bigg(\i C_4 - \i C_2(k - C_2^2)^{-1/2}
\arcsinh\Big((-1 + C_1^2(\xi^{-1} + \tilde C_3)^2/(k - C_2^2))^{-1/2}\Big)\bigg) ,
\\&
\quad
C_2^2 < k,
\quad
C_1>0,
\quad
n=1,
\end{aligned}
\label{Trans-ODE-dil-U-e-sol1}
\end{equation}
\begin{equation}
\begin{aligned}
U = & \left((C_2^2 - k)\xi^2/C_1 + C_1(1 + \tilde C_3\xi)^2\right)^{1/2}
\\&\times
\exp\bigg(\i C_4 - \i C_2(C_2^2 - k)^{-1/2}
\arcsin\Big((1 + C_1^2(\xi^{-1} + \tilde C_3)^2/(C_2^2 - k))^{-1/2}\Big)\bigg) ,
\\&
\quad
C_2^2 > k,
\quad
C_1>0,
\quad
n=1,
\end{aligned}
\label{Trans-ODE-dil-U-e-sol2}
\end{equation}
\begin{equation}
\begin{aligned}
U =  & \left((C_2^2 - k)\xi^2/C_1 + C_1(1 + \tilde C_3\xi)^2\right)^{1/2}
\\&\times
\exp\bigg(\i C_4 - \i C_2(k - C_2^2)^{-1/2}
\arccosh\Big((1 + C_1^2(\xi^{-1} + \tilde C_3)^2/(C_2^2 - k))^{-1/2}\Big)\bigg) ,
\\&
\quad
C_2^2 < k,
\quad
C_1<0,
\quad
n=1,
\end{aligned}
\label{Trans-ODE-dil-U-e-sol3}
\end{equation}
\begin{equation}
\begin{aligned}
& U = (4(k - C_2^2))^{1/4} (C_3\xi^2 \pm \xi)^{1/2}
\exp\left(\i \tilde C_4 + \tfrac{1}{2} \i C_2 (k - C_2^2)^{-1/2}\ln |C_3\pm \xi^{-1}|\right) ,
\\&
\quad
C_2^2 < k,
\quad
n=1,
\end{aligned}
\label{Trans-ODE-dil-U-c-sol}
\end{equation}
\begin{equation}
\begin{aligned}
U = C_1^{1/2} |1 + \tilde C_3\xi|
\exp\left(\i C_4 \mp \i (k^{1/2}/C_1)|\xi^{-1} + \tilde C_3|^{-1}\right) ,
\quad
C_1>0, 
\quad
n=1,
\end{aligned}
\label{Trans-ODE-dil-U-e-sol4}
\end{equation}
with real constants $C_1$, $C_2$, $\tilde C_3$, $C_4$, $\tilde C_4$.
\end{prop}

\subsubsection{Quadrature for $p=-8$}

From case \eqref{Trans-ODE-dil-y-quadratures-c}, we have
$s=4$, $\tilde s=2$, $C_1=0$ and $C_2\neq0$.
The quadratures \eqref{V-quadrature} are thus given by
\begin{equation}\label{Trans-ODE-dil-z-quadr-d}
\pm z +C_3 =
(1/4) \int\frac{dy}{\sqrt{3k - C_2^2 y}} ,
\quad
y=A^4
\end{equation}
and
\begin{equation}\label{Trans-ODE-dil-Phi-quadr-d}
\Phi =
C_4 + (C_2/2)\int\frac{dy}{\sqrt{3k - C_2^2 y^2}},
\quad
y=A^2.
\end{equation}
Quadrature \eqref{Trans-ODE-dil-z-quadr-d} can
be matched to the form \eqref{special-integrals-sqrt},
which yields
\begin{equation}\label{Trans-ODE-dil-A-d-sol}
A^4 = 3k/C_2^2 - 4C_2^2(\pm z + C_3)^2,
\quad
k>0.
\end{equation}
Quadrature \eqref{Trans-ODE-dil-Phi-quadr-d} can be matched to
the form \eqref{special-integrals-sin}.
This yields
\begin{equation}\label{Trans-ODE-dil-Phi-d-sol}
\Phi
= C_4 + (1/2)\arcsin\left((3k)^{-1/2} C_2 A^2\right),
\quad
k>0.
\end{equation}
Hence we obtain a solution of ODE \eqref{Trans-ODE-dil-coords}
for $p = -8$:
\begin{align}
V =
&
\left(3k/C_2^2 - 4C_2^2(\pm z + C_3)^2\right)^{1/4}
\exp\left(\i C_4 + \tfrac{1}{2} \i
\arcsin((1 - 4C_2^4(\pm z + C_3)^2/(3k))^{1/2})\right),
\nonumber\\&
k>0.
\label{Trans-ODE-dil-V-d-sol}
\end{align}
This solution \eqref{Trans-ODE-dil-V-d-sol} gives the following result
after we change variables \eqref{Trans+Phas-dil-zV}.

\begin{prop}\label{prop:Trans-ODE-dil-U-case-d}
For $p=-8$, 
the translation-group ODE \eqref{Trans-ODE} has a solution
\begin{equation}
\begin{aligned}
U = &
\left((3k/C_2^2) \xi^{4/3} - 4C_2^2(\xi^{1/3} + \tilde C_3\xi^{2/3})^2\right)^{1/4}
\\&\times
\exp\bigg(\i C_4 + \tfrac{1}{2} \i
\arcsin\Big((1 - 4C_2^4(\xi^{-1/3} + \tilde C_3)^2/(3k))^{1/2}\Big)\bigg),
\quad
k>0,
\quad
n=5/3,
\end{aligned}
\label{Trans-ODE-dil-U-d-sol}
\end{equation}
with real constants $C_2$, $\tilde C_3$, $C_4$.
\end{prop}

\subsection{Conditional-symmetry quadratures for the translation-group ODE}

From Proposition \ref{prop:hiddensymms:Trans+Phas},
the hidden conditional symmetries \eqref{symm3:Trans+Phas}--\eqref{symm6:Trans+Phas}
with $\nu=0$ can be naturally split up into two types:
symmetries \eqref{symm3:Trans+Phas}, \eqref{symm4:Trans+Phas}
hold for $p=-4$, $C_1\neq 0$ and have the form \eqref{A-symm}
to which Lemmas~\ref{lemma:B-firstintegr} and \ref{lemma:B-invsoln} 
can be applied;
symmetries \eqref{symm5:Trans+Phas}, \eqref{symm6:Trans+Phas}
hold for $p=1$, $C_1=0$ and have a slightly more general form
\begin{equation}\label{A-symm-shift}
\tilde\Y= \zeta(\xi)\partial/\partial\xi + (\Upsilon_1(\xi)A+\Upsilon_0(\xi))\partial/\partial A,
\end{equation}
for which the methods that underlie
Lemmas~\ref{lemma:B-firstintegr} and \ref{lemma:B-invsoln} still can be used.

\subsubsection{Quadratures for $p=-4$, $C_1\neq 0$}

Point symmetry \eqref{symm3:Trans+Phas} of
the level-set ODE \eqref{A-ODE:Trans+Phas}
has the canonical coordinates
\begin{align}\label{Trans+Phas-hidden3-zB}
z = (-3/4)\ln ({C_1}^2/k - \xi^{2/3}),
\quad
B = ({C_1}^2/k - \xi^{2/3})^{-3/4} A
\end{align}
with $n=4/3$, $\nu=0$.
Hence the Lagrangian \eqref{A-L:Trans+Phas} for ODE \eqref{A-ODE:Trans+Phas}
gets transformed into
\begin{equation}\label{Trans+Phas-hidden3-B-L}
\hat L = \exp(-(2/3)z)( -(1/2){B'}^2 -(1/6)B^2 +2kB^{-2} ),
\end{equation}
yielding the transformed ODE
\begin{equation}\label{Trans+Phas-hidden3-B-ODE}
\tfrac{1}{2}B'' - \tfrac{1}{3}B' - \tfrac{1}{6}B - 2kB^{-3} = 0.
\end{equation}
Since the coefficient of $B'$ is non-zero,
the reduction shown in Lemma \ref{lemma:B-firstintegr} 
cannot be applied to ODE \eqref{Trans+Phas-hidden3-B-ODE}.
Instead, from Lemma \ref{lemma:B-invsoln},
an invariant solution can be derived.
This yields
\begin{equation}\label{Trans+Phas-hidden3-B-invsol}
B = (-12k)^{1/4}.
\end{equation}
Substituting this solution into expressions \eqref{A-invsoln}--\eqref{Phi-invsoln}, we obtain
\begin{equation}\label{Trans+Phas-hidden3-APhi-invsol}
A =
(-12k)^{1/4}({C_1}^2/k - \xi^{2/3})^{3/4},
\quad
\Phi =
(\sqrt{3}C_1/2)(k\xi^{2/3} - {C_1}^2)^{-1/2} + C_2,
\end{equation}
which is the corresponding invariant solution of
the system \eqref{A-EL-ODE}--\eqref{Phi-EL-ODE}
with $p=-4$, $n=4/3$, $\nu=0$.

Point symmetry \eqref{symm4:Trans+Phas} of
the level-set ODE \eqref{A-ODE:Trans+Phas}
has the canonical coordinates
\begin{align}\label{Trans+Phas-hidden4-zB}
z = (1/(4k))\ln(\xi^2/({C_1}^2\xi^2 - k)),
\quad
B = \xi^{-1/2}({C_1}^2\xi^2 - k)^{-3/4} A
\end{align}
with $n=0$, $\nu=0$.
Hence the Lagrangian \eqref{A-L:Trans+Phas} for ODE \eqref{A-ODE:Trans+Phas}
gets transformed into
\begin{equation}\label{Trans+Phas-hidden4-B-L}
\hat L = \exp(-2k z)( (1/2){B'}^2 +(3k^2/2)B^2 -2B^{-2} ).
\end{equation}
This yields the transformed ODE
\begin{equation}\label{Trans+Phas-hidden4-B-ODE}
-\tfrac{1}{2}B'' + kB' + \tfrac{3}{2}k^2B + 2B^{-3}=0,
\end{equation}
whose coefficient of $B'$ is non-zero.
Again, while the reduction shown in Lemma \ref{lemma:B-firstintegr} 
cannot be applied,
an invariant solution to ODE \eqref{Trans+Phas-hidden4-B-ODE} can be derived
from Lemma \ref{lemma:B-invsoln},
yielding
\begin{equation}\label{Trans+Phas-hidden4-B-invsol}
B = (-3k^2/4)^{-1/4}.
\end{equation}
Substituting this solution into expressions \eqref{A-invsoln}--\eqref{Phi-invsoln}, we obtain
\begin{equation}\label{Trans+Phas-hidden4-APhi-invsol}
A =
(3k^2/4)^{-1/4}\xi^{1/2}(k - {C_1}^2\xi^2)^{3/4},
\quad
\Phi =
-(\sqrt{3}C_1/2)(k\xi^{-2} - {C_1}^2)^{-1/2} + C_2,
\end{equation}
which is the corresponding invariant solution of
the system \eqref{A-EL-ODE}--\eqref{Phi-EL-ODE}
with $p=-4$, $n=0$, $\nu=0$.

These two solutions \eqref{Trans+Phas-hidden3-APhi-invsol}
and \eqref{Trans+Phas-hidden4-APhi-invsol}
give the following result.

\begin{prop}
For $p=-4$,
the translation-group ODE \eqref{Trans-ODE} has solutions
\begin{equation}\label{Trans+Phas-hidden-U-sol-a}
\begin{aligned}
& U = 
(-12k)^{1/4} (\tilde C_1/k - \xi^{2/3})^{3/4}
\exp\big(\i(\sqrt{3}/2)((k/\tilde C_1)\xi^{2/3} - 1)^{-1/2} + \i C_2\big),
\\&
\quad
k<0,
\quad
n=4/3, 
\end{aligned}
\end{equation}
\begin{equation}\label{Trans+Phas-hidden-U-sol-b}
\begin{aligned}
& U =
(4k/3)^{1/4}\xi^{2}( \xi^{-2} - \tilde C_1/k )^{3/4}
\exp\big(-\i(\sqrt{3}/2)((k/\tilde C_1)\xi^{-2} - 1)^{-1/2} + \i C_2\big),
\\&
\quad
n=0,
\end{aligned}
\end{equation}
with real constants $\tilde C_1$, $C_2$.
\end{prop}

\subsubsection{Quadratures for $p=1$, $C_1=0$}

Point symmetry \eqref{symm5:Trans+Phas} of the level-set ODE \eqref{A-ODE:Trans+Phas}
has the canonical coordinates
\begin{align}\label{Trans+Phas-hidden5-zB}
z = -(1/2)\xi^{-2},
\quad
B = \xi^6 A - (24/k)\xi^4
\end{align}
with $n=16$, $C_1=0$, $\nu=0$.
Hence the ODE \eqref{A-ODE:Trans+Phas} gets transformed into
\begin{align}\label{Trans+Phas-hidden5-B-ODE}
B'' + kB^2 = 0.
\end{align}
This ODE \eqref{Trans+Phas-hidden5-B-ODE} admits the integrating factor $B'$,
leading to the quadrature
\begin{equation}\label{B-quadrature-hidden5}
\int\frac{dB}{\sqrt{C_2 - (2k/3)B^3}} = \pm z + C_3,
\end{equation}
which is an elliptic integral when $C_2\neq 0$.
(This solution will be worked out in detail elsewhere \cite{AncFen}.)
In the case $C_2=0$, the integral \eqref{B-quadrature-hidden5}
can be evaluated explicitly, yielding
\begin{equation}\label{B-hidden5}
B = (-6/k)(\pm z + C_3)^{-2}.
\end{equation}
Hence, from the canonical coordinates \eqref{Trans+Phas-hidden5-zB}
and the first integral \eqref{Phi-integral},
we obtain
\begin{equation}\label{Trans+Phas-hidden5-APhi-sol}
A = (96C_3/k)(C_3\xi^2 \pm 1)/(2C_3\xi^2 \pm 1)^2,
\quad
\Phi=C_4,
\end{equation}
which is the solution of the system \eqref{A-EL-ODE}--\eqref{Phi-EL-ODE}
with $p=1$, $n=16$, $\nu=0$, and $C_1=C_2=0$.
Note that the invariant solution of the ODE \eqref{Trans+Phas-hidden5-B-ODE},
given by $B'=0$, is simply $B=0$.
This solution also arises as the special limit $C_3\rightarrow\infty$
in the solution \eqref{B-hidden5}.
The corresponding invariant solution of system \eqref{A-EL-ODE}--\eqref{Phi-EL-ODE}
is given by
\begin{equation}\label{Trans+Phas-hidden5-APhi-invsol}
A = 24/(k\xi^2), 
\quad
\Phi = C_4,
\end{equation}
which is non-trivial due to the shift term in the change of variables
\eqref{Trans+Phas-hidden5-zB} for $B$ in terms of $A$.

Point symmetry \eqref{symm6:Trans+Phas} of the level-set ODE \eqref{A-ODE:Trans+Phas}
has the canonical coordinates
\begin{align}\label{Trans+Phas-hidden6-zB}
z = (1/3)\xi^{1/3},
\quad
B = \xi^{4/3}A - (2/(3k))\xi^{-2/3}
\end{align}
with $n=13/3$, $\nu=0$, $C_1=0$.
Hence the ODE \eqref{A-ODE:Trans+Phas} gets transformed into
\begin{align}\label{Trans+Phas-hidden6-B-ODE}
\tfrac{1}{81}B'' + k B^2 = 0.
\end{align}
This ODE \eqref{Trans+Phas-hidden6-B-ODE} admits the integrating factor $B'$,
leading to the quadrature
\begin{equation}\label{B-quadrature-d}
\int\frac{dB}{\sqrt{81C_2 - 54kB^3}} = \pm z + C_3.
\end{equation}
When $C_2\neq 0$, this quadrature is an elliptic integral.
(The resulting solution will be worked out in detail elsewhere \cite{AncFen}.)
In the case $C_2=0$, the integral \eqref{B-quadrature-d}
can be evaluated explicitly,
yielding
\begin{equation}\label{B-hidden6}
B = -(2/27k)(\pm z + C_3)^{-2}.
\end{equation}
The canonical coordinates \eqref{Trans+Phas-hidden5-zB}
and the first integral \eqref{Phi-integral} then give
\begin{equation}\label{Trans+Phas-hidden6-APhi-sol}
A = (2C_3/k)\xi^{-2}(3C_3 \pm 2\xi^{1/3})/(3C_3 \pm \xi^{1/3})^2,
\quad
\Phi=C_4,
\end{equation}
which is the solution of the system \eqref{A-EL-ODE}--\eqref{Phi-EL-ODE} with $p=1$, $n=13/3$, $\nu=0$, and $C_1=C_2=0$.
Note that the invariant solution of the ODE \eqref{Trans+Phas-hidden6-B-ODE}
is again $B=0$ which also arises as the special limit $C_3\rightarrow\infty$
in the solution \eqref{B-hidden6}.
The corresponding invariant solution of system \eqref{A-EL-ODE}--\eqref{Phi-EL-ODE}
is given by
\begin{equation}\label{Trans+Phas-hidden6-APhi-invsol}
A = 2/(3k\xi^2), 
\quad
\Phi = C_4.
\end{equation}

The solutions
\eqref{Trans+Phas-hidden5-APhi-sol}, \eqref{Trans+Phas-hidden5-APhi-invsol},
\eqref{Trans+Phas-hidden6-APhi-sol}, \eqref{Trans+Phas-hidden6-APhi-invsol}
give the following result.

\begin{prop}\label{prop:Trans-ODE-hidden-U-case-b}
For $p=1$, 
the translation-group ODE \eqref{Trans-ODE} has solutions
\begin{equation}
\begin{aligned}
U = &(24/k)\xi^{-2}\exp(\i C_4), 
\quad 
n=16,
\end{aligned}
\label{Trans+Phas-hidden-U-c-sol1}
\end{equation}
\begin{equation}
\begin{aligned}
U = &(96/k)(\xi^2+ \tilde C_3)(2\xi^2 + \tilde C_3)^{-2}\exp(\i C_4), 
\quad 
n=16,
\end{aligned}
\label{Trans+Phas-hidden-U-c-sol2}
\end{equation}
\begin{equation}
\begin{aligned}
U =& (2/(3k))\xi^{-2}\exp(\i C_4), 
\quad 
n=13/3,
\end{aligned}
\label{Trans+Phas-hidden-U-d-sol1}
\end{equation}
\begin{equation}
\begin{aligned}
U = &(2/k)\xi^{-2}(2\tilde C_3\xi^{1/3} + 3)(\tilde C_3\xi^{1/3} + 3)^{-2}\exp(\i C_4), 
\quad 
n=13/3,
\end{aligned}
\label{Trans+Phas-hidden-U-d-sol2}
\end{equation}
with real constants $\tilde C_3$, $C_4$.
\end{prop}

\subsection{Linearization of the translation-group ODE}

When $C_1=0$ and $p=-1$,
the level-set ODE \eqref{A-ODE:Trans+Phas} for $A(\xi)$ becomes linear,
\begin{equation}\label{Trans+Phas-linear}
A'' + (n-1)\xi^{-1} A' + \nu A + k =0.
\end{equation}
The general solution of this ODE splits into three cases.

For the case $\nu> 0$,
we can use a change of variable
$A=\xi^{1-n/2} B$, $z=\sqrt{\nu}\xi$
which transforms the ODE \eqref{Trans+Phas-linear}
to an inhomogeneous form of Bessel's equation
\begin{equation}
z^2 B''+ z B'+( z^2 - (\tfrac{1}{2}n-1)^2 )B + k\nu^{-(2+n)/4} z^{1+n/2}=0.
\end{equation}
This equation has the general solution
\begin{equation}\label{Trans+Phas-linear-sol-B-JY}
B = C_2 J_{|n-2|/2}(z) + C_3 Y_{|n-2|/2}(z)- k\nu^{-(2+n)/4} z^{|n-2|/2},
\end{equation}
where $J_{\mu}(z)$ and  $Y_{\mu}(z)$ are
the Bessel functions of the first and second kinds, respectively.
Hence the general solution of ODE \eqref{Trans+Phas-linear} for $A(\xi)$
with $\nu\neq0$ is given by
\begin{equation}\label{Trans+Phas-linear-sol-A-JY}
A = \xi^{1-n/2} (C_2 J_{|n-2|/2}(\sqrt{\nu}\xi) + C_3 Y_{|n-2|/2}(\sqrt{\nu}\xi)) - k/\nu.
\end{equation}
For the case $\nu< 0$,
we can use a change of variable
$A=\xi^{1-n/2} B$, $z=\sqrt{-\nu}\xi$
which transforms the ODE \eqref{Trans+Phas-linear}
to an inhomogeneous form of a modified Bessel equation
\begin{equation}
z^2 B''+ z B' - ( z^2 + (\tfrac{1}{2}n-1)^2 )B + k(-\nu)^{-(2+n)/4} z^{1+n/2}=0.
\end{equation}
This equation has the general solution
\begin{equation}\label{Trans+Phas-linear-sol-B-IK}
B = C_2 I_{|n-2|/2}(z) + C_3 K_{|n-2|/2}(z) + k(-\nu)^{-(2+n)/4} z^{|n-2|/2},
\end{equation}
where $I_{\mu}(z)$ and  $K_{\mu}(z)$ are
the modified Bessel functions of the first and second kinds, respectively.
Hence the general solution of ODE \eqref{Trans+Phas-linear} for $A(\xi)$
with $\nu\neq0$ is given by
\begin{equation}\label{Trans+Phas-linear-sol-A-IK}
A = \xi^{1-n/2} \left(C_2 I_{|n-2|/2}(\sqrt{-\nu}\xi) + C_3 K_{|n-2|/2}(\sqrt{-\nu}\xi)\right) - k/\nu.
\end{equation}
Since $C_1=0$, the first integral \eqref{Phi-integral} for $\Phi(\xi)$
reduces to $\Phi'=0$, which yields
\begin{equation}\label{Trans+Phas-linear-sol-Phi}
\Phi=\const.
\end{equation}

For the case $\nu=0$,
the ODE \eqref{Trans+Phas-linear} can be integrated directly
to get
\begin{align}
& A = -(k/(2n))\xi^2 + C_3 \xi^{2-n} + C_2,
\quad
n\neq 0,2,
\label{Trans-linear-sol1-A}
\\
& A = -(k/4) \xi^2 + C_3 \ln \xi + C_2,
\quad
n=2,
\label{Trans-linear-sol2-A}
\\
& A = (-k/2) \xi^2\ln \xi + C_3 \xi^2 + C_2,
\quad
n=0.
\label{Trans-linear-sol3-A}
\end{align}
From the first integral \eqref{Phi-integral} for $\Phi(\xi)$,
we again have
\begin{equation}\label{Trans-linear-sol-Phi}
\Phi=\const.
\end{equation}

Hence we obtain the following result.

\begin{prop}\label{prop:Trans+Phas-linear-U}
For $p=-1$,
the translation-group ODE \eqref{Trans+Phas-ODE} has solutions
\begin{equation}\label{Trans+Phas-linear-U-sol1}
\begin{aligned}
U = \big( \xi^{1-n/2} (C_2 J_{|n-2|/2}(\sqrt{\nu}\xi) + C_3 Y_{|n-2|/2}(\sqrt{\nu}\xi)) - k/\nu\big)\exp(\i C_4),
\quad
\nu> 0, 
\end{aligned}
\end{equation}
\begin{equation}\label{Trans+Phas-linear-U-sol2}
\begin{aligned}
U = \big( \xi^{1-n/2} (C_2 I_{|n-2|/2}(\sqrt{-\nu}\xi) + C_3 K_{|n-2|/2}(\sqrt{-\nu}\xi)) - k/\nu\big)\exp(\i C_4),
\quad
\nu< 0, 
\end{aligned}
\end{equation}
\begin{equation}\label{Trans-linear-U-sol1}
\begin{aligned}
U = \big( -(k/(2n))\xi^2 + C_3 \xi^{2-n} + C_2\big)\exp(\i C_4),
\quad 
\nu=0, 
\quad
n\neq 0,2
\end{aligned}
\end{equation}
\begin{equation}
\label{Trans-linear-U-sol2}
\begin{aligned}
U = \big(-(k/4) \xi^2 + C_3 \ln \xi + C_2\big)\exp(\i C_4), 
\quad
\nu =0, 
\quad
n= 2, 
\end{aligned}
\end{equation}
\begin{equation}\label{Trans-linear-U-sol3}
\begin{aligned}
U = \big((-k/2) \xi^2\ln \xi + C_3 \xi^2 + C_2\big)\exp(\i C_4), 
\quad
\nu=0, 
\quad
n= 0, 
\end{aligned}
\end{equation}
with real constants $C_2$, $C_3$, $C_4$.
\end{prop}

\section{\large Solutions to the optimal scaling-group ODE}
\label{scaling-group}

Next the reduction of order methods from section~\ref{methods}
will be applied to the scaling-group ODE \eqref{Scal+Phas-ODE}
arising by the reduction of the radial Schr\"odinger equation
under its optimal subgroup of point symmetries \eqref{optimal-ScalPhas}.

To proceed,
we use polar variables $U=A\exp(\i\Phi)$ to convert
this $\U(1)$-invariant ODE \eqref{Scal+Phas-ODE}
into a semilinear system of real ODEs
\begin{subequations}\label{APhi-ODEs:Scal+Phas}
\begin{align}
&
4\xi^2 A'' - 4\xi^2 A \Phi'^2 - 2(n-4-4/p)\xi A' +
(1+4\mu \xi) A\Phi' \\\nonumber&\qquad
- (\mu^2 -(4-2n)/p-4/p^2)A +kA^{1+p}=0,
\\
&
4\xi^2\Phi'' + 8\xi^2 A^{-1} A'\Phi' -(1+4\mu \xi)A^{-1} A'
- 2 (n-4-4/p) \xi \Phi' + \mu (n-2-4/p)=0.
\end{align}
\end{subequations}
As shown by the results stated in Proposition~\ref{prop:U-ODE-symms},
the point symmetries of this system \eqref{APhi-ODEs:Scal+Phas}
consist of only phase rotations
\begin{equation}
\Y_{\phas}  = \partial/\partial \Phi.
\label{APhi-phas:Scal+Phas}
\end{equation}

In the case of the pseudo-conformal power $p=4/n$,
the ODEs in the polar system \eqref{APhi-ODEs:Scal+Phas}
are the respective Euler--Lagrange equations
$\delta L/\delta A=0$ and $A^{-1}\delta L/\delta\Phi=0$
of the $\U(1)$-invariant Lagrangian \eqref{L:Scal+Phas}
expressed in polar variables,
\begin{equation}\label{APhi-L:Scal+Phas}
L =
-4 \xi^2 ( A'^2 + A^2(\Phi' -(\tfrac{1}{8}\xi^{-2}+\tfrac{1}{2}\mu\xi^{-1}))^2 )
+(\tfrac{1}{16}\xi^{-2}+\tfrac{1}{2}\mu\xi^{-1} +4\tfrac{p-1}{p^2})A^2
+ \tfrac{2}{p+2}kA^{p+2},
\quad
p=4/n.
\end{equation}
Invariance of the polar Lagrangian \eqref{APhi-L:Scal+Phas}
under $\Y_{\phas}$ produces a first integral \eqref{Phi-integral},
yielding
\begin{equation}\label{Phi-integral:Scal+Phas}
\Phi' = \tfrac{1}{8} \xi^{-2}(C_1 A^{-2}+1+4\mu\xi),
\quad
p=4/n.
\end{equation}
The polar system \eqref{APhi-ODEs:Scal+Phas} thereby reduces to
a single real semilinear ODE
\begin{equation}\label{A-ODE:Scal+Phas}
4\xi^2 A'' +8\xi A' - \tfrac{1}{4}C_1^2 \xi^{-2} A^{-3}
+(\tfrac{1}{16}\xi^{-2}+\tfrac{1}{2}\mu\xi^{-1} +4\tfrac{p-1}{p^2})A
+kA^{1+p}=0,
\quad
p=4/n,
\end{equation}
which is the Euler--Lagrange equation
$\delta \tilde L/\delta A=0$ of a modified Lagrangian \eqref{A-Lagr},
given by
\begin{equation}\label{A-L:Scal+Phas}
L = -4 \xi^2 A'^2 + \tfrac{1}{4}C_1^2 \xi^{-2} A^{-2}
+(\tfrac{1}{16}\xi^{-2}+\tfrac{1}{2}\mu\xi^{-1} +4\tfrac{p-1}{p^2})A^2
+ \tfrac{2}{p+2}kA^{p+2},
\quad
p=4/n.
\end{equation}
Solutions of the ODE \eqref{A-ODE:Scal+Phas} for $A(\xi)$
represent the level set $C_1=\const$ of solutions $(A(\xi),\Phi(\xi))$
to the polar system \eqref{APhi-ODEs:Scal+Phas},
or equivalently the level set
\begin{equation}\label{levelset:Scal+Phas}
C_1= 2\i \xi^2 (U\bar U'- U'\bar U) -(\tfrac{1}{2}+2\mu\xi)|U|^2
=\const
\end{equation}
of solutions $U(\xi)$ to the scaling-group ODE \eqref{Scal+Phas-ODE}.

Note that the level-set ODE \eqref{A-ODE:Scal+Phas} will be linear iff
$C_1=0$ and $p=-1$.

\begin{prop}\label{prop:hiddensymms:Scal+Phas}
In the nonlinear case $C_1 \neq 0$ or $p \neq -1$,
the level-set ODE \eqref{A-ODE:Scal+Phas} admits no point symmetries.
\end{prop}

Since both the level-set ODE \eqref{A-ODE:Scal+Phas}
and the polar system \eqref{APhi-ODEs:Scal+Phas}
have no point symmetries,
only the second reduction method (cf section~\ref{method-2}) is applicable
by using the linearization of ODE \eqref{A-ODE:Scal+Phas}
which holds when $C_1=0$, $p=-1$.

\subsection{Linearization of the scaling-group ODE}

In the case $C_1=0$ and $p=-1$ ($n=-4$),
the level-set ODE \eqref{A-ODE:Scal+Phas} for $A(\xi)$
becomes linear,
\begin{equation}\label{Scal+Phas-linear}
4\xi^2 A'' +8\xi A' + (\tfrac{1}{16}\xi^{-2}+\tfrac{1}{2}\mu\xi^{-1} -8)A + k=0.
\end{equation}
The general solution of this ODE is given by
\begin{equation}\label{Scal+Phas-linear-sol-A}
\begin{aligned}
A = &
C_2 M_{-\i \mu/2,3/2}(\tfrac{1}{4}\i\xi^{-1})+C_3 M_{-\i \mu/2,-3/2}(\tfrac{1}{4}\i\xi^{-1})
\\&\quad
+ (\i k/3)\big(
M_{-\i \mu/2,-3/2}(\tfrac{1}{4}\i\xi^{-1})\int M_{-\i \mu/2,3/2}(\tfrac{1}{4}\i\xi^{-1})\,d\xi
\\&\qquad
- M_{-\i \mu/2,3/2}(\tfrac{1}{4}\i\xi^{-1})\int M_{-\i \mu/2,-3/2}(\tfrac{1}{4}\i\xi^{-1})\,d\xi
\big),
\end{aligned}
\end{equation}
where $M_{\lambda,\nu}(z)$ is the Whittaker function.

Since $C_1=0$, the first integral \eqref{Phi-integral} for $\Phi(\xi)$
reduces to $\Phi'=\tfrac{1}{8}\xi^{-2}+\tfrac{1}{2}\mu\xi^{-1}$, which yields
\begin{equation}\label{Scal+Phas-linear-sol-Phi}
\Phi=-\tfrac{1}{8}\xi^{-1} + \tfrac{1}{2}\mu\ln \xi + C_4.
\end{equation}

Hence we obtain the following result.

\begin{prop}\label{prop:Scal+Phas-linear-U}
For $p=-1$, 
the scaling-group ODE \eqref{Scal+Phas-ODE} has a solution
\begin{equation}\label{Scal+Phas-linear-U-sol}
\begin{aligned}
U = & \Big(
C_2 M_{-\i \mu/2,3/2}(\tfrac{1}{4}\i\xi^{-1})+C_3 M_{-\i \mu/2,-3/2}(\tfrac{1}{4}\i\xi^{-1})
\\&\quad
+ (\i k/3)\big(
M_{-\i \mu/2,-3/2}(\tfrac{1}{4}\i\xi^{-1})\int M_{-\i \mu/2,3/2}(\tfrac{1}{4}\i\xi^{-1})\,d\xi
\\&\qquad
- M_{-\i \mu/2,3/2}(\tfrac{1}{4}\i\xi^{-1})\int M_{-\i \mu/2,-3/2}(\tfrac{1}{4}\i\xi^{-1})\,d\xi
\big) \Big)
\exp(-\tfrac{1}{8}\i \xi^{-1} + \tfrac{1}{2}\i \mu\ln \xi + \i C_4),
\\&
\quad
n=-4, 
\end{aligned}
\end{equation}
with real constants $C_2$, $C_3$, $C_4$.
\end{prop}

\section{\large Solutions to the optimal pseudo-conformal-group ODE}
\label{conformal-group}

The reduction of order methods from section~\ref{methods}
will be considered last for the pseudo-conformal-group ODE \eqref{Trans+Inver+Phas-ODE}
arising by the reduction of the radial Schr\"odinger equation
under its optimal subgroup of point symmetries \eqref{optimal-TransInverPhas}.

We begin by using polar variables $U=A\exp(\i\Phi)$ to convert
this $\U(1)$-invariant ODE \eqref{Trans+Inver+Phas-ODE}
into a semilinear system of real ODEs
\begin{subequations}\label{APhi-ODEs:Trans+Inver+Phas}
\begin{align}
&
4\xi^2 A'' - 4\xi^2 A\Phi'^2 + 8\xi A'
+ (\kappa\xi^{-1}-\tfrac{1}{4}\xi^{-2}+n(1-\tfrac{1}{4}n))A + kA^{1+4/n}=0,
\\
& \Phi'' + 2A^{-1}A'\Phi' + 2\xi^{-1}\Phi'=0.
\end{align}
\end{subequations}
The ODEs in this system \eqref{APhi-ODEs:Trans+Inver+Phas}
are the respective Euler--Lagrange equations
$\delta L/\delta A=0$ and $A^{-1}\delta L/\delta\Phi=0$
of the $\U(1)$-invariant Lagrangian \eqref{L:Trans+Inver+Phas}
expressed in polar variables,
\begin{equation}\label{APhi-L:Trans+Inver+Phas}
L =
-4\xi^2 A'^2 - 4\xi^2A^2\Phi'^2 +
\left(\kappa\xi^{-1} - \tfrac{1}{4}\xi^{-2} + n(1-\tfrac{1}{4}n)\right)A^2
+ \tfrac{n}{n+2}kA^{2+4/n}.
\end{equation}

As shown by the results stated in Proposition~\ref{prop:U-ODE-symms},
the point symmetries of the polar system \eqref{APhi-ODEs:Trans+Inver+Phas}
consist of only phase rotations
\begin{equation}
\Y_{\phas}  = \partial/\partial \Phi.
\label{APhi-phas:Trans+Inver+Phas}
\end{equation}
Invariance of the polar Lagrangian \eqref{APhi-L:Trans+Inver+Phas}
under $\Y_{\phas}$ produces a first integral \eqref{Phi-integral},
yielding
\begin{equation}\label{Phi-integral:Trans+Inver+Phas}
\Phi' = \tfrac{1}{4}C_1 \xi^{-2} A^{-2}.
\end{equation}
The polar system \eqref{APhi-ODEs:Trans+Inver+Phas} thereby reduces to
a single real semilinear ODE
\begin{equation}\label{A-ODE:Trans+Inver+Phas}
4\xi^2 A'' + 8\xi A' - \tfrac{1}{4}C_1^2\xi^{-2} A^{-3}
+\left(\kappa\xi^{-1} - \tfrac{1}{4}\xi^{-2} + n(1-\tfrac{1}{4}n)\right)A
+ kA^{1+4/n}=0,
\end{equation}
which is the Euler--Lagrange equation
$\delta \tilde L/\delta A=0$ of a modified Lagrangian \eqref{A-Lagr},
given by
\begin{equation}\label{A-L:Trans+Inver+Phas}
\tilde L = -4\xi^2A'^2 + \tfrac{1}{4}C_1^2\xi^{-2}A^{-2}
+\left(\kappa\xi^{-1} - \tfrac{1}{4}\xi^{-2} + n(1-\tfrac{1}{4}n)\right)A^2
+ \tfrac{n}{n+2}kA^{2+4/n}.
\end{equation}
Solutions of the ODE \eqref{A-ODE:Trans+Inver+Phas} for $A(\xi)$
represent the level set $C_1=\const$ of solutions $(A(\xi),\Phi(\xi))$
to the polar system \eqref{APhi-ODEs:Trans+Inver+Phas},
or equivalently the level set
\begin{equation}\label{levelset:Trans+Inver+Phas}
C_1= 2\i \xi^{-2} (U\bar U'- U'\bar U)
=\const
\end{equation}
of solutions $U(\xi)$ to the pseudo-conformal-group ODE \eqref{Trans+Inver+Phas-ODE}.

Note that the level-set ODE \eqref{A-ODE:Trans+Inver+Phas} will be linear iff
$C_1=0$ and $p=-1$ ($n=-4$).

\begin{prop}\label{prop:hiddensymms:Trans+Inver+Phas}
In the nonlinear case $C_1 \neq 0$ or $p \neq -1$ ($n\neq -4$),
the level-set ODE \eqref{A-ODE:Trans+Inver+Phas} admits no point symmetries.
\end{prop}

Since both the level-set ODE \eqref{A-ODE:Trans+Inver+Phas}
and the polar system \eqref{APhi-ODEs:Trans+Inver+Phas}
have no point symmetries,
only the second reduction method (cf section~\ref{method-2}) is applicable
by using the linearization of ODE \eqref{A-ODE:Trans+Inver+Phas}
which holds when $C_1=0$, $p=-1$ ($n=-4$).

\subsection{Linearization of the pseudo-conformal-group ODE}

In the case $C_1=0$ and $p=-1$ ($n=-4$),
the level-set ODE \eqref{A-ODE:Trans+Inver+Phas} for $A(\xi)$
becomes linear,
\begin{equation}\label{Trans+Inver+Phas-linear}
4\xi^2 A'' + 8\xi A' +( \kappa\xi^{-1} -\tfrac{1}{4}\xi^{-2} - 8 )A + k=0.
\end{equation}
The general solution of this ODE is given by
\begin{equation}\label{Trans+Inver+Phas-linear-sol-A}
\begin{aligned}
A = &
C_2 M_{\kappa/2,3/2}(\tfrac{1}{2}\xi^{-1})+C_3 M_{\kappa/2,-3/2}(\tfrac{1}{2}\xi^{-1})
\\&\quad
+(k/6)\big(
M_{\kappa/2,3/2}(\tfrac{1}{2}\xi^{-1})\int M_{\kappa/2,-3/2}(\tfrac{1}{2}\xi^{-1})\,d\xi
-M_{\kappa/2,-3/2}(\tfrac{1}{2}\xi^{-1})\int M_{\kappa/2,3/2}(\tfrac{1}{2}\xi^{-1})\,d\xi
\big),
\end{aligned}
\end{equation}
where $M_{\lambda,\nu}(z)$ is the Whittaker function.

Since $C_1=0$, the first integral \eqref{Phi-integral} for $\Phi(\xi)$
reduces to $\Phi'=0$, which yields
\begin{equation}\label{Trans+Inver+Phas-linear-sol-Phi}
\Phi=\const.
\end{equation}

Hence we obtain the following result.

\begin{prop}\label{prop:Trans+Inver+Phas-linear-U}
For $p=-1$, 
the pseudo-conformal-group ODE \eqref{Trans+Inver+Phas-ODE} has a solution
\begin{equation}\label{Trans+Inver+Phas-linear-U-sol}
\begin{aligned}
U = & \Big(
C_2 M_{\kappa/2,3/2}(\tfrac{1}{2}\xi^{-1})+C_3 M_{\kappa/2,-3/2}(\tfrac{1}{2}\xi^{-1})
\\&\quad
+(k/6)\big(
M_{\kappa/2,3/2}(\tfrac{1}{2}\xi^{-1})\int M_{\kappa/2,-3/2}(\tfrac{1}{2}\xi^{-1})\,d\xi
\\&\qquad
-M_{\kappa/2,-3/2}(\tfrac{1}{2}\xi^{-1})\int M_{\kappa/2,3/2}(\tfrac{1}{2}\xi^{-1})\,d\xi
\big) \Big)\exp(\i C_4),
\quad
n=-4, 
\end{aligned}
\end{equation}
with real constants $C_2$, $C_3$, $C_4$.
\end{prop}

\section{\large Group-invariant Radial Solutions}
\label{solns}

All of the group-invariant solutions derived in 
sections~\ref{translation-group}, ~\ref{scaling-group}, and~\ref{conformal-group}
through the optimal subgroups of point symmetries 
\eqref{optimal-TransPhas}, \eqref{optimal-ScalPhas}, and \eqref{optimal-TransInverPhas}
will now be written out in the form $u=f(t,r)$
for the radial Schr\"odinger equation \eqref{radial-nls} 
and its alternative two-dimensional formulation \eqref{2D-radial-nls}. 
Wherever possible, solutions are merged into families that do not overlap.

\subsection{Solutions from the optimal translation-symmetry subgroup}

\begin{thm}\label{thm:Trans-inv-solns}
The radial Schr\"odinger equations \eqref{radial-nls} and \eqref{2D-radial-nls}
have the following group-invariant solutions 
with respect to time translations \eqref{Trans-group}:
{\allowdisplaybreaks
\begin{align}
& \begin{aligned}
u=\big(2(p(n-2)-2)/(kp^2)\big)^{1/p} r^{-2/p}\exp(\i c_1),
\quad
p\neq 2/(n-2), k/(p(n-2)-2)>0,
\label{radial-trans-sol-0}
\end{aligned}\\
& \begin{aligned}
u = &
\left(\pm 8(p+2)/(kp^2)\right)^{1/p}(c_2 r^2 \pm 1/c_2)^{-2/p}\exp(\i c_1),
\\&
\quad
p=4/(n-2), \pm k(1-2/n)>0, n\neq 2,
\label{radial-trans-sol-1}
\end{aligned}\\
& \begin{aligned}
u = &
(-\tfrac{1}{8}kp^2(p + 2))^{-1/p}\big(r + c_2 r^{3-n}\big)^{-2/p}\exp(\i c_1),
\\&
\quad
p=2(n-3)/(2-n), k/(n-2)<0, n\neq 2,3,
\label{radial-trans-sol-16}
\end{aligned}\\
& \begin{aligned}
u =
\big( -\tfrac{1}{2}(k/n)r^2 + c_3 r^{2-n} + c_2\big)\exp(\i c_1),
\quad
p=-1, n\neq 0,2,
\label{radial-sol-31}
\end{aligned}\\
& \begin{aligned}
u =
\big(-\tfrac{1}{4}k r^2 + c_3 \ln r + c_2\big)\exp(\i c_1),
\quad
p=-1, n=2,
\label{radial-sol-32}
\end{aligned}\\
& \begin{aligned}
u =
(96/k)(r^2 + c_2)(2 r^2 + c_2)^{-2}\exp(\i c_1),
\quad
p=1,n=16,
\label{radial-sol-27}
\end{aligned}\\
& \begin{aligned}
u =
\big( -\tfrac{1}{2}k r^2\ln r + c_3 r^2 + c_2\big)\exp(\i c_1),
\quad
p=-1, n=0,
\label{radial-sol-33}
\end{aligned}\\
& \begin{aligned}
u = &
(4k/3)^{1/4}r^2 \big(r^{-2} - c_2/k\big)^{3/4}
\exp\big(\i c_1 - \i\tfrac{\sqrt{3}}{2}((k/c_2)r^{-2} - 1)^{-1/2}\big),
\\&
\quad
p=-4,n=0,k>0,
\label{radial-sol-25}
\end{aligned}\\
& \begin{aligned}
u = &
(-12k)^{1/4} (c_2/k - r^{2/3})^{3/4}
\exp\big(\i c_1 + \i\tfrac{\sqrt{3}}{2}((k/c_2)r^{2/3} - 1)^{-1/2}\big),
\\&
\quad
p=-4,n=4/3,k<0,
\label{radial-sol-24}
\end{aligned}\\
& \begin{aligned}
u = &
\big( (3k/{c_3}^2) r^{4/3} - 4{c_3}^2(r^{1/3} + c_2 r^{2/3})^2 \big)^{1/4}
\\&\qquad\times
\exp\bigg(\i c_1 + \tfrac{1}{2} \i\arcsin\Big(\big(1 - \tfrac{4}{3}({c_3}^4/k)(r^{-1/3} + c_2)^2\big)^{1/2}\Big)\bigg),
\\&
\quad
p=-8, n=5/3,k>0,
\label{radial-trans-sol-23}
\end{aligned}\\
& \begin{aligned}
u =
(2/k)r^{-2}(2c_2r^{1/3} + 3)(c_2 r^{1/3} + 3)^{-2}\exp(\i c_1),
\quad
p=1,n=13/3,
\label{radial-sol-29}
\end{aligned}\\
& \begin{aligned}
u =
\big(-2c_3 r + (\pm(k - {c_3}^2))^{1/2}(c_2 r^2 \mp 1/c_2)\big)^{1/2}\exp(\i c_1),
\quad
p=-4, n=1, {c_3}^2\neq k,
\label{radial-trans-sol-5}
\end{aligned}\\
& \begin{aligned}
u = &
\big(-2c_4 r + (k - {c_3}^2 - {c_4}^2)^{1/2}(c_2 r^2 - 1/c_2)\big)^{1/2}
\\&\qquad\times
\exp\bigg(\i c_1 - \tfrac{1}{2}\i (k/{c_3}^2 -1)^{-1/2}
\arcsinh\Big(\frac{2(k - {c_3}^2 - {c_4}^2)^{1/2}r + c_4(c_2 r^2 -1/c_2 )}{-2c_4 r + (k - {c_3}^2 - {c_4}^2)^{1/2}(c_2 r^2 -1/c_2)}\Big)\bigg) ,
\\&
\quad
p=-4, n=1, {c_3}^2 + {c_4}^2 < k,
\label{radial-trans-sol-8}
\end{aligned}\\
& \begin{aligned}
u = &
\big(-2c_4 r + ({c_3}^2 + {c_4}^2 - k\big)^{1/2}
(c_2 r^2 + 1/c_2))^{1/2}
\\&\qquad\times
\exp\bigg(\i c_1 - \tfrac{1}{2}\i (k/{c_3}^2 - 1)^{-1/2}
\arccosh\Big(\frac{2({c_3}^2 + {c_4}^2 - k)^{1/2}r - c_4 (c_2 r^2 + 1/c_2)}
{2c_4 r - ({c_3}^2 + {c_4}^2 - k)^{1/2}(c_2 r^2 + 1/c_2)}\Big)\bigg) ,
\\&
\quad
p=-4, n=1, {c_3}^2 < k < {c_3}^2 + {c_4}^2,
\label{radial-trans-sol-9}
\end{aligned}\\
& \begin{aligned}
u = &
\big(-2c_4 r + ({c_3}^2 + {c_4}^2 - k)^{1/2}(c_2 r^2 + 1/c_2)\big)^{1/2}
\\&\qquad\times
\exp\bigg(\i c_1 - \tfrac{1}{2}\i (1 - k/{c_3}^2)^{-1/2}
\arcsin\Big(\frac{2({c_3}^2 + {c_4}^2 - k)^{1/2}r - c_4(c_2 r^2 + 1/c_2)}
{-2c_4 r + ({c_3}^2 + {c_4}^2 - k)^{1/2}(c_2 r^2 + 1/c_2 )}\Big)\bigg) ,
\\&
\quad
p=-4, n=1, {c_3}^2 > k,
\label{radial-trans-sol-10}
\end{aligned}\\
& \begin{aligned}
u = &
(-2c_3 r + c_2 r^2)^{1/2}
\exp\big(\i c_1 \pm\i\tfrac{1}{2}(k/{c_3}^2-1)^{1/2}\ln|c_2-2c_3/r|\big) ,
\quad
p=-4, n=1,
\label{radial-trans-sol-11}
\end{aligned}\\
& \begin{aligned}
u = &
(-2c_3 r + c_2)^{1/2}
\exp\big(\i c_1 \pm \i\tfrac{1}{2}(k/{c_3}^2-1)^{1/2}\ln|c_2-2c_3 r|\big) ,
\quad
p=-4, n=1,
\label{radial-trans-sol-12}
\end{aligned}\\
& \begin{aligned}
u =
(c_3/c_2)^{1/2} (c_2 r - 1)
\exp\Big(\i c_1 - \tfrac{1}{2}\i(k^{1/2}/c_3)
\Big(\frac{c_2 r + 1}{c_2 r - 1}\Big)\Big) ,
\quad
p=-4, n=1, k>0,
\label{radial-trans-sol-13}
\end{aligned}\\
& \begin{aligned}
u =
c_2 r\exp\big(\i c_1 - \i (k^{1/2}/{c_2}^2) r^{-1}\big) ,
\quad
p=-4, n=1, k>0,
\label{radial-trans-sol-14}
\end{aligned}\\
& \begin{aligned}
u =
c_2 \exp\big(\i c_1 - \i (k^{1/2} /{c_2}^2) r\big) ,
\quad
p=-4, n=1, k>0,
\label{radial-trans-sol-15}
\end{aligned}\\
& \begin{aligned}
u = &
\big((({c_3}^2 - k)/c_4)r^2 + c_4(1 + c_2 r)^2\big)^{1/2}
\\&\qquad\times
\exp\bigg(\i c_1 - \i c_3(k - {c_3}^2)^{-1/2}
\arcsinh\Big(\big(-1 + ({c_4}^2/(k - {c_3}^2))(r^{-1} + c_2)^2\big)^{-1/2}\Big)\bigg) ,
\\&
\quad
p=-4, n=1, {c_3}^2 < k, c_4>0,
\label{radial-trans-sol-18}
\end{aligned}\\
& \begin{aligned}
u = &
\big((({c_3}^2 - k)/c_4)r^2 + c_4(1 + c_2 r)^2\big)^{1/2}
\\&\qquad\times
\exp\bigg(\i c_1 - \i c_3({c_3}^2 - k)^{-1/2}
\arcsin\Big(\big(1 + ({c_4}^2/({c_3}^2 - k))(r^{-1} + c_2)^2\big)^{-1/2}\Big)\bigg) ,
\\&
\quad
p=-4, n=1, {c_3}^2 > k, c_4>0,
\label{radial-trans-sol-19}
\end{aligned}\\
& \begin{aligned}
u = &
\big((({c_3}^2 - k)/c_4)r^2 + c_4(1 + c_2 r)^2\big)^{1/2}
\\&\qquad\times
\exp\bigg(\i c_1 - \i c_3(k - {c_3}^2)^{-1/2}
\arccosh\Big(\big(1 + ({c_4}^2/({c_3}^2 - k))(r^{-1} + c_2)^2\big)^{-1/2}\Big)\bigg),
\\&
\quad
p=-4, n=1, {c_3}^2 < k, c_4<0,
\label{radial-trans-sol-20}
\end{aligned}\\
& \begin{aligned}
u = &
\big(4(k - {c_3}^2)\big)^{1/4} (c_2 r^2 \pm r)^{1/2}
\exp\Big(\i c_1 + \tfrac{1}{2} \i c_3 (k - {c_3}^2)^{-1/2}\ln |c_2 \pm r^{-1}|\Big),
\\&
\quad
p=-4, n=1, {c_3}^2 < k,
\label{radial-trans-sol-21}
\end{aligned}\\
& \begin{aligned}
u =
c_3 (1 + c_2 r) \exp\big(\i c_1 \mp \i (k^{1/2}/{c_3}^2)(r^{-1} + c_2)^{-1}\big),
\quad
p=-4, n=1.
\label{radial-trans-sol-22}
\end{aligned}
\end{align}
\endallowdisplaybreaks}
\end{thm}

\begin{thm}\label{thm:TransPhas-inv-solns}
The radial Schr\"odinger equations \eqref{radial-nls} and \eqref{2D-radial-nls}
have the following group-invariant solutions 
with respect to time translations combined with phase rotations \eqref{TransPhas-group}:
{\allowdisplaybreaks
\begin{align}
& \begin{aligned}
u =
\Big( r^{1-n/2}\big(c_2 J_{|n-2|/2}(\sqrt{\nu}r) + c_3 Y_{|n-2|/2}(\sqrt{\nu}r)\big) - k/\nu\Big)\exp(\i c_1+ \i\nu t),
\quad
p=-1,
\label{radial-sol-30}
\end{aligned}\\
& \begin{aligned}
u =
\Big( r^{1-n/2} \big(c_2 I_{|n-2|/2}(\sqrt{-\nu}r) + c_3 K_{|n-2|/2}(\sqrt{-\nu}r)\big) - k/\nu\Big)\exp(\i c_1+ \i\nu t),
\quad
p=-1.
\label{radial-sol-36}
\end{aligned}
\end{align}
\endallowdisplaybreaks}
\end{thm}

\subsection{Solutions from the optimal scaling-symmetry and pseudo-conformal-symmetry subgroups}

\begin{thm}\label{thm:ScalPhas-inv-solns}
The radial Schr\"odinger equations \eqref{radial-nls} and \eqref{2D-radial-nls}
have the following group-invariant solutions 
with respect to scalings combined with phase rotations \eqref{ScalPhas-group}:
{\allowdisplaybreaks
\begin{align}
& \begin{aligned}
u = &
\tfrac{1}{3}\i k r^2 \Big(
M_{-\i \mu/2,-3/2}(\tfrac{1}{4}\i r^2/t)\int^{t/r^2}_{c_2} M_{-\i \mu/2,3/2}(\tfrac{1}{4}\i\xi^{-1})\,d\xi
\\&\qquad\quad
- M_{-\i \mu/2,3/2}(\tfrac{1}{4}\i r^2/t)\int^{t/r^2}_{c_3} M_{-\i \mu/2,-3/2}(\tfrac{1}{4}\i\xi^{-1})\,d\xi
\Big)
\exp\left(\i c_1 - \tfrac{1}{8}\i (r^2/t) + \tfrac{1}{2}\i \mu\ln t\right),
\\&
\quad
p=-1, n=-4.
\label{radial-sol-34}
\end{aligned}
\end{align}
\endallowdisplaybreaks}
\end{thm}

\begin{thm}\label{thm:TransInverPhas-inv-solns}
The radial Schr\"odinger equations \eqref{radial-nls} and \eqref{2D-radial-nls}
have the following group-invariant solutions 
with respect to time translations and inversions
combined with phase rotations \eqref{TransInverPhas-group}:
{\allowdisplaybreaks
\begin{align}
& \begin{aligned}
u = &
\tfrac{1}{6}kr^2 \Big(
M_{\kappa/2,3/2}(\tfrac{1}{2}r^2/(1+t^2))\int^{(1+t^2)/r^2}_{c_2} M_{\kappa/2,-3/2}(\tfrac{1}{2}\xi^{-1})\,d\xi
\\&\qquad\quad
- M_{\kappa/2,-3/2}(\tfrac{1}{2}r^2/(1+t^2))\int^{(1+t^2)/r^2}_{c_3} M_{\kappa/2,3/2}(\tfrac{1}{2}\xi^{-1})\,d\xi
\Big)
\\&\qquad
\times \exp\left(\i c_1 - \i\kappa\arctan(1/t) - \tfrac{1}{4}\i r^2t/(1+t^2)\right),
\\&
\quad
p=-1, n=-4.
\label{radial-sol-35}
\end{aligned}
\end{align}
\endallowdisplaybreaks}
\end{thm}

\subsection{Solutions from the full symmetry group}

From Theorem~\ref{thm:radial-pointsymms},
the full group of point symmetries admitted by
the radial Schr\"odinger equation \eqref{radial-nls}
and its alternative two-dimensional formulation \eqref{2D-radial-nls} 
can be applied to each of the solutions $u=f(t,r)$ listed in
Theorems~\ref{thm:Trans-inv-solns}, \ref{thm:TransPhas-inv-solns},
\ref{thm:ScalPhas-inv-solns}, and~\ref{thm:TransInverPhas-inv-solns}.
Phase rotations \eqref{radial-Phas} and scalings \eqref{radial-Scal}
change only the constants appearing in these solutions,
while time translations \eqref{radial-Trans} at most
shift $t$ by a new constant.
In contrast, inversions \eqref{radial-Inver} produce additional new solutions,
which are listed in the next theorem.

\begin{thm}\label{thm:conformal-inv-solns}
In the case of the pseudo-conformal power $p=4/n$,
the radial Schr\"odinger equations \eqref{radial-nls} and \eqref{2D-radial-nls}
have the following additional group-invariant solutions:
{\allowdisplaybreaks
\begin{align}
& \begin{aligned}
u = &
(n(n-4)/(4k))^{n/4} r^{-n/2} \exp\big(\i c_1- \i c_2 r^2/(4(1+c_2 t))\big) ,
\\&
\quad
n\neq 4, k/(n(n-4))>0,
\label{radial-trans-sol-0-apply-inver}
\end{aligned}\\
& \begin{aligned}
u = &
(-n^3/(2k(2n+4)))^{n/4} r^{-n/2} \big(1 + c_2 (r/(1+c_3 t))^{2-4/p}\big)^{-2/p}
\exp\big(\i c_1- \i c_3 r^2/(4(1+c_3 t))\big),
\\&
\quad
n^2-n+4=0,
\label{radial-trans-sol-16-apply-inver}
\end{aligned}\\
& \begin{aligned}
u = &
\big( \tfrac{1}{8}k r^2 + c_3 r^6(1+c_4t)^{-4} + c_2(1+c_4t)^2\big)
\exp\big(\i c_1 - \i c_4r^2/(1+c_4t)\big),
\\&
\quad
p=-1, n=-4,
\label{radial-sol-31-apply-inver}
\end{aligned}\\
& \begin{aligned}
u = &
\Big( r^3(1+c_4t)^{-1}
\big(c_2 J_{3}(\sqrt{\nu}r/(1+c_4t)) + c_3 Y_{3}(\sqrt{\nu}r/(1+c_4t))\big)
- (k/\nu)(1+c_4t)^2\Big)
\\&\qquad\times
\exp\big(\i c_1+ \i\nu t/(1+c_4t) - \i c_4r^2/(4(1+c_4t))\big),
\\&
\quad
p=-1,n=-4,\nu>0,
\label{radial-sol-30-apply-inver}
\end{aligned}\\
& \begin{aligned}
u = &
\Big( r^3(1+c_4t)^{-1}
\big(c_2 I_{3}(\sqrt{-\nu}r/(1+c_4t)) + c_3 K_{3}(\sqrt{-\nu}r/(1+c_4t))\big)
- (k/\nu)(1+c_4t)^2\Big)
\\&\qquad\times
\exp\big(\i c_1+ \i\nu t/(1+c_4t) - \i c_4r^2/(4(1+c_4t))\big),
\\&
\quad
p=-1,n=-4,\nu<0,
\label{radial-sol-36-apply-inver}
\end{aligned}\\
& \begin{aligned}
u = &
\tfrac{1}{3}\i kr^2 \Big(
M_{-\i \mu/2,-3/2}(\tfrac{1}{4}\i r^2/(t(1 + c_4t)))\int^{t(1+c_4t)/r^2}_{c_2} M_{-\i \mu/2,3/2}(\tfrac{1}{4}\i\xi^{-1})\,d\xi
\\&\qquad\quad
- M_{-\i \mu/2,3/2}(\tfrac{1}{4}\i r^2/(t(1 + c_4t)))\int^{t(1+c_4t)/r^2}_{c_3} M_{-\i \mu/2,-3/2}(\tfrac{1}{4}\i\xi^{-1})\,d\xi
\Big)
\\&\qquad
\times\exp\left(\i c_1 - \tfrac{1}{2}\i \mu\ln (c_4 + 1/t) -\tfrac{1}{8}\i r^2(1+2c_4t)/(t(1+c_4t)) \right),
\\&
\quad
p=-1, n=-4,
\label{radial-sol-34-apply-inver}
\end{aligned}\\
& \begin{aligned}
u =  &
\tfrac{1}{6}k r^2 \Big(
M_{\kappa/2,3/2}(\tfrac{1}{2}r^2/(t^2+(1+c_4t)^2))\int^{(t^2+(1+c_4t)^2)/r^2}_{c_2} M_{\kappa/2,-3/2}(\tfrac{1}{2}\xi^{-1})\,d\xi
\\&\qquad\quad
- M_{\kappa/2,-3/2}(\tfrac{1}{2}r^2/(t^2+(1+c_4t)^2))\int^{(t^2+(1+c_4t)^2)/r^2}_{c_3} M_{\kappa/2,3/2}(\tfrac{1}{2}\xi^{-1})\,d\xi
\Big)
\\&\qquad
\times \exp\left(\i c_1 - \i\kappa\arctan(c_4+1/t) - \tfrac{1}{4}\i r^2(c_4+t+{c_4}^2t)/(t^2+ (1+c_4t)^2)\right),
\\&
\quad
p=-1, n=-4.
\label{radial-sol-35-apply-inver}
\end{aligned}
\end{align}
\endallowdisplaybreaks}
\end{thm}

\subsection{Analytical behaviour of group-invariant solutions}

The radial solutions listed in
Theorems~\ref{thm:Trans-inv-solns} to~\ref{thm:conformal-inv-solns}
exhibit several types of interesting behaviour describing
(1) standing waves;
(2) static and dynamic monopoles;
(3) static ``bright solitons'';
(4) static and dynamic ``dark solitons''.

A function $u(t,r)$ describes a {\em radial monopole} if
it is smooth on $0<r<\infty$ such that
$|u|\rightarrow 0$ as $r\rightarrow\infty$
and $|u|\rightarrow \infty$ as $r\rightarrow 0$.
Solution \eqref{radial-trans-sol-0} for $p>0$
is a $n$-dimensional static monopole;
solution \eqref{radial-sol-27} for $c_2=0$
is a $16$-dimensional static monopole;
solutions \eqref{radial-trans-sol-0} for $n\neq 1,2,\ldots$,
\eqref{radial-trans-sol-16} for $2<n<3$,
and \eqref{radial-sol-29}
are planar static monopoles.
Solution \eqref{radial-trans-sol-0-apply-inver}
is a $n$-dimensional monopole with a dynamic phase;
solution \eqref{radial-trans-sol-16-apply-inver}
is a planar dynamic monopole.

A function $u(t,r)$ describes a {\em radial standing wave}
if it is smooth on $0\leq r<\infty$ such that
$u=U(r)\exp(\i\omega t)$ with $\omega\neq 0$ and $|U|$ bounded as $r\rightarrow\infty$.
Solution \eqref{radial-sol-30}
and solution \eqref{radial-sol-36} in the case $c_2=0$
are $n$-dimensional standing waves.

A function $u(t,r)$ describes a {\em ``bright radial soliton''}
or a {\em ``dark radial soliton''}
if it is smooth on $0\leq r<\infty$ such that
$|u|\rightarrow 0$ as $r\rightarrow\infty$
or $|u|\rightarrow A\neq 0$ as $r\rightarrow\infty$, respectively.
Solution \eqref{radial-trans-sol-1} for $n>2$ in the ``$+$'' case
is a $n$-dimensional static bright soliton;
solution \eqref{radial-sol-27} in the case $c_2=0$
is a $16$-dimensional static bright soliton.
Solution \eqref{radial-sol-36-apply-inver} in the case $c_2=0$
is a planar dynamic dark soliton exhibiting blow-up.

More details about these physically interesting radial solutions,
including their $L^2$ norms and conserved energies,
will be discussed in a separate paper \cite{Anc}.
The remaining radial solutions found in Theorems~\ref{thm:Trans-inv-solns} to~\ref{thm:conformal-inv-solns}
all have unphysical behaviour, in particular $|u|$ is unbounded
as $r\rightarrow\infty$.

\section{\large Concluding Remarks}
\label{concl}

In this paper,
all explicit group-invariant solutions given by elementary functions
have been derived
(cf Theorems~\ref{thm:Trans-inv-solns} to~\ref{thm:conformal-inv-solns})
for the class of semilinear radial Schr\"odinger equations \eqref{radial-nls}
with a power nonlinearity $p\neq 0$ in multi-dimensions $n\neq 1$.
Among these solutions $u(t,r)$,
a few describe $n$-dimensional radial standing waves, radial monopoles,
and static radial ``bright solitons'',
which have some physical interest.

Several solutions exist, surprisingly, only for non-integer values of $n$.
In such cases the radial Schr\"odinger equation \eqref{radial-nls}
is shown to have an alternative interpretation as
a planar (\ie/ $2$-dimensional) radial equation \eqref{2D-radial-nls}
containing an extra modulation term $m u_r/r$ that describes 
a point-source disturbance at the origin $r=0$, with $m=2-n$.
Some of these planar solutions are physically interesting
dynamic radial monopoles and dynamic radial ``dark solitons''.

However, no $n$-dimensional radial solutions are obtained
in the analytically important cases $p\geq 4/n$ 
relevant for blow-up behaviour when $n\geq 2$.
In particular, it is rigorously known \cite{Cav,Sul} that 
some radial solutions exhibit a finite time blow-up such that 
$|u(t,r)|\rightarrow \infty$ as $t\rightarrow T<\infty$
(with the energy and $L^2$ norm of $u(t,r)$ being finite). 

For the critical case $p=4/n$, 
a special class of radial blow-up solutions can be shown to have 
an exact group-invariant form \cite{Sul}
\begin{equation}\label{crit-blowup}
u(t,r) = (T-t)^{-n/2} U(\xi) \exp(\i(\omega+r^2/4)/(T-t)),
\quad
\xi = r/(T-t),
\quad
\omega\neq 0,
\end{equation}
where $U(\xi)$ satisfies the complex nonlinear 2nd order ODE 
\begin{equation}\label{crit-blowup-ode}
U'' + (n-1)\xi^{-1} U' +\omega U +k|U|^{4/n}U =0
\end{equation}
which is given by reduction of the radial Schr\"odinger equation \eqref{radial-nls}
under a pseudo-conformal symmetry group generated by 
\begin{equation}\label{crit-blowup-X}
T^2\X_{\trans}-T\X_{\scal}+\X_{\inver} +\omega \X_{\phas} . 
\end{equation}
Such pseudo-conformal blow-up solutions \eqref{crit-blowup} 
are related by a certain symmetry transformation to 
standing wave solutions \eqref{u:Trans+Phas}--\eqref{Trans+Phas-ODE}, 
which arise from the optimal translation group \eqref{TransPhas-group}
given by the generator \eqref{optimal-TransPhas}.
Specifically, if $u=f(t,r)$ has a standing-wave form 
with frequency $\nu\neq 0$ when $p=4/n$, 
then 
\begin{equation}
u=f(t/(T-t),r/(T-t)) (T-t)^{-2/p} \exp\big(\i r^2/(4(T-t))\big)
\end{equation}
(modulo a constant phase rotation) has a blow-up form with $\omega=T\nu\neq0$, 
where 
\begin{equation}
t\rightarrow t/(1-t/T),
\quad 
r\rightarrow r/(T-t),
\quad
u\rightarrow (T-t)^{2/p} \exp\big(-\i r^2/(4(T-t))\big)u
\end{equation}
is an inversion transformation combined with a scaling transformation
acting on $(t,r,u)$.
Hence the ODEs \eqref{Trans+Phas-ODE} and \eqref{crit-blowup-ode}
are equivalent up to at most a point transformation on $(\xi,U,\bar U)$. 
From the results 
in \propref{prop:U-ODE-symms}(ii) and \propref{prop:hiddensymms:Trans+Phas}, 
we conclude that, since $\omega\neq0$, 
the blow-up ODE \eqref{crit-blowup-ode} has too few point symmetries 
to allow it to be reduced to quadratures by means of first integrals
(except in the case $p\neq -1$, $n=4/p=-4$ when special solutions can be found
in terms of Bessel functions). 
The only obvious explicit solution, by inspection, is $U=(-\omega/k)^{n/4}$, 
but this solution has infinite energy and infinite $L^2$ norm. 

In the supercritical case $p>4/n$, 
numerical evidence \cite{Sul} suggests that a general class of blow-up solutions
for the radial Schr\"odinger equation \eqref{radial-nls}
asymptotically approach an exact similarity form 
\begin{equation}\label{supercrit-blowup}
u(t,r) = (T-t)^{-1/p} U(\xi) \exp(\i\omega\ln((T-t)/T)),
\quad
\xi = r/\sqrt{T-t} , 
\quad
\omega\neq 0
\end{equation}
where $U(\xi)$ satisfies a more complicated complex nonlinear 2nd order ODE 
\begin{equation}\label{supercrit-blowup-ode}
U'' + ((n-1)\xi^{-1} -\tfrac{1}{2}\i\xi) U' -(\omega +\i/p) U +k|U|^{p}U =0,
\end{equation}
which is given by reduction with respect to a scaling symmetry group 
generated by 
\begin{equation}\label{supercrit-blowup-X}
T\X_{\trans}-\tfrac{1}{2}\X_{\scal} -\omega \X_{\phas} . 
\end{equation}
Modulo a time translation symmetry transformation $t\rightarrow t+T$, 
these blow-up solutions \eqref{supercrit-blowup} are the same as 
the similarity solutions \eqref{u:Scal+Phas}--\eqref{Scal+Phas-ODE}, 
which arise from the optimal scaling group \eqref{ScalPhas-group}
given by the generator \eqref{optimal-ScalPhas}.
Since the ODEs \eqref{Scal+Phas-ODE} and \eqref{supercrit-blowup-ode}
are thereby equivalent up to at most a point transformation on $(\xi,U,\bar U)$,
the results in \propref{prop:U-ODE-symms}(i) and \propref{prop:hiddensymms:Scal+Phas} show that the blow-up ODE \eqref{supercrit-blowup-ode} 
has too few point symmetries 
to allow it to be reduced to quadratures by means of first integrals
(except, again, in the case $p\neq -1$, $n=4/p=-4$ when special solutions 
can be found in terms of Whittaker functions). 

Consequently, symmetry reduction methods are unable to yield 
any explicit $n$-dimensional radial blow-up solutions 
\eqref{crit-blowup} and \eqref{supercrit-blowup}. 
To look for such solutions,
we plan to apply the method of group-foliation reduction \cite{AncAliWol2},
which has been successfully used in previous work \cite{AncLiu,AncAliWol}
to obtain blow-up and dispersive radial solutions
to semilinear wave equations and semilinear heat conduction equations
with power nonlinearities in multi-dimensions.

\subsection*{Acknowledgements}
S.~Anco is supported by an NSERC research grant.
W.~Feng is indebted to the China Scholarship Council for financial support 
to work as a visiting scholar at Brock University 
and thanks the Department of Mathematics for support
during the period when this paper was written.
The referees are thanked for valuable comments 
which have improved this paper.


\begin{thebibliography}{99}

\bibitem{Sul}
C. Sulem and P.-L. Sulem,
{\it The Nonlinear Schr\"odinger Equation},
Applied Math. Sci. Volume 139
(Springer, New York) 1999.

\bibitem{Cav}
T. Cazenave,
{\it Semilinear Schr\"odinger Equations},
Courant Lecture Notes 10
(American Mathematical Society, Providence) 2003.

\bibitem{PolZai}
A. D. Polyanin and V. F. Zaitsev,
{\it Handbook of Nonlinear Partial Differential Equations (2nd edition)},
CRC (Chapman and Hall) 2011.

\bibitem{Olv}
P.J. Olver,
{\it Applications of Lie Groups to Differential Equations},
(Springer, New York) 1986.

\bibitem{BluAnc}
G. Bluman and S.C. Anco,
{\it Symmetry and Integration Methods for Differential Equations},
Applied Math. Sci. Volume 154
(Springer, New York) 2002.

\bibitem{NikPop}
A.G. Nikitin and R.O. Popovych,
Ukr. Math. J. 53 (2001), no. 8, 1255--1265.

\bibitem{FusSer}
W.I. Fushchich and N.I. Serov,
J. Phys. A: Math. Gen. 20 (1987) L929--L933.

\bibitem{GagWin}
L. Gagnon and P. Winternitz,
J. Phys. A 21 (1988), 1493--1511;
ibid.
J. Phys. A 22 (1989), 469--497;
ibid.
J. Phys. A 22 (1989), 499--509.

\bibitem{GagWin2}
L. Gagnon and P. Winternitz,
Phys. Rev. A 39 (1989), 296--306.

\bibitem{PopKunEsh}
R.O. Popovych, M. Kunzinger, H. Eshraghi,
Acta Appl. Math. 109 (2010), 315--359.

\bibitem{WinPat}
J. Patera and P. Winternitz,
J. Math. Phys. 18 (1977) 1449--1455.

\bibitem{AncFen}
S.C. Anco and W. Feng,
In preparation. 

\bibitem{Anc}
S.C. Anco,
In preparation.

\bibitem{AncAliWol2}
S.C. Anco, S. Ali, T. Wolf,
SIGMA 7 (2011) 066 (10 pages).

\bibitem{AncLiu}
S.C. Anco and S. Liu,
J. Math. Anal. Appl. 297 (2004), 317--342.

\bibitem{AncAliWol}
S.C. Anco, S. Ali, T. Wolf,
J. Math. Anal. Appl. 379 (2011), 748--763.


\end{thebibliography}
\end{document}